\documentclass[prl,aps,superscriptaddress,
floatfix,tightenlines,twocolumn,pdflatex,longbibliography]{revtex4-1}
\usepackage{amsmath}
\usepackage{amssymb}
\usepackage{amsfonts}
\usepackage{epsfig}
\usepackage{subfigure}
\usepackage[dvips]{color}
\usepackage[colorlinks,bookmarks=false,citecolor=blue,linkcolor=blue,urlcolor=blue]{hyperref}
\usepackage{bm}
\usepackage{amsthm}
\usepackage{graphicx}
\usepackage{multirow}
\usepackage{mathrsfs}
\usepackage{times}
\theoremstyle{plain}

\theoremstyle{definition}

\theoremstyle{remark}

\begin{document}

\title{Characterization of 
 topological  states via dual multipartite entanglement}
\author{Yu-Ran Zhang}
\affiliation{Beijing Computational Science Research Center, Beijing 100094, China}
\affiliation{Theoretical Quantum Physics Laboratory, RIKEN, Saitama 351-0198, Japan}
\author{Yu Zeng}
\affiliation{Institute of Physics, Chinese Academy of Sciences, Beijing 100190, China}
\author{Heng Fan}
\affiliation{Institute of Physics, Chinese Academy of Sciences, Beijing 100190, China}
\affiliation{CAS Central of Excellence for Topological Quantum Computation, University of Chinese Academy of Sciences, Beijing 100190, China}
\author{J. Q. You}
\email{jqyou@zju.edu.cn}
\affiliation{Beijing Computational Science Research Center, Beijing 100094, China}
\affiliation{Department of Physics, Zhejiang University, Hangzhou 310027, China}
\author{Franco Nori}
\email{fnori@riken.jp}
\affiliation{Theoretical Quantum Physics Laboratory, RIKEN, Saitama 351-0198, Japan}
\affiliation{Physics Department, University of Michigan, Ann Arbor, Michigan 48109-1040, USA}


\begin{abstract}
We demonstrate that multipartite entanglement is able to characterize one-dimensional
 symmetry-protected topological order, which is witnessed by the scaling behavior
of the quantum Fisher information of the ground state with respect to the spin operators defined in the dual lattice.
We investigate an extended Kitaev chain
with a $\mathbf{Z}$ symmetry identified equivalently by winding
numbers and paired Majorana zero modes at each end.
The topological phases with high winding numbers are detected by the scaling coefficient of the quantum Fisher information density
 with respect to generators in different dual lattices.
Containing richer properties and more complex structures than bipartite entanglement, the dual
multipartite entanglement of the topological state has promising applications in
robust quantum computation and  quantum metrology, and can be generalized to 
identify topological order in the Kitaev honeycomb model.
\end{abstract}

\maketitle
\emph{Introduction.---}%
In recent years, quantum topological phases \cite{Zeng2015} in extended systems have become
of great significance in modern physics due to its promise for both topological quantum
computation \cite{Mong2014,Kraus2013,Sau2010,Alicea2011,Alicea2012,Kitaev2003} and
condensed matter physics \cite{Xia2009,Albrecht2016}. Topological
phase transitions, beyond the Landau symmetry-breaking theory, are described by the change of 
its topological order or symmetry-protected topological (SPT) order \cite{Zeng2015}.
Topological order \cite{Wen1990}, e.g. quantum Hall states or spin liquids  \cite{Wen2004},
cannot be described by local order parameters
\cite{Amico2008,Eisert2010} but can be characterized by the long-range entanglement encoded in
the states of the systems, such as the topological entanglement entropy \cite{Kitaev2006,Levin2006}
and entanglement spectrum \cite{Li2008}. Further enriched by symmetries,
SPT phases, corresponding to short-range entangled phases with symmetry-protected edge modes  \cite{Gu2009,Chen2011,Scaffidi2017,Bliokh2015},
are theoretically proposed and experimentally discovered in topological insulators and superconductors
\cite{Bliokh2015,Konig2007,Fu2008,Hsieh2008,Hasan2010,Qi2011,Sato2017}.
These characteristics make topological states robust against local noise, which has emerged as one of the
most exciting approaches to realizing topologically protected quantum information processing and
fault-tolerant quantum computing \cite{Nayak2008}. The simplest realization would be the Majorana
zero modes (MZMs) at the edges of low-dimensional systems
\cite{Sarma2015,Elliott2015,You2010,You2014,Akzyanov2015,Akzyanov2016,Zhang2016}, e.g., extended Kitaev models
\cite{Vodola2014,Vodola2015,Alecce2017,Lepori2017}, which have recently been
observed in various experimental platforms including nanowire devices \cite{Mourik2012,Nadj-Perge2014}
and quantum spin liquids \cite{Do2017}.

In addition to the fruitful results from bipartite entanglement \cite{Kitaev2006,Levin2006,Li2008}, multipartite
entanglement \cite{Pezze2009,Hyllus2012,Toth2012,Strobel2014} (witnessed by the quantum Fisher information (QFI)
\cite{BRAUNSTEIN1994,Giovannetti2006,Giovannetti2011} with respect to nonlocal operators \cite{Pezze2016})
displays much richer properties of complex structures of topological states and deserves
further investigation. The QFI quantifies useful multipartite entanglement for quantum metrology,
which is confirmed by quantum parameter estimation with sub-shot-noise sensitivity
\cite{Strobel2014,BRAUNSTEIN1994,Giovannetti2006,Giovannetti2011,Gross2010,Lucke2011}.  Recently, it was shown that the scaling behavior of the QFI with
respect to spin operators in the original lattice is sensitive for detecting the topologically nontrivial
phases with low winding numbers $\nu=\pm1,\pm\frac{1}{2}$ \cite{Pezze2017}.
However, we find that topological phases with higher winding numbers cannot be characterized by
 the QFI with respect to these operators.

In this Letter, we provide a general method to characterize 1D SPT
order with higher winding numbers by multipartite entanglement defined in the dual lattice.
We focus on an extended Kitaev fermion chain
with $p$-wave superconductivity and a chiral symmetry belonging to the $\textbf{Z}$-type
{BDI} class \cite{Chiu2016,Li2016, Tewari2012} identified equivalently by high winding numbers  and
boundary MZMs from the Bogoliubov-de Gennes (BdG) Hamiltonian. Dual multipartite
entanglement is signaled by the scaling behavior of the QFI density of the ground state with
respect to spin operators by the duality transformation \cite{FRADKIN1978,Feng2007,Qin2017,Smacchia2011}.
By exploiting the duality of the model, we find that the QFI density in dual lattices, written in terms of string correlation
functions (SCFs) \cite{Smacchia2011,Venuti2005,Cui2013}, has a linear
scaling behavior versus system size in SPT phases and detects 1D quantum SPT  phase transitions. Therefore, together with \cite{Pezze2017},
dual multipartite entanglement 
can be used to identify SPT order.
We also extend our investigation to
the Kitaev honeycomb model \cite{Kitaev2006a}, indicating that our results can be generalized to
2D systems with topological order. Our work reveals the possibility of promising
applications of topologically protected multipartite entanglement in robust quantum computation
and quantum metrology.

\emph{Winding numbers, Majorana zero modes, and topological phase transitions.---}%
 We study the extended Kitaev fermion chain with extensive pairing and hopping terms \cite{Alecce2017},
\begin{align}
H=&\sum_{n=1}^{N_f}\sum_{j=1}^{L}\left(\frac{{J}_n^+}{2}c^\dag_jc_{j+n}+\frac{{J}_n^-}{2}c^\dag_jc_{j+n}^\dag+\textrm{h.c.}\right)\nonumber\\
&-\sum_{j=1}^L\mu\left(c^\dag_jc_j-\frac{1}{2}\right),\label{1dK}
\end{align}
where  $L$ (assumed even)  is the total number of sites, $N_f$ denotes the farthest pairing and hopping distance,
and the antiperiodic conditions $c_{j+L}=-c_j$ are assumed.
The hopping and pairing parameters are all chosen as real to make the Hamiltonian preserve
time-reversal symmetry and belong to the {BDI} class ($\mathbf{Z}$ type) characterized by
a winding number \cite{Chiu2016,Li2016}. Through the Jordan-Wigner transformation
$c_1=-\sigma_1^+$, $c_j=-\sigma_j^+\prod_{i=1}^{j-1}\sigma^z_i$,
this spinless fermion model corresponds to the extended Ising model
\cite{Suzuki1971,Niu2012,Zhang2015,Zhang2017a,Zhang2017,SM}
\begin{equation}
H=\sum_{n=1}^{N_f}\sum_{j=1}^L\left(\!\frac{{J}_n^x}{2}\sigma_j^x\sigma_{j+n}^x\!+\frac{{J}_n^y}{2}\sigma_j^y\sigma_{j+n}^y\!\right)\!\!\prod_{l=j+1}^{j+n-1}\!\!\!\sigma_l^z+\!\sum_{j=1}^L\frac{\mu}{2}\sigma_j^z,
\end{equation}
with ${J}_{x,y}\equiv ({J}_n^+\pm {J}_n^-)/2$.
In the thermodynamic limit $L\gg N_f\geq1$, the Hamiltonian (\ref{1dK}) can be diagonalized by a Fourier-Bogoliubov
transformation with energy spectrum $\epsilon_q=\pm\frac{1}{2}\sqrt{y(q)^2+z(q)^2}$, where
$y(q)=\sum_{n=1}^{N_f}{J}_n^-\sin(nq)$, $z(q)=\sum_{n=1}^{N_f}{J}_n^+\cos(nq)-\mu$, with
$q$ the wavevector \cite{SM}.

\begin{figure}[t]
\centering
\includegraphics[width=0.46\textwidth]{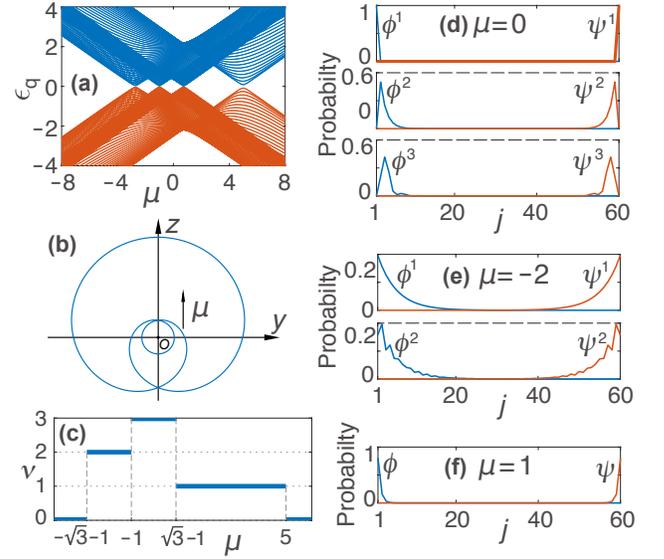}\\
\caption{(color online) (a) Energy spectrum for $L=200$ sites, (b) trajectory of the winding
vector $\bm{r}(q)=(0,y(q),z(q))$, (c-e)  probability distributions [blue (red) curve is for
 left (right) modes] of MZMs for $L=60$ sites given different
values of a chemical potential $\mu$ for the extended Kitaev chain with $N_f=3$ and
${J}_1^\pm=1$, ${J}_{2}^\pm=2$, ${J}_{3}^\pm=2$.
(c) The phase diagram characterized by the winding number.
(d) For $\mu=0$,
the winding number $\nu=3$ and we have three pairs of non-degenerate MZMs
exponentially localized at the domain wall. (e) For $\mu=-2$,
$\nu=2$ and there are two pairs of MZMs.  (f) When $\mu=1$,
$\nu=1$ which leads to one pair of MZMs.
}\label{fig:1}
\end{figure}

 As a $\textbf{Z}$ topological invariant
\cite{Chiu2016,Zhang2015}, the winding number of the closed loop with the vector
$\bm{r}(q)=(0,y(q),z(q))$ in the auxiliary $y$-$z$ plane around the origin can be written
as $\nu=({1}/{2\pi})\oint(ydz-zdy)/|\bm{r}|^2$. Substituting $\zeta(q)\equiv\exp(iq)$, for
$y(q)\equiv Y(\zeta)$ and $z(q)\equiv Z(\zeta)$, we can define a complex characteristic
function $g(\zeta)\equiv Z(\zeta)+iY(\zeta)$ and obtain the winding number by
calculating the logarithmic residue of $g(\zeta)$ in accordance with the Cauchy's
argument principle \cite{Ahlfors1953}
$\nu=({1}/{2\pi i})\oint_{{}_{|\zeta|=1}}\!\!\!\!\!\!\!\!\!\!d\zeta\:\:{g'(\zeta)}/{g(\zeta)}=\mathcal{N}-\mathcal{P}$,
where in the complex region $|\zeta|<1$,  $\mathcal{N}$ is the number of zeros 
and $\mathcal{P}$ is the number of poles. 
Moreover, topological
phase transitions are characterized by the change of winding numbers at the critical points
that can be calculated by solving $g(\zeta)=0$ on the contour $|\zeta|=1$ \cite{SM}.
Similarly, the topologically nontrivial phases for the model (\ref{1dK}) are also identified
by the existence of paired boundary MZMs of which the properties are obtained from the
solution of the BdG Hamiltonian with open boundary conditions
\cite{Fendley2012,Niu2012,Viyuela2016}. This can also be transformed to calculating
zeros of $g(\zeta)$ in $|\zeta|<1$, such that the number of MZMs at each end of the open
chain, defined as $\mathcal{M}_{0}$, equals the absolute value of the winding number:
$\mathcal{M}_{0}=|\mathcal{N}-\mathcal{P}|=|\nu|$. 
Therefore, these two approaches [($i$) by winding numbers from the geometric topology
in the 2D auxiliary space, and ($ii$) by MZMs from BdG equations to characterize
topological phases] in the extended Kitaev chain in Eq.~(\ref{1dK}) are 
equivalent \cite{SM} (see, e.g.,  Fig.~\ref{fig:1}).

\emph{Multipartite entanglement and QFI density.---}%
Multipartite entanglement \cite{Hyllus2012,Toth2012} plays a key role in quantum physics and quantum metrology, and
moreover, it is central to understanding quantum many-body systems. QFI, similar as
quantum spin squeezing \cite{KITAGAWA1993,Ma2011},
is a significant quantity in both large-scale multipartite entanglement detection and
high-precision quantum metrology \cite{Strobel2014,BRAUNSTEIN1994,Giovannetti2006,Giovannetti2011,Gross2010,Lucke2011}. Given a generator
$\mathcal{O}$ and a mixed state $\rho=\sum_{i}p_i|i\rangle\langle i|$, with
$\langle i|j\rangle=\delta_{ij}$, the QFI of a state
$\rho(t)=\exp({-it\mathcal{O}})\rho \exp({it\mathcal{O}})$ with respect to a parameter $t$
is  \cite{BRAUNSTEIN1994}
$F_Q[\mathcal{O},\rho]=\sum_{p_i+p_j\neq0}\frac{2(p_i-p_j)^2}{p_i+p_j}|\langle i|\mathcal{O}| j\rangle|^2$.
For a pure state $|\psi\rangle$, the QFI can be simplified as
$F_Q[\mathcal{O},|\psi\rangle]=4(\Delta_\psi\mathcal{O})^2$, where the variance of the generator is
$(\Delta_\psi\mathcal{O})^2\equiv\langle\mathcal{O}^2\rangle_{\psi}-\langle\mathcal{O}\rangle_{\psi}^2$.
The QFI relates to dynamic susceptibilities \cite{Hauke2016} that are routinely measured
in laboratory experiments. Furthermore, the scaling of the QFI
with respect to nonlocal operators \cite{Pezze2016} would be sensitive to topological quantum phase
transitions \cite{Pezze2017}. For critical systems with $L$ sites, we consider a QFI density with form
$f_Q=F_Q/L$, and the violation of the inequality $f_Q\leq\kappa$ signals $(\kappa+1)$-partite
entanglement ($1\leq\kappa\leq L-1$) \cite{Pezze2009}.

To detect a topological phase of an extended Kitaev chain with a winding number $\nu=\pm1$,
the generators in terms of spin operators in the $x,y$ directions through the
Jordan-Wigner transformation are chosen as \cite{Pezze2017}
$\mathcal{O}_{\nu=\pm1}=\sum_{j=1}^L\sigma_j^{x,y}/2$,
and staggered operators as
${\mathcal{O}}^{(\textrm{st})}_{\nu=\pm1}=\sum_{j=1}^L(-)^j\sigma_j^{x,y}/2$.
 Then, the QFI density for the ground state $|\mathcal{G}\rangle$ becomes
${f}_Q[{\mathcal{O}}_{\nu=\pm1},|\mathcal{G}\rangle]=1+\sum_{r=1}^{L-1}C_{\nu=\pm1}(r)$ and
${f}_Q[{\mathcal{O}}^{(\textrm{st})}_{\nu=\pm1},|\mathcal{G}\rangle]=1+\sum_{r=1}^{L-1}(-)^rC_{\nu=\pm1}(r)$,
where the spin-spin correlation functions are
$C_{\nu=\pm1}(r)\equiv\langle\sigma_j^{x,y}\sigma_{j+r}^{x,y}\rangle_{\mathcal{G}}$,
with $\langle\cdots\rangle_{\mathcal{G}}$ the average of the ground state $|\mathcal{G}\rangle$.
A topological phase with a low winding number can be characterized by power-law diverging finite-size
scaling of the QFI density, $f_Q\propto L$, as discussed in \cite{Pezze2017}.

\begin{figure}[t]
\centering
\includegraphics[width=0.44\textwidth]{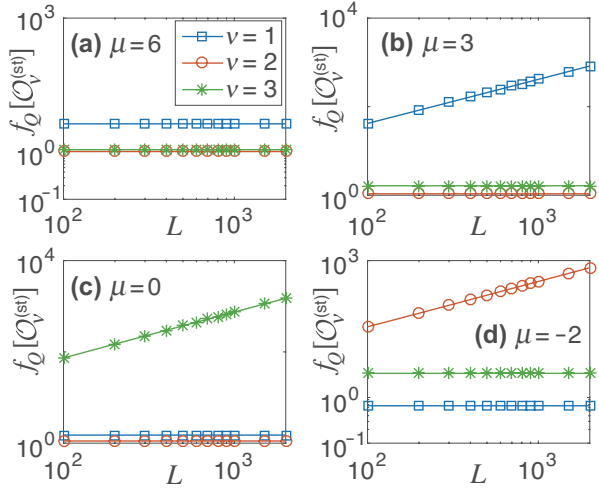}\\
\caption{(color online) Dual QFI density $f_Q[\mathcal{O}^{(\textrm{st})}_\nu,|\mathcal{G}\rangle]$ of the ground state $|\mathcal{G}\rangle$ versus $L$ for the extended Kitaev chain with $N_f=3$ and nonzero parameters (${J}_1^\pm=1$, ${J}_{2}^\pm=2$, ${J}_{3}^\pm=2$) in different topological phases.
(a) For $\mu=6$, the winding
number $\nu=0$. (b) For $\mu=3$, $\nu=1$, and the fitting nontrivial scaling topological index
$\lambda_{1}^{(\textrm{st})}=0.9965$. (c) For $\mu=0$, $\nu=3$, and $\lambda^{(\textrm{st})}_{3}=1.0047$. (d) For
$\mu=-2$, $\nu=2$, and $\lambda^{(\textrm{st})}_{2}=0.9957$.
}\label{fig:2}
\end{figure}

\begin{figure}[t]
\centering
\includegraphics[width=0.45\textwidth]{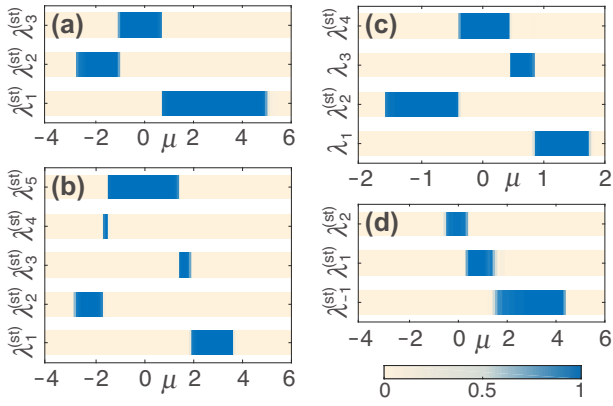}\\
\caption{(color online) Scaling topological index $\lambda_\nu$ and $\lambda_\nu^{(\textrm{st})}$  
of the dual QFI density $f_Q[\mathcal{O}_\nu,|\mathcal{G}\rangle]$ and $f_Q[\mathcal{O}_\nu^{(\textrm{st})},|\mathcal{G}\rangle]$, respectively, versus system size $L$ up to 1200.
The extended Kitaev chain in Eq.~(\ref{1dK}) has the following nonzero parameters: (a)
${J}_1^\pm=1$, ${J}_{2}^\pm=2$, ${J}_{3}^\pm=2$ ($N_f=3$);
(b) ${J}_1^\pm=0.1$, ${J}_{2}^\pm=0.21$, ${J}_{3}^\pm=0.44$, ${J}_4^\pm=0.9$, ${J}_{5}^\pm=2$ ($N_f=5$);
(c) ${J}_1^\pm=0.1$, ${J}_{2}^\pm=0.21$, ${J}_{3}^\pm=-0.74$, ${J}_4^\pm=0.9$ ($N_f=4$);
and (d) ${J}_{2}^\pm=2.4$, $J_3^\pm=\pm2$ ($N_f=3$).}\label{fig:3}
\end{figure}

\emph{Characterization of topological phases by multipartite entanglement in the dual lattice.---}%
 Duality in physics provides different but equivalent mathematical descriptions of a system
and provides an overall understanding of the same physical phenomena from different angles \cite{Qin2017}.
For example, an Ising chain with an external field $h$ has a self-dual symmetry, mapping between
the ordered and disordered phases, expressed as $H_{\textrm{Ising}}=\sum_j(\sigma_j^x\sigma_{j+1}^x+h\sigma_j^z)=h\sum_j(s_j^xs_{j+1}^x+h^{-1}s_j^z)$,
with the duality transformation $s_j^x=\prod_{k\leq j}\sigma_k^z$,
$s_j^z=\sigma_j^x\sigma_{j+1}^x$, and $s_j^y=-is_j^zs_j^x$ \cite{Suzuki2012}. Here both $\sigma$ and $s$ satisfy the
same algebra. Furthermore, the \emph{nonlocal} SCF \cite{Smacchia2011,Venuti2005,Cui2013}, characterizing
SPT order by the $\mathbf{Z}_2\times\mathbf{Z}_2$ symmetry in the cluster Ising model \cite{Zeng2015} with Hamiltonian
$H_{\textrm{cluster}}=\sum_j(\sigma_{j-1}^x\sigma_j^z\sigma_{j+1}^x+h\sigma_j^z)$,  can
be written as a \emph{local} correlator $(-)^r\langle s_j^ys_{j+r}^y\rangle_{\mathcal{G}}$ in the dual
lattice of the Ising model \cite{Venuti2005,Cui2013}.
Through the Jordan-Wigner transformation (also regarded as a duality transformation using a
bond-algebraic approach \cite{Cobanera2011}), the self-duality properties of a spin-$\frac{1}{2}$
model can help to study topological phases and multipartite entanglement in the extended
Kitaev chain (\ref{1dK}).

To detect a SPT phase with a positive integer winding number $\nu=n\geq2$, we
consider the duality transformation of an extended Ising model
$H=\sum_j(\sigma_j^{x}\sigma_{j+n-1}^{x}\prod_{l=1}^{n-2}\sigma_{j+l}^z+h\sigma_j^z)=h\sum_j(\hat{s}_j^{x}\hat{s}_{j+n-1}^{x}\prod_{l=1}^{n-2}\hat{s}_{j+l}^z+h^{-1}\hat{s}_j^z)$,
corresponding to an extended Kitaev chain with $\nu=n-1$.
We can define the dual operator $\tau^{(\nu=n)}_j\equiv \hat{s}^y_j=-i\hat{s}_j^z\hat{s}_j^x$. For a negative winding number, $\nu=-n$, we consider another extended Ising model by transforming $x\rightarrow y$:
$H=\sum_j(\sigma_j^{y}\sigma_{j+n-1}^{y}\prod_{l=1}^{n-2}\sigma_{j+l}^z+h\sigma_j^z)=h\sum_j(\tilde{s}_j^{y}\tilde{s}_{j+n-1}^{y}\prod_{l=1}^{n-2}\tilde{s}_{j+l}^z+h^{-1}\tilde{s}_j^z)$
and obtain the dual spin operator $\tau^{(\nu=-n)}_j\equiv \tilde{s}^x_j=i\tilde{s}_j^z\tilde{s}_j^y$.
The expressions of the dual spin operators $\tau_j^{(\nu)}$ differ
according to the parity of the winding number $\nu$ \cite{Fidkowski2011}.
Explicitly with $p\geq1$, we have \cite{SM} for even winding numbers,
\begin{align}
\tau_{j}^{(\nu\pm2p)}=-\left(\prod_{k=1}^{j-1}\sigma_k^z\right)\left(\prod_{l=1}^{p}\sigma_{j+2l-2}^{y,x}\sigma_{j+2l-1}^{x,y}\right),
\end{align}
and for odd winding numbers,
\begin{align}
\tau_{j}^{(\nu=\pm(2p+1))}=\sigma_j^{x,y}\left(\prod_{l=1}^{p}\sigma_{j+2l-1}^{y,x}\sigma_{j+2l}^{x,y}\right).
\end{align}
The SCF \cite{Venuti2005,Cui2013} equals the spin correlation
function from site $j$ to $(j+r)$ in the dual lattice:
\begin{eqnarray}
C_{\nu}(r)\equiv\langle\tau_j^{(\nu)}\tau_{j+r}^{(\nu)}\rangle_{\mathcal{G}}
=\left\langle\prod_{l=j}^{j+r-1}\!\!\!\Big(\sigma_l^{\alpha}\sigma_{l+|\nu|}^{\alpha}\!\!\prod_{k=1}^{|\nu|-1}\!\!\sigma_{l+k}^z\Big)\!\right\rangle_{\!\!\!\mathcal{G}}\!,
\end{eqnarray}
where $\alpha=x$ (or $y$) for a positive (or negative) $\nu$. It is clearer to write the SCF, in
terms of Majorana fermion operators 
$a_j=c_j^\dag+c_j$ and 
$b_j=i(c^\dag_j-c_j)$, as 
\begin{align}
C_{\nu}(r)=\left\langle\prod_{l=j}^{j+r}(- ib_{l}a_{l+\nu})\!\right\rangle_{\!\!\!\mathcal{G}}\!=\left\langle\prod_{l=j}^{j+r}(1-2d_{l,\nu}^\dag d_{l,\nu})\!\right\rangle_{\!\!\!\mathcal{G}},
\end{align}
where we define $d_{l,\nu}=(b_l+ia_{l+\nu})/{2}$ and $d_{l,\nu}^\dag=(b_l-ia_{l+\nu})/{2}$
as Dirac fermion operators \cite{Viyuela2016}. Therefore, the SCF can also be regarded as the
ground-state average of $\mathbf{Z}$ type Majorana parity \cite{Fidkowski2011}, and in
particular, $\Delta_{\nu}\equiv\lim_{r\rightarrow\infty}C_\nu(r)$ and $\Delta_{\nu}^{\textrm{(st)}}\equiv\lim_{r\rightarrow\infty}(-)^rC_\nu(r)$ are the
string order parameters \cite{Cui2013,Smacchia2011}, capturing hidden SPT order.

The generators of the dual QFI density are defined in the dual lattice as 
$\mathcal{O}_\nu=\sum_{j=1}^{M}\tau_{j}^{(\nu)}$, and
$\mathcal{O}_\nu^{(\textrm{st})}=\sum_{j=1}^{M}(-)^j\tau_{j}^{(\nu)}$,
with $M\equiv L-|\nu|+1$,  where the choice of dual generators
depends on the sign of the direct interaction between the Majorana fermions at chain ends \cite{Kitaev2001,SM}. The  operator $\mathcal{O}_\nu$ applies for the positive interaction, and the staggered operator $\mathcal{O}_\nu^{(\textrm{st})}$ is for the negative one.
Then, we obtain the dual QFI density of the ground state for $L\gg N_f\geq1$ as 
$f_Q[\mathcal{O}_\nu,|\mathcal{G}\rangle]\simeq1+\sum_{r=1}^{M-1}\!C_{\nu}(r)$, and $f_Q[\mathcal{O}^{(\textrm{st})}_\nu,|\mathcal{G}\rangle]\simeq1+\sum_{r=1}^{M-1}(-)^rC_{\nu}(r)$,
where we have used $(\tau_j^{(\nu)})^2=\mathbb{I}$, with $\mathbb{I}$ the identity.
Using Wick's theorem, the dual QFI density can be expressed in terms of fermion correlators and
may be measured in many-body systems using experimentally mature techniques,
such as Bragg spectroscopy \cite{Stoferle2004,Ernst2010} or neutron scattering
\cite{Shirane2002}.

The SCF has a similar scaling behavior in the topologically nontrivial phase with a higher winding number as
the spin correlator used in \cite{Pezze2017} (see, e.g., \cite{SM}). Thus, we find that the dual QFI density as a function of $L$ also
follows an asymptotic power law scaling in the thermodynamic limit as
${f}_{Q}[\mathcal{O}_\nu, |\mathcal{G}\rangle]=1+\gamma_\nu L^{\lambda_\nu}$, and
${f}_{Q}[\mathcal{O}^{(\textrm{st})}_\nu, |\mathcal{G}\rangle]=1+\gamma_\nu^{(\textrm{st})} L^{\lambda_\nu^{(\textrm{st})}}$,
where the scaling coefficients $\gamma$ and $\lambda$  depend
on the choice of the dual generators and the parameters of the Hamiltonian (\ref{1dK}). For a topological phase with a definite
winding number $\nu$, we could find that $\lambda_\nu$ or $\lambda_\nu^{(\textrm{st})}$
is equal to 1 ($F_Q\propto L^2$), and the scaling coefficients $\lambda_{{\omega}}$ and $\lambda^{(\textrm{st})}_{{\omega}}$ for other integer winding numbers, $\omega\neq\nu$, are approximately
zero (see, e.g., Fig.~\ref{fig:2}). Thus, the scaling topological index
$\lambda_\nu$ or $\lambda_\nu^{(\textrm{st})}$, relating directly to the SCF, characterizes the features of the
topological phase with a winding number $\nu$ of the extended Kitaev model. In Fig.~\ref{fig:3}, we consider four different
types of extended Kitaev chain models and plot the fitting scaling coefficients $\lambda_\nu$ or $\lambda_\nu^{(\textrm{st})}$
of the QFI density versus system size $L$ up to $1200$, and also versus the chemical potential
$\mu$, which clearly show the topological phase diagrams.
Therefore, we conclude that by choosing the generators in different dual lattices, the
scaling behavior of the QFI density, a witness of multipartite entanglement, can detect 1D SPT phase transitions.
In the topologically nontrivial phase with integer winding number,
the quadratic growth of the QFI  can also be broadly applicable to practical quantum metrology
\cite{Giovannetti2006,Giovannetti2011,BRAUNSTEIN1994,Strobel2014,Gross2010,Lucke2011}.
The scaling coefficients of the QFI density in phases with half-integer
winding numbers or on the critical boundary between two topological phases would be
complicated \cite{Pezze2017} and deserve further investigations, of which more simulations
and discussions are given in \cite{SM}.

\begin{figure}[t]
\centering
\includegraphics[width=0.45\textwidth]{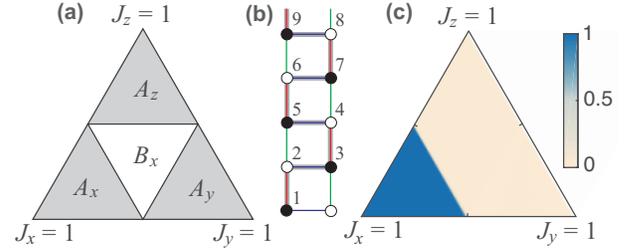}\\
\caption{(color online) (a) The phase diagram of the Kitaev honeycomb model on the
$J_x+J_y+J_z=1$ plane.  In the region 
$J_x\leq J_y+J_z$, $J_y\leq J_z+J_x$, and $J_z\leq J_x+J_y$, there is a gapless phase $B$ with non-Abelian
excitations, and in other regions, there are three gapped phases $A_{x,y,z}$ with Abelian anyon
excitations. (b) A single-chain representation
of the two-leg spin ladder of the Kitaev model. (c) The scaling topological index $\lambda_x^{(\textrm{st})}$ of
the dual QFI density $f_Q[{\mathcal{O}}_x^{(\textrm{st})},|\mathcal{G}\rangle]$
for different values of $J_{x,y,z}$ versus system size $2L$
up to 400.
}\label{fig:4}
\end{figure}

\emph{Dual multipartite entanglement in the Kitaev honeycomb model.---}
The Kitaev honeycomb model \cite{Kitaev2006a}, on a hexagonal lattice
with topological order at zero temperature, has been widely investigated using a variety of quantum-information
methods \cite{Yang2008,Abasto2009,Shi2009,Chen2016}.
The Hamiltonian is
$H_{\textrm{hc}}=-\sum_{\alpha=x,y,z}\!J_\alpha\!\sum_{\langle ij\rangle_\alpha}\sigma_i^\alpha\sigma_j^\alpha$,
where $\langle ij\rangle_\alpha$ denotes the nearest neighbor bonds in the $\alpha$-direction.
We consider positive bonds, $J_{x,y,z}>0$, and focus on the
$J_x+J_y+J_z=1$ parametric plane. The phase diagram is shown in Fig.~\ref{fig:4}(a).

Here, through the two-leg spin ladder \cite{Feng2007} of the Kitaev honeycomb model, we find that
the quantum phase  with hidden topological order can also be characterized by  dual multipartite entanglement.
As shown in Fig.~\ref{fig:4}(b), we relabel all the sites along a special path and rewrite the Hamiltonian with 
third-nearest-neighbor couplings \cite{Feng2007}:
$H_{\textrm{2l}}=-\sum_{j=1}^L(J_x\sigma_{2j-1}^x\sigma_{2j}^x+J_y\sigma_{2j}^y\sigma_{2j+3}^y+J_z\sigma_{2j}^z\sigma_{2j+1}^z)$.
With the duality transformation
$\check{s}_j^x=\prod_{k=1}^{j}\sigma^x_k$, $\check{s}_j^z=\sigma^z_{j}\sigma^z_{j+1}$, and $\check{s}_j^y=-i\check{s}_j^z\check{s}_j^x$,
we obtain an anisotropic $XY$ spin chain with a transverse field in the dual space
\begin{align}
H_{\textrm{2l}}=-\sum_{j=1}^L(J_x\check{s}^x_{2j}\check{s}^x_{2j+2}\!+J_y{W}_{j}\check{s}^y_{2j}\check{s}^y_{2j+2}\!+J_z\check{s}^z_{2j}),
\end{align}
where ${W}_{j}\equiv\check{s}^x_{2j-1}\check{s}^z_{2j+1}\check{s}^x_{2j+3}$ is the plaquette operator in the dual lattice
(a good quantum number \cite{Feng2007})  and has ${W}_{j}=-1$ ($\pi$-flux phase \cite{Lieb1994}) for
the ground state.  Then, with respect to the dual generator
${\mathcal{O}}_x^{(\textrm{st})}=\sum_{j=1}^{L}(-)^j\check{s}^x_{2j}$,
the QFI density is $f_Q[{\mathcal{O}}_x^{(\textrm{st})},|\mathcal{G}\rangle]\equiv1+\sum_{r=1}^{L-1}(-)^r{C}_x(r)\simeq 1+\gamma_x^{(\textrm{st})} L^{\lambda_x^{(\textrm{st})}}$,
where the staggered SCF is  $(-)^r{C}_x(r)\equiv(-)^r\langle \check{s}^x_{2j}\check{s}^x_{2j+2r}\rangle_{\mathcal{G}}=(-)^r\langle\prod_{k=1}^{2r}\!\sigma_{2j+k}^x\rangle_{\mathcal{G}}$.
The dual QFI density is linear versus $L$ in the gapped phase $A_x$ ($J_x\geq J_y+J_z$) and
  constant in other regions [see Fig.~\ref{fig:4}(c)].
The other two gapped phases $A_y$ and $A_z$ [see Fig.~\ref{fig:4}(a)]
can be obtained by the substitutions $J_x\rightarrow J_{y,z}\rightarrow J_{z,y}\rightarrow J_x$, respectively.
Moreover, when considering the equivalent brick-wall lattice \cite{Feng2007} of the Kitaev honeycomb model,
these results can also be extended to the general 2D
lattice by transforming the second index of site to the momentum space \cite{Feng2007,SM}.

\emph{Conclusions.---}Recent work \cite{Pezze2017} shows that 1D
SPT order with winding numbers $\nu=\pm1$ can be characterized by a super-extensive QFI
 with respect to the spin operator, $ F_Q\propto L^2$. By introducing the above duality, we have
shown that 1D SPT order with higher winding numbers can be characterized by the scaling
behavior of multipartite entanglement with respect to the spin generators in the dual lattice.
By choosing the generators in different dual lattices, the scaling coefficients 
$\lambda_\nu$ and $\lambda_\nu^{(\textrm{st})}$ of the dual QFI density, as a 
witness of multipartite entanglement \cite{Hyllus2012,Toth2012},  effectively identify different
nontrivial topological phases with high winding numbers.
Moreover, further investigations on the  Kitaev honeycomb model have shown that
our results for detecting 1D SPT order could be well generalized to characterize topological order in 2D systems
(e.g., the toric code model \cite{Kitaev2006a} and fractional quantum Hall states \cite{Stormer1999})
and SPT order in non-Hermitian systems \cite{Leykam2017}. This work paves the way to characterizing  
topological phases
using multipartite entanglement of the ground state, and also the
detection of topologically protected multipartite entanglement, with promising applications in
both quantum computation and quantum metrology.

\begin{acknowledgements}
Y.R.Z. would like to thank Xian-Xin Wu, Wei Qin, Tao Liu, Zhou Li and Tian-Si Han for useful discussions.
H.F. was partially supported by Ministry of Technology of China (grants No. 2016YFA0302104
and 2016YFA0300600), National Natural Science Foundation of China  (grant No. 11774406) and Chinese Academy of Sciences (grants No. XDPB-0803). J.Q.Y. was partially supported
by the National Key Research and Development Program
of China (grant No. 2016YFA0301200), the NSFC (grant No. 11774022), 
and the NSAF (grant No. U1530401). F.N. was partially supported by the
MURI Center for Dynamic Magneto-Optics via the AFOSR Award No. FA9550-14-1-0040,
the Japan Society for the Promotion of Science (KAKENHI), the IMPACT program of JST,
CREST Grant No. JPMJCR1676,  RIKEN-AIST Challenge Research Fund,
JSPS-RFBR Grant No. 17-52-50023, and the Sir John Templeton Foundation.
\end{acknowledgements}

\bibliography{manuscript_rev}

\begin{thebibliography}{88}%
\makeatletter
\providecommand \@ifxundefined [1]{%
 \@ifx{#1\undefined}
}%
\providecommand \@ifnum [1]{%
 \ifnum #1\expandafter \@firstoftwo
 \else \expandafter \@secondoftwo
 \fi
}%
\providecommand \@ifx [1]{%
 \ifx #1\expandafter \@firstoftwo
 \else \expandafter \@secondoftwo
 \fi
}%
\providecommand \natexlab [1]{#1}%
\providecommand \enquote  [1]{``#1''}%
\providecommand \bibnamefont  [1]{#1}%
\providecommand \bibfnamefont [1]{#1}%
\providecommand \citenamefont [1]{#1}%
\providecommand \href@noop [0]{\@secondoftwo}%
\providecommand \href [0]{\begingroup \@sanitize@url \@href}%
\providecommand \@href[1]{\@@startlink{#1}\@@href}%
\providecommand \@@href[1]{\endgroup#1\@@endlink}%
\providecommand \@sanitize@url [0]{\catcode `\\12\catcode `\$12\catcode
  `\&12\catcode `\#12\catcode `\^12\catcode `\_12\catcode `\%12\relax}%
\providecommand \@@startlink[1]{}%
\providecommand \@@endlink[0]{}%
\providecommand \url  [0]{\begingroup\@sanitize@url \@url }%
\providecommand \@url [1]{\endgroup\@href {#1}{\urlprefix }}%
\providecommand \urlprefix  [0]{URL }%
\providecommand \Eprint [0]{\href }%
\providecommand \doibase [0]{http://dx.doi.org/}%
\providecommand \selectlanguage [0]{\@gobble}%
\providecommand \bibinfo  [0]{\@secondoftwo}%
\providecommand \bibfield  [0]{\@secondoftwo}%
\providecommand \translation [1]{[#1]}%
\providecommand \BibitemOpen [0]{}%
\providecommand \bibitemStop [0]{}%
\providecommand \bibitemNoStop [0]{.\EOS\space}%
\providecommand \EOS [0]{\spacefactor3000\relax}%
\providecommand \BibitemShut  [1]{\csname bibitem#1\endcsname}%
\let\auto@bib@innerbib\@empty
\bibitem [{\citenamefont {Zeng}\ \emph {et~al.}(2015)\citenamefont {Zeng},
  \citenamefont {Chen}, \citenamefont {Zhou},\ and\ \citenamefont
  {Wen}}]{Zeng2015}%
  \BibitemOpen
  \bibfield  {author} {\bibinfo {author} {\bibfnamefont {B.}~\bibnamefont
  {Zeng}}, \bibinfo {author} {\bibfnamefont {X.}~\bibnamefont {Chen}}, \bibinfo
  {author} {\bibfnamefont {D.~L.}\ \bibnamefont {Zhou}}, \ and\ \bibinfo
  {author} {\bibfnamefont {X.~G.}\ \bibnamefont {Wen}},\ }\bibfield  {title}
  {\enquote {\bibinfo {title} {Quantum information meets quantum matter--{F}rom
  quantum entanglement to topological phase in many-body systems},}\
  }\href@noop {} {\bibfield  {journal} {\bibinfo  {journal} {arXiv:1508.02595}\
  } (\bibinfo {year} {2015})}\BibitemShut {NoStop}%
\bibitem [{\citenamefont {Mong}\ \emph {et~al.}(2014)\citenamefont {Mong},
  \citenamefont {Clarke}, \citenamefont {Alicea}, \citenamefont {Lindner},
  \citenamefont {Fendley}, \citenamefont {Nayak}, \citenamefont {Oreg},
  \citenamefont {Stern}, \citenamefont {Berg}, \citenamefont {Shtengel},\ and\
  \citenamefont {Fisher}}]{Mong2014}%
  \BibitemOpen
  \bibfield  {author} {\bibinfo {author} {\bibfnamefont {R.~S.~K.}\
  \bibnamefont {Mong}}, \bibinfo {author} {\bibfnamefont {D.~J.}\ \bibnamefont
  {Clarke}}, \bibinfo {author} {\bibfnamefont {J.}~\bibnamefont {Alicea}},
  \bibinfo {author} {\bibfnamefont {N.~H.}\ \bibnamefont {Lindner}}, \bibinfo
  {author} {\bibfnamefont {P.}~\bibnamefont {Fendley}}, \bibinfo {author}
  {\bibfnamefont {C.}~\bibnamefont {Nayak}}, \bibinfo {author} {\bibfnamefont
  {Y.}~\bibnamefont {Oreg}}, \bibinfo {author} {\bibfnamefont {A.}~\bibnamefont
  {Stern}}, \bibinfo {author} {\bibfnamefont {E.}~\bibnamefont {Berg}},
  \bibinfo {author} {\bibfnamefont {K.}~\bibnamefont {Shtengel}}, \ and\
  \bibinfo {author} {\bibfnamefont {M.~P.~A.}\ \bibnamefont {Fisher}},\
  }\bibfield  {title} {\enquote {\bibinfo {title} {Universal topological
  quantum computation from a superconductor-$\textrm{A}$belian quantum
  $\textrm{H}$all heterostructure},}\ }\href {\doibase
  10.1103/PhysRevX.4.011036} {\bibfield  {journal} {\bibinfo  {journal} {Phys.
  Rev. X}\ }\textbf {\bibinfo {volume} {4}},\ \bibinfo {pages} {011036}
  (\bibinfo {year} {2014})}\BibitemShut {NoStop}%
\bibitem [{\citenamefont {Kraus}\ \emph {et~al.}(2013)\citenamefont {Kraus},
  \citenamefont {Zoller},\ and\ \citenamefont {Baranov}}]{Kraus2013}%
  \BibitemOpen
  \bibfield  {author} {\bibinfo {author} {\bibfnamefont {C.~V.}\ \bibnamefont
  {Kraus}}, \bibinfo {author} {\bibfnamefont {P.}~\bibnamefont {Zoller}}, \
  and\ \bibinfo {author} {\bibfnamefont {M.~A.}\ \bibnamefont {Baranov}},\
  }\bibfield  {title} {\enquote {\bibinfo {title} {Braiding of atomic
  $\textrm{M}$ajorana fermions in wire networks and implementation of the
  $\textrm{D}$eutsch-$\textrm{J}$ozsa algorithm},}\ }\href {\doibase
  10.1103/PhysRevLett.111.203001} {\bibfield  {journal} {\bibinfo  {journal}
  {Phys. Rev. Lett.}\ }\textbf {\bibinfo {volume} {111}},\ \bibinfo {pages}
  {203001} (\bibinfo {year} {2013})}\BibitemShut {NoStop}%
\bibitem [{\citenamefont {Sau}\ \emph {et~al.}(2010)\citenamefont {Sau},
  \citenamefont {Lutchyn}, \citenamefont {Tewari},\ and\ \citenamefont
  {Das~Sarma}}]{Sau2010}%
  \BibitemOpen
  \bibfield  {author} {\bibinfo {author} {\bibfnamefont {J.~D.}\ \bibnamefont
  {Sau}}, \bibinfo {author} {\bibfnamefont {R.~M.}\ \bibnamefont {Lutchyn}},
  \bibinfo {author} {\bibfnamefont {S.}~\bibnamefont {Tewari}}, \ and\ \bibinfo
  {author} {\bibfnamefont {S.}~\bibnamefont {Das~Sarma}},\ }\bibfield  {title}
  {\enquote {\bibinfo {title} {Generic new platform for topological quantum
  computation using semiconductor heterostructures},}\ }\href {\doibase
  10.1103/PhysRevLett.104.040502} {\bibfield  {journal} {\bibinfo  {journal}
  {Phys. Rev. Lett.}\ }\textbf {\bibinfo {volume} {104}},\ \bibinfo {pages}
  {040502} (\bibinfo {year} {2010})}\BibitemShut {NoStop}%
\bibitem [{\citenamefont {Alicea}\ \emph {et~al.}(2011)\citenamefont {Alicea},
  \citenamefont {Oreg}, \citenamefont {Refael}, \citenamefont {von Oppen},\
  and\ \citenamefont {Fisher}}]{Alicea2011}%
  \BibitemOpen
  \bibfield  {author} {\bibinfo {author} {\bibfnamefont {J.}~\bibnamefont
  {Alicea}}, \bibinfo {author} {\bibfnamefont {Y.}~\bibnamefont {Oreg}},
  \bibinfo {author} {\bibfnamefont {G.}~\bibnamefont {Refael}}, \bibinfo
  {author} {\bibfnamefont {F.}~\bibnamefont {von Oppen}}, \ and\ \bibinfo
  {author} {\bibfnamefont {M.~P.~A.}\ \bibnamefont {Fisher}},\ }\bibfield
  {title} {\enquote {\bibinfo {title} {Non-$\textrm{A}$belian statistics and
  topological quantum information processing in 1$\textrm{D}$ wire networks},}\
  }\href {\doibase 10.1038/NPHYS1915} {\bibfield  {journal} {\bibinfo
  {journal} {Nat. Phys.}\ }\textbf {\bibinfo {volume} {7}},\ \bibinfo {pages}
  {412--417} (\bibinfo {year} {2011})}\BibitemShut {NoStop}%
\bibitem [{\citenamefont {Alicea}(2012)}]{Alicea2012}%
  \BibitemOpen
  \bibfield  {author} {\bibinfo {author} {\bibfnamefont {J.}~\bibnamefont
  {Alicea}},\ }\bibfield  {title} {\enquote {\bibinfo {title} {New directions
  in the pursuit of $\textrm{M}$ajorana fermions in solid state systems},}\
  }\href {\doibase 10.1088/0034-4885/75/7/076501} {\bibfield  {journal}
  {\bibinfo  {journal} {Rep. Prog. Phys.}\ }\textbf {\bibinfo {volume} {75}},\
  \bibinfo {pages} {076501} (\bibinfo {year} {2012})}\BibitemShut {NoStop}%
\bibitem [{\citenamefont {Kitaev}(2003)}]{Kitaev2003}%
  \BibitemOpen
  \bibfield  {author} {\bibinfo {author} {\bibfnamefont {A.~Y.}\ \bibnamefont
  {Kitaev}},\ }\bibfield  {title} {\enquote {\bibinfo {title} {Fault-tolerant
  quantum computation by anyons},}\ }\href {\doibase
  10.1016/S0003-4916(02)00018-0} {\bibfield  {journal} {\bibinfo  {journal}
  {Ann. Phys.}\ }\textbf {\bibinfo {volume} {303}},\ \bibinfo {pages} {2--30}
  (\bibinfo {year} {2003})}\BibitemShut {NoStop}%
\bibitem [{\citenamefont {Xia}\ \emph {et~al.}(2009)\citenamefont {Xia},
  \citenamefont {Qian}, \citenamefont {Hsieh}, \citenamefont {Wray},
  \citenamefont {Pal}, \citenamefont {Lin}, \citenamefont {Bansil},
  \citenamefont {Grauer}, \citenamefont {Hor}, \citenamefont {Cava},\ and\
  \citenamefont {Hasan}}]{Xia2009}%
  \BibitemOpen
  \bibfield  {author} {\bibinfo {author} {\bibfnamefont {Y.}~\bibnamefont
  {Xia}}, \bibinfo {author} {\bibfnamefont {D.}~\bibnamefont {Qian}}, \bibinfo
  {author} {\bibfnamefont {D.}~\bibnamefont {Hsieh}}, \bibinfo {author}
  {\bibfnamefont {L.}~\bibnamefont {Wray}}, \bibinfo {author} {\bibfnamefont
  {A.}~\bibnamefont {Pal}}, \bibinfo {author} {\bibfnamefont {H.}~\bibnamefont
  {Lin}}, \bibinfo {author} {\bibfnamefont {A.}~\bibnamefont {Bansil}},
  \bibinfo {author} {\bibfnamefont {D.}~\bibnamefont {Grauer}}, \bibinfo
  {author} {\bibfnamefont {Y.~S.}\ \bibnamefont {Hor}}, \bibinfo {author}
  {\bibfnamefont {R.~J.}\ \bibnamefont {Cava}}, \ and\ \bibinfo {author}
  {\bibfnamefont {M.~Z.}\ \bibnamefont {Hasan}},\ }\bibfield  {title} {\enquote
  {\bibinfo {title} {Observation of a large-gap topological-insulator class
  with a single $\textrm{D}$irac cone on the surface},}\ }\href {\doibase
  10.1038/NPHYS1274} {\bibfield  {journal} {\bibinfo  {journal} {Nat. Phys.}\
  }\textbf {\bibinfo {volume} {5}},\ \bibinfo {pages} {398--402} (\bibinfo
  {year} {2009})}\BibitemShut {NoStop}%
\bibitem [{\citenamefont {Albrecht}\ \emph {et~al.}(2016)\citenamefont
  {Albrecht}, \citenamefont {Higginbotham}, \citenamefont {Madsen},
  \citenamefont {Kuemmeth}, \citenamefont {Jespersen}, \citenamefont {Nygard},
  \citenamefont {Krogstrup},\ and\ \citenamefont {Marcus}}]{Albrecht2016}%
  \BibitemOpen
  \bibfield  {author} {\bibinfo {author} {\bibfnamefont {S.~M.}\ \bibnamefont
  {Albrecht}}, \bibinfo {author} {\bibfnamefont {A.~P.}\ \bibnamefont
  {Higginbotham}}, \bibinfo {author} {\bibfnamefont {M.}~\bibnamefont
  {Madsen}}, \bibinfo {author} {\bibfnamefont {F.}~\bibnamefont {Kuemmeth}},
  \bibinfo {author} {\bibfnamefont {T.~S.}\ \bibnamefont {Jespersen}}, \bibinfo
  {author} {\bibfnamefont {J.}~\bibnamefont {Nygard}}, \bibinfo {author}
  {\bibfnamefont {P.}~\bibnamefont {Krogstrup}}, \ and\ \bibinfo {author}
  {\bibfnamefont {C.~M.}\ \bibnamefont {Marcus}},\ }\bibfield  {title}
  {\enquote {\bibinfo {title} {Exponential protection of zero modes in
  $\textrm{M}$ajorana islands},}\ }\href {\doibase 10.1038/nature17162}
  {\bibfield  {journal} {\bibinfo  {journal} {Nature}\ }\textbf {\bibinfo
  {volume} {531}},\ \bibinfo {pages} {206--209} (\bibinfo {year}
  {2016})}\BibitemShut {NoStop}%
\bibitem [{\citenamefont {Wen}\ and\ \citenamefont {Niu}(1990)}]{Wen1990}%
  \BibitemOpen
  \bibfield  {author} {\bibinfo {author} {\bibfnamefont {X.~G.}\ \bibnamefont
  {Wen}}\ and\ \bibinfo {author} {\bibfnamefont {Q.}~\bibnamefont {Niu}},\
  }\bibfield  {title} {\enquote {\bibinfo {title} {Ground-state degeneracy of
  the fractional quantum $\textrm{H}$all states in the presence of a random
  potential and on high-genus $\textrm{R}$iemann surfaces},}\ }\href {\doibase
  10.1103/PhysRevB.41.9377} {\bibfield  {journal} {\bibinfo  {journal} {Phys.
  Rev. B}\ }\textbf {\bibinfo {volume} {41}},\ \bibinfo {pages} {9377--9396}
  (\bibinfo {year} {1990})}\BibitemShut {NoStop}%
\bibitem [{\citenamefont {Wen}(2004)}]{Wen2004}%
  \BibitemOpen
  \bibfield  {author} {\bibinfo {author} {\bibfnamefont {X.~G.}\ \bibnamefont
  {Wen}},\ }\href@noop {} {\emph {\bibinfo {title} {Quantum field theory of
  many-body systems: from the origin of sound to an origin of light and
  electrons}}}\ (\bibinfo  {publisher} {Oxford University Press on Demand},\
  \bibinfo {year} {2004})\BibitemShut {NoStop}%
\bibitem [{\citenamefont {Amico}\ \emph {et~al.}(2008)\citenamefont {Amico},
  \citenamefont {Fazio}, \citenamefont {Osterloh},\ and\ \citenamefont
  {Vedral}}]{Amico2008}%
  \BibitemOpen
  \bibfield  {author} {\bibinfo {author} {\bibfnamefont {L.}~\bibnamefont
  {Amico}}, \bibinfo {author} {\bibfnamefont {R.}~\bibnamefont {Fazio}},
  \bibinfo {author} {\bibfnamefont {A.}~\bibnamefont {Osterloh}}, \ and\
  \bibinfo {author} {\bibfnamefont {V.}~\bibnamefont {Vedral}},\ }\bibfield
  {title} {\enquote {\bibinfo {title} {Entanglement in many-body systems},}\
  }\href {\doibase 10.1103/RevModPhys.80.517} {\bibfield  {journal} {\bibinfo
  {journal} {Rev. Mod. Phys.}\ }\textbf {\bibinfo {volume} {80}},\ \bibinfo
  {pages} {517--576} (\bibinfo {year} {2008})}\BibitemShut {NoStop}%
\bibitem [{\citenamefont {Eisert}\ \emph {et~al.}(2010)\citenamefont {Eisert},
  \citenamefont {Cramer},\ and\ \citenamefont {Plenio}}]{Eisert2010}%
  \BibitemOpen
  \bibfield  {author} {\bibinfo {author} {\bibfnamefont {J.}~\bibnamefont
  {Eisert}}, \bibinfo {author} {\bibfnamefont {M.}~\bibnamefont {Cramer}}, \
  and\ \bibinfo {author} {\bibfnamefont {M.~B.}\ \bibnamefont {Plenio}},\
  }\bibfield  {title} {\enquote {\bibinfo {title} {Colloquium: $\textrm{A}$rea
  laws for the entanglement entropy},}\ }\href {\doibase
  10.1103/RevModPhys.82.277} {\bibfield  {journal} {\bibinfo  {journal} {Rev.
  Mod. Phys.}\ }\textbf {\bibinfo {volume} {82}},\ \bibinfo {pages} {277--306}
  (\bibinfo {year} {2010})}\BibitemShut {NoStop}%
\bibitem [{\citenamefont {Kitaev}\ and\ \citenamefont
  {Preskill}(2006)}]{Kitaev2006}%
  \BibitemOpen
  \bibfield  {author} {\bibinfo {author} {\bibfnamefont {A.}~\bibnamefont
  {Kitaev}}\ and\ \bibinfo {author} {\bibfnamefont {J.}~\bibnamefont
  {Preskill}},\ }\bibfield  {title} {\enquote {\bibinfo {title} {Topological
  entanglement entropy},}\ }\href {\doibase 10.1103/PhysRevLett.96.110404}
  {\bibfield  {journal} {\bibinfo  {journal} {Phys. Rev. Lett.}\ }\textbf
  {\bibinfo {volume} {96}},\ \bibinfo {pages} {110404} (\bibinfo {year}
  {2006})}\BibitemShut {NoStop}%
\bibitem [{\citenamefont {Levin}\ and\ \citenamefont {Wen}(2006)}]{Levin2006}%
  \BibitemOpen
  \bibfield  {author} {\bibinfo {author} {\bibfnamefont {M.}~\bibnamefont
  {Levin}}\ and\ \bibinfo {author} {\bibfnamefont {X.~G.}\ \bibnamefont
  {Wen}},\ }\bibfield  {title} {\enquote {\bibinfo {title} {Detecting
  topological order in a ground state wave function},}\ }\href {\doibase
  10.1103/PhysRevLett.96.110405} {\bibfield  {journal} {\bibinfo  {journal}
  {Phys. Rev. Lett.}\ }\textbf {\bibinfo {volume} {96}},\ \bibinfo {pages}
  {110405} (\bibinfo {year} {2006})}\BibitemShut {NoStop}%
\bibitem [{\citenamefont {Li}\ and\ \citenamefont {Haldane}(2008)}]{Li2008}%
  \BibitemOpen
  \bibfield  {author} {\bibinfo {author} {\bibfnamefont {H.}~\bibnamefont
  {Li}}\ and\ \bibinfo {author} {\bibfnamefont {F.~D.~M.}\ \bibnamefont
  {Haldane}},\ }\bibfield  {title} {\enquote {\bibinfo {title} {Entanglement
  spectrum as a generalization of entanglement entropy: Identification of
  topological order in non-$\textrm{A}$belian fractional quantum
  $\textrm{H}$all effect states},}\ }\href {\doibase
  10.1103/PhysRevLett.101.010504} {\bibfield  {journal} {\bibinfo  {journal}
  {Phys. Rev. Lett.}\ }\textbf {\bibinfo {volume} {101}},\ \bibinfo {pages}
  {010504} (\bibinfo {year} {2008})}\BibitemShut {NoStop}%
\bibitem [{\citenamefont {Gu}\ and\ \citenamefont {Wen}(2009)}]{Gu2009}%
  \BibitemOpen
  \bibfield  {author} {\bibinfo {author} {\bibfnamefont {Z.~C.}\ \bibnamefont
  {Gu}}\ and\ \bibinfo {author} {\bibfnamefont {X.~G.}\ \bibnamefont {Wen}},\
  }\bibfield  {title} {\enquote {\bibinfo {title}
  {Tensor-entanglement-filtering renormalization approach and
  symmetry-protected topological order},}\ }\href {\doibase
  10.1103/PhysRevB.80.155131} {\bibfield  {journal} {\bibinfo  {journal} {Phys.
  Rev. B}\ }\textbf {\bibinfo {volume} {80}},\ \bibinfo {pages} {155131}
  (\bibinfo {year} {2009})}\BibitemShut {NoStop}%
\bibitem [{\citenamefont {Chen}\ \emph {et~al.}(2011)\citenamefont {Chen},
  \citenamefont {Gu},\ and\ \citenamefont {Wen}}]{Chen2011}%
  \BibitemOpen
  \bibfield  {author} {\bibinfo {author} {\bibfnamefont {X.}~\bibnamefont
  {Chen}}, \bibinfo {author} {\bibfnamefont {Z.~C.}\ \bibnamefont {Gu}}, \ and\
  \bibinfo {author} {\bibfnamefont {X.~G.}\ \bibnamefont {Wen}},\ }\bibfield
  {title} {\enquote {\bibinfo {title} {Classification of gapped symmetric
  phases in one-dimensional spin systems},}\ }\href {\doibase
  10.1103/PhysRevB.83.035107} {\bibfield  {journal} {\bibinfo  {journal} {Phys.
  Rev. B}\ }\textbf {\bibinfo {volume} {83}},\ \bibinfo {pages} {035107}
  (\bibinfo {year} {2011})}\BibitemShut {NoStop}%
\bibitem [{\citenamefont {Scaffidi}\ \emph {et~al.}(2017)\citenamefont
  {Scaffidi}, \citenamefont {Parker},\ and\ \citenamefont
  {Vasseur}}]{Scaffidi2017}%
  \BibitemOpen
  \bibfield  {author} {\bibinfo {author} {\bibfnamefont {T.}~\bibnamefont
  {Scaffidi}}, \bibinfo {author} {\bibfnamefont {D.~E.}\ \bibnamefont
  {Parker}}, \ and\ \bibinfo {author} {\bibfnamefont {R.}~\bibnamefont
  {Vasseur}},\ }\bibfield  {title} {\enquote {\bibinfo {title} {Gapless
  symmetry-protected topological order},}\ }\href {\doibase
  10.1103/PhysRevX.7.041048} {\bibfield  {journal} {\bibinfo  {journal} {Phys.
  Rev. X}\ }\textbf {\bibinfo {volume} {7}},\ \bibinfo {pages} {041048}
  (\bibinfo {year} {2017})}\BibitemShut {NoStop}%
\bibitem [{\citenamefont {Bliokh}\ \emph {et~al.}(2015)\citenamefont {Bliokh},
  \citenamefont {Smirnova},\ and\ \citenamefont {Nori}}]{Bliokh2015}%
  \BibitemOpen
  \bibfield  {author} {\bibinfo {author} {\bibfnamefont {K.~Y.}\ \bibnamefont
  {Bliokh}}, \bibinfo {author} {\bibfnamefont {D.}~\bibnamefont {Smirnova}}, \
  and\ \bibinfo {author} {\bibfnamefont {F.}~\bibnamefont {Nori}},\ }\bibfield
  {title} {\enquote {\bibinfo {title} {Quantum spin {H}all effect of light},}\
  }\href {\doibase 10.1126/science.aaa9519} {\bibfield  {journal} {\bibinfo
  {journal} {Science}\ }\textbf {\bibinfo {volume} {348}},\ \bibinfo {pages}
  {1448--1451} (\bibinfo {year} {2015})}\BibitemShut {NoStop}%
\bibitem [{\citenamefont {Konig}\ \emph {et~al.}(2007)\citenamefont {Konig},
  \citenamefont {Wiedmann}, \citenamefont {Brune}, \citenamefont {Roth},
  \citenamefont {Buhmann}, \citenamefont {Molenkamp}, \citenamefont {Qi},\ and\
  \citenamefont {Zhang}}]{Konig2007}%
  \BibitemOpen
  \bibfield  {author} {\bibinfo {author} {\bibfnamefont {M.}~\bibnamefont
  {Konig}}, \bibinfo {author} {\bibfnamefont {S.}~\bibnamefont {Wiedmann}},
  \bibinfo {author} {\bibfnamefont {C.}~\bibnamefont {Brune}}, \bibinfo
  {author} {\bibfnamefont {A.}~\bibnamefont {Roth}}, \bibinfo {author}
  {\bibfnamefont {H.}~\bibnamefont {Buhmann}}, \bibinfo {author} {\bibfnamefont
  {L.~W.}\ \bibnamefont {Molenkamp}}, \bibinfo {author} {\bibfnamefont {X.~L.}\
  \bibnamefont {Qi}}, \ and\ \bibinfo {author} {\bibfnamefont {S.~C.}\
  \bibnamefont {Zhang}},\ }\bibfield  {title} {\enquote {\bibinfo {title}
  {Quantum spin {H}all insulator state in {H}g{T}e quantum wells},}\ }\href
  {\doibase 10.1126/science.1148047} {\bibfield  {journal} {\bibinfo  {journal}
  {Science}\ }\textbf {\bibinfo {volume} {318}},\ \bibinfo {pages} {766--770}
  (\bibinfo {year} {2007})}\BibitemShut {NoStop}%
\bibitem [{\citenamefont {Fu}\ and\ \citenamefont {Kane}(2008)}]{Fu2008}%
  \BibitemOpen
  \bibfield  {author} {\bibinfo {author} {\bibfnamefont {L.}~\bibnamefont
  {Fu}}\ and\ \bibinfo {author} {\bibfnamefont {C.~L.}\ \bibnamefont {Kane}},\
  }\bibfield  {title} {\enquote {\bibinfo {title} {Superconducting proximity
  effect and $\textrm{M}$ajorana fermions at the surface of a topological
  insulator},}\ }\href {\doibase 10.1103/PhysRevLett.100.096407} {\bibfield
  {journal} {\bibinfo  {journal} {Phys. Rev. Lett.}\ }\textbf {\bibinfo
  {volume} {100}},\ \bibinfo {pages} {096407} (\bibinfo {year}
  {2008})}\BibitemShut {NoStop}%
\bibitem [{\citenamefont {Hsieh}\ \emph {et~al.}(2008)\citenamefont {Hsieh},
  \citenamefont {Qian}, \citenamefont {Wray}, \citenamefont {Xia},
  \citenamefont {Hor}, \citenamefont {Cava},\ and\ \citenamefont
  {Hasan}}]{Hsieh2008}%
  \BibitemOpen
  \bibfield  {author} {\bibinfo {author} {\bibfnamefont {D.}~\bibnamefont
  {Hsieh}}, \bibinfo {author} {\bibfnamefont {D.}~\bibnamefont {Qian}},
  \bibinfo {author} {\bibfnamefont {L.}~\bibnamefont {Wray}}, \bibinfo {author}
  {\bibfnamefont {Y.}~\bibnamefont {Xia}}, \bibinfo {author} {\bibfnamefont
  {Y.~S.}\ \bibnamefont {Hor}}, \bibinfo {author} {\bibfnamefont {R.~J.}\
  \bibnamefont {Cava}}, \ and\ \bibinfo {author} {\bibfnamefont {M.~Z.}\
  \bibnamefont {Hasan}},\ }\bibfield  {title} {\enquote {\bibinfo {title} {A
  topological {D}irac insulator in a quantum spin {H}all phase},}\ }\href
  {\doibase 10.1038/nature06843} {\bibfield  {journal} {\bibinfo  {journal}
  {Nature}\ }\textbf {\bibinfo {volume} {452}},\ \bibinfo {pages} {970--975}
  (\bibinfo {year} {2008})}\BibitemShut {NoStop}%
\bibitem [{\citenamefont {Hasan}\ and\ \citenamefont {Kane}(2010)}]{Hasan2010}%
  \BibitemOpen
  \bibfield  {author} {\bibinfo {author} {\bibfnamefont {M.~Z.}\ \bibnamefont
  {Hasan}}\ and\ \bibinfo {author} {\bibfnamefont {C.~L.}\ \bibnamefont
  {Kane}},\ }\bibfield  {title} {\enquote {\bibinfo {title} {Colloquium:
  $\textrm{T}$opological insulators},}\ }\href {\doibase
  10.1103/RevModPhys.82.3045} {\bibfield  {journal} {\bibinfo  {journal} {Rev.
  Mod. Phys.}\ }\textbf {\bibinfo {volume} {82}},\ \bibinfo {pages}
  {3045--3067} (\bibinfo {year} {2010})}\BibitemShut {NoStop}%
\bibitem [{\citenamefont {Qi}\ and\ \citenamefont {Zhang}(2011)}]{Qi2011}%
  \BibitemOpen
  \bibfield  {author} {\bibinfo {author} {\bibfnamefont {X.~L.}\ \bibnamefont
  {Qi}}\ and\ \bibinfo {author} {\bibfnamefont {S.~C.}\ \bibnamefont {Zhang}},\
  }\bibfield  {title} {\enquote {\bibinfo {title} {Topological insulators and
  superconductors},}\ }\href {\doibase 10.1103/RevModPhys.83.1057} {\bibfield
  {journal} {\bibinfo  {journal} {Rev. Mod. Phys.}\ }\textbf {\bibinfo {volume}
  {83}},\ \bibinfo {pages} {1057--1110} (\bibinfo {year} {2011})}\BibitemShut
  {NoStop}%
\bibitem [{\citenamefont {Sato}\ and\ \citenamefont {Ando}(2017)}]{Sato2017}%
  \BibitemOpen
  \bibfield  {author} {\bibinfo {author} {\bibfnamefont {M.}~\bibnamefont
  {Sato}}\ and\ \bibinfo {author} {\bibfnamefont {Y.}~\bibnamefont {Ando}},\
  }\bibfield  {title} {\enquote {\bibinfo {title} {Topological superconductors:
  a review},}\ }\href {\doibase 10.1088/1361-6633/aa6ac7} {\bibfield  {journal}
  {\bibinfo  {journal} {Rep. Prog. Phys.}\ }\textbf {\bibinfo {volume} {80}},\
  \bibinfo {pages} {076501} (\bibinfo {year} {2017})}\BibitemShut {NoStop}%
\bibitem [{\citenamefont {Nayak}\ \emph {et~al.}(2008)\citenamefont {Nayak},
  \citenamefont {Simon}, \citenamefont {Stern}, \citenamefont {Freedman},\ and\
  \citenamefont {Das~Sarma}}]{Nayak2008}%
  \BibitemOpen
  \bibfield  {author} {\bibinfo {author} {\bibfnamefont {C.}~\bibnamefont
  {Nayak}}, \bibinfo {author} {\bibfnamefont {S.~H.}\ \bibnamefont {Simon}},
  \bibinfo {author} {\bibfnamefont {A.}~\bibnamefont {Stern}}, \bibinfo
  {author} {\bibfnamefont {M.}~\bibnamefont {Freedman}}, \ and\ \bibinfo
  {author} {\bibfnamefont {S.}~\bibnamefont {Das~Sarma}},\ }\bibfield  {title}
  {\enquote {\bibinfo {title} {Non-$\textrm{A}$belian anyons and topological
  quantum computation},}\ }\href {\doibase 10.1103/RevModPhys.80.1083}
  {\bibfield  {journal} {\bibinfo  {journal} {Rev. Mod. Phys.}\ }\textbf
  {\bibinfo {volume} {80}},\ \bibinfo {pages} {1083--1159} (\bibinfo {year}
  {2008})}\BibitemShut {NoStop}%
\bibitem [{\citenamefont {Sarma}\ \emph {et~al.}(2015)\citenamefont {Sarma},
  \citenamefont {Freedman},\ and\ \citenamefont {Nayak}}]{Sarma2015}%
  \BibitemOpen
  \bibfield  {author} {\bibinfo {author} {\bibfnamefont {S.~D.}\ \bibnamefont
  {Sarma}}, \bibinfo {author} {\bibfnamefont {M.}~\bibnamefont {Freedman}}, \
  and\ \bibinfo {author} {\bibfnamefont {C.}~\bibnamefont {Nayak}},\ }\bibfield
   {title} {\enquote {\bibinfo {title} {Majorana zero modes and topological
  quantum computation},}\ }\href {\doibase 10.1038/npjqi.2015.1} {\bibfield
  {journal} {\bibinfo  {journal} {npj Quantum Inf.}\ }\textbf {\bibinfo
  {volume} {1}},\ \bibinfo {pages} {15001} (\bibinfo {year}
  {2015})}\BibitemShut {NoStop}%
\bibitem [{\citenamefont {Elliott}\ and\ \citenamefont
  {Franz}(2015)}]{Elliott2015}%
  \BibitemOpen
  \bibfield  {author} {\bibinfo {author} {\bibfnamefont {S.~R.}\ \bibnamefont
  {Elliott}}\ and\ \bibinfo {author} {\bibfnamefont {M.}~\bibnamefont
  {Franz}},\ }\bibfield  {title} {\enquote {\bibinfo {title} {Colloquium:
  $\textrm{M}$ajorana fermions in nuclear, particle, and solid-state
  physics},}\ }\href {\doibase 10.1103/RevModPhys.87.137} {\bibfield  {journal}
  {\bibinfo  {journal} {Rev. Mod. Phys.}\ }\textbf {\bibinfo {volume} {87}},\
  \bibinfo {pages} {137--163} (\bibinfo {year} {2015})}\BibitemShut {NoStop}%
\bibitem [{\citenamefont {You}\ \emph {et~al.}(2010)\citenamefont {You},
  \citenamefont {Shi}, \citenamefont {Hu},\ and\ \citenamefont
  {Nori}}]{You2010}%
  \BibitemOpen
  \bibfield  {author} {\bibinfo {author} {\bibfnamefont {J.~Q.}\ \bibnamefont
  {You}}, \bibinfo {author} {\bibfnamefont {X.~F.}\ \bibnamefont {Shi}},
  \bibinfo {author} {\bibfnamefont {X.~D.}\ \bibnamefont {Hu}}, \ and\ \bibinfo
  {author} {\bibfnamefont {F.}~\bibnamefont {Nori}},\ }\bibfield  {title}
  {\enquote {\bibinfo {title} {Quantum emulation of a spin system with
  topologically protected ground states using superconducting quantum
  circuits},}\ }\href {\doibase 10.1103/PhysRevB.81.014505} {\bibfield
  {journal} {\bibinfo  {journal} {Phys. Rev. B}\ }\textbf {\bibinfo {volume}
  {81}},\ \bibinfo {pages} {014505} (\bibinfo {year} {2010})}\BibitemShut
  {NoStop}%
\bibitem [{\citenamefont {You}\ \emph {et~al.}(2014)\citenamefont {You},
  \citenamefont {Wang}, \citenamefont {Zhang},\ and\ \citenamefont
  {Nori}}]{You2014}%
  \BibitemOpen
  \bibfield  {author} {\bibinfo {author} {\bibfnamefont {J.~Q.}\ \bibnamefont
  {You}}, \bibinfo {author} {\bibfnamefont {Z.~D.}\ \bibnamefont {Wang}},
  \bibinfo {author} {\bibfnamefont {W.~X.}\ \bibnamefont {Zhang}}, \ and\
  \bibinfo {author} {\bibfnamefont {F.}~\bibnamefont {Nori}},\ }\bibfield
  {title} {\enquote {\bibinfo {title} {Encoding a qubit with
  $\textrm{M}$ajorana modes in superconducting circuits},}\ }\href {\doibase
  10.1038/srep05535} {\bibfield  {journal} {\bibinfo  {journal} {Sci. Rep.}\
  }\textbf {\bibinfo {volume} {4}},\ \bibinfo {pages} {5535} (\bibinfo {year}
  {2014})}\BibitemShut {NoStop}%
\bibitem [{\citenamefont {Akzyanov}\ \emph {et~al.}(2015)\citenamefont
  {Akzyanov}, \citenamefont {Rakhmanov}, \citenamefont {Rozhkov},\ and\
  \citenamefont {Nori}}]{Akzyanov2015}%
  \BibitemOpen
  \bibfield  {author} {\bibinfo {author} {\bibfnamefont {R.~S.}\ \bibnamefont
  {Akzyanov}}, \bibinfo {author} {\bibfnamefont {A.~L.}\ \bibnamefont
  {Rakhmanov}}, \bibinfo {author} {\bibfnamefont {A.~V.}\ \bibnamefont
  {Rozhkov}}, \ and\ \bibinfo {author} {\bibfnamefont {F.}~\bibnamefont
  {Nori}},\ }\bibfield  {title} {\enquote {\bibinfo {title} {Majorana fermions
  at the edge of superconducting islands},}\ }\href {\doibase
  10.1103/PhysRevB.92.075432} {\bibfield  {journal} {\bibinfo  {journal} {Phys.
  Rev. B}\ }\textbf {\bibinfo {volume} {92}},\ \bibinfo {pages} {075432}
  (\bibinfo {year} {2015})}\BibitemShut {NoStop}%
\bibitem [{\citenamefont {Akzyanov}\ \emph {et~al.}(2016)\citenamefont
  {Akzyanov}, \citenamefont {Rakhmanov}, \citenamefont {Rozhkov},\ and\
  \citenamefont {Nori}}]{Akzyanov2016}%
  \BibitemOpen
  \bibfield  {author} {\bibinfo {author} {\bibfnamefont {R.~S.}\ \bibnamefont
  {Akzyanov}}, \bibinfo {author} {\bibfnamefont {A.~L.}\ \bibnamefont
  {Rakhmanov}}, \bibinfo {author} {\bibfnamefont {A.~V.}\ \bibnamefont
  {Rozhkov}}, \ and\ \bibinfo {author} {\bibfnamefont {F.}~\bibnamefont
  {Nori}},\ }\bibfield  {title} {\enquote {\bibinfo {title} {Tunable {M}ajorana
  fermion from {L}andau quantization in 2{D} topological superconductors},}\
  }\href {\doibase 10.1103/PhysRevB.94.125428} {\bibfield  {journal} {\bibinfo
  {journal} {Phys. Rev. B}\ }\textbf {\bibinfo {volume} {94}},\ \bibinfo
  {pages} {125428} (\bibinfo {year} {2016})}\BibitemShut {NoStop}%
\bibitem [{\citenamefont {Zhang}\ and\ \citenamefont {Nori}(2016)}]{Zhang2016}%
  \BibitemOpen
  \bibfield  {author} {\bibinfo {author} {\bibfnamefont {P.}~\bibnamefont
  {Zhang}}\ and\ \bibinfo {author} {\bibfnamefont {F.}~\bibnamefont {Nori}},\
  }\bibfield  {title} {\enquote {\bibinfo {title} {Majorana bound states in a
  disordered quantum dot chain},}\ }\href {\doibase
  10.1088/1367-2630/18/4/043033} {\bibfield  {journal} {\bibinfo  {journal}
  {New J. Phys.}\ }\textbf {\bibinfo {volume} {18}},\ \bibinfo {pages} {043033}
  (\bibinfo {year} {2016})}\BibitemShut {NoStop}%
\bibitem [{\citenamefont {Vodola}\ \emph {et~al.}(2014)\citenamefont {Vodola},
  \citenamefont {Lepori}, \citenamefont {Ercolessi}, \citenamefont {Gorshkov},\
  and\ \citenamefont {Pupillo}}]{Vodola2014}%
  \BibitemOpen
  \bibfield  {author} {\bibinfo {author} {\bibfnamefont {D.}~\bibnamefont
  {Vodola}}, \bibinfo {author} {\bibfnamefont {L.}~\bibnamefont {Lepori}},
  \bibinfo {author} {\bibfnamefont {E.}~\bibnamefont {Ercolessi}}, \bibinfo
  {author} {\bibfnamefont {A.~V.}\ \bibnamefont {Gorshkov}}, \ and\ \bibinfo
  {author} {\bibfnamefont {G.}~\bibnamefont {Pupillo}},\ }\bibfield  {title}
  {\enquote {\bibinfo {title} {Kitaev chains with long-range pairing},}\ }\href
  {\doibase 10.1103/PhysRevLett.113.156402} {\bibfield  {journal} {\bibinfo
  {journal} {Phys. Rev. Lett.}\ }\textbf {\bibinfo {volume} {113}},\ \bibinfo
  {pages} {156402} (\bibinfo {year} {2014})}\BibitemShut {NoStop}%
\bibitem [{\citenamefont {Vodola}\ \emph {et~al.}(2015)\citenamefont {Vodola},
  \citenamefont {Lepori}, \citenamefont {Ercolessi},\ and\ \citenamefont
  {Pupillo}}]{Vodola2015}%
  \BibitemOpen
  \bibfield  {author} {\bibinfo {author} {\bibfnamefont {D.}~\bibnamefont
  {Vodola}}, \bibinfo {author} {\bibfnamefont {L.}~\bibnamefont {Lepori}},
  \bibinfo {author} {\bibfnamefont {E.}~\bibnamefont {Ercolessi}}, \ and\
  \bibinfo {author} {\bibfnamefont {G.}~\bibnamefont {Pupillo}},\ }\bibfield
  {title} {\enquote {\bibinfo {title} {Long-range $\textrm{I}$sing and
  $\textrm{K}$itaev models: phases, correlations and edge modes},}\ }\href
  {\doibase 10.1088/1367-2630/18/1/015001} {\bibfield  {journal} {\bibinfo
  {journal} {New J. Phys.}\ }\textbf {\bibinfo {volume} {18}},\ \bibinfo
  {pages} {015001} (\bibinfo {year} {2015})}\BibitemShut {NoStop}%
\bibitem [{\citenamefont {Alecce}\ and\ \citenamefont
  {Dell'Anna}(2017)}]{Alecce2017}%
  \BibitemOpen
  \bibfield  {author} {\bibinfo {author} {\bibfnamefont {A.}~\bibnamefont
  {Alecce}}\ and\ \bibinfo {author} {\bibfnamefont {L.}~\bibnamefont
  {Dell'Anna}},\ }\bibfield  {title} {\enquote {\bibinfo {title} {Extended
  $\textrm{K}$itaev chain with longer-range hopping and pairing},}\ }\href
  {\doibase 10.1103/PhysRevB.95.195160} {\bibfield  {journal} {\bibinfo
  {journal} {Phys. Rev. B}\ }\textbf {\bibinfo {volume} {95}},\ \bibinfo
  {pages} {195160} (\bibinfo {year} {2017})}\BibitemShut {NoStop}%
\bibitem [{\citenamefont {Lepori}\ and\ \citenamefont
  {Dell'Anna}(2017)}]{Lepori2017}%
  \BibitemOpen
  \bibfield  {author} {\bibinfo {author} {\bibfnamefont {L.}~\bibnamefont
  {Lepori}}\ and\ \bibinfo {author} {\bibfnamefont {L.}~\bibnamefont
  {Dell'Anna}},\ }\bibfield  {title} {\enquote {\bibinfo {title} {Long-range
  topological insulators and weakened bulk-boundary correspondence},}\ }\href
  {\doibase https://doi.org/10.1088/1367-2630/aa84d0} {\bibfield  {journal}
  {\bibinfo  {journal} {New J. Phys.}\ }\textbf {\bibinfo {volume} {19}},\
  \bibinfo {pages} {103030} (\bibinfo {year} {2017})}\BibitemShut {NoStop}%
\bibitem [{\citenamefont {Mourik}\ \emph {et~al.}(2012)\citenamefont {Mourik},
  \citenamefont {Zuo}, \citenamefont {Frolov}, \citenamefont {Plissard},
  \citenamefont {Bakkers},\ and\ \citenamefont {Kouwenhoven}}]{Mourik2012}%
  \BibitemOpen
  \bibfield  {author} {\bibinfo {author} {\bibfnamefont {V.}~\bibnamefont
  {Mourik}}, \bibinfo {author} {\bibfnamefont {K.}~\bibnamefont {Zuo}},
  \bibinfo {author} {\bibfnamefont {S.~M.}\ \bibnamefont {Frolov}}, \bibinfo
  {author} {\bibfnamefont {S.~R.}\ \bibnamefont {Plissard}}, \bibinfo {author}
  {\bibfnamefont {E.P.A.M.}\ \bibnamefont {Bakkers}}, \ and\ \bibinfo {author}
  {\bibfnamefont {L.~P.}\ \bibnamefont {Kouwenhoven}},\ }\bibfield  {title}
  {\enquote {\bibinfo {title} {Signatures of $\textrm{M}$ajorana fermions in
  hybrid superconductor-semiconductor nanowire devices},}\ }\href {\doibase
  10.1126/science.1222360} {\bibfield  {journal} {\bibinfo  {journal}
  {Science}\ }\textbf {\bibinfo {volume} {336}},\ \bibinfo {pages} {1003--1007}
  (\bibinfo {year} {2012})}\BibitemShut {NoStop}%
\bibitem [{\citenamefont {Nadj-Perge}\ \emph {et~al.}(2014)\citenamefont
  {Nadj-Perge}, \citenamefont {Drozdov}, \citenamefont {Li}, \citenamefont
  {Chen}, \citenamefont {Jeon}, \citenamefont {Seo}, \citenamefont {MacDonald},
  \citenamefont {Bernevig},\ and\ \citenamefont {Yazdani}}]{Nadj-Perge2014}%
  \BibitemOpen
  \bibfield  {author} {\bibinfo {author} {\bibfnamefont {S.}~\bibnamefont
  {Nadj-Perge}}, \bibinfo {author} {\bibfnamefont {I.~K.}\ \bibnamefont
  {Drozdov}}, \bibinfo {author} {\bibfnamefont {J.}~\bibnamefont {Li}},
  \bibinfo {author} {\bibfnamefont {H.}~\bibnamefont {Chen}}, \bibinfo {author}
  {\bibfnamefont {S.}~\bibnamefont {Jeon}}, \bibinfo {author} {\bibfnamefont
  {J.}~\bibnamefont {Seo}}, \bibinfo {author} {\bibfnamefont {A.~H.}\
  \bibnamefont {MacDonald}}, \bibinfo {author} {\bibfnamefont {B.~A.}\
  \bibnamefont {Bernevig}}, \ and\ \bibinfo {author} {\bibfnamefont
  {A.}~\bibnamefont {Yazdani}},\ }\bibfield  {title} {\enquote {\bibinfo
  {title} {Observation of $\textrm{M}$ajorana fermions in ferromagnetic atomic
  chains on a superconductor},}\ }\href {\doibase 10.1126/science.1259327}
  {\bibfield  {journal} {\bibinfo  {journal} {Science}\ }\textbf {\bibinfo
  {volume} {346}},\ \bibinfo {pages} {602--607} (\bibinfo {year}
  {2014})}\BibitemShut {NoStop}%
\bibitem [{\citenamefont {Do}\ \emph {et~al.}(2017)\citenamefont {Do},
  \citenamefont {Park}, \citenamefont {Yoshitake}, \citenamefont {Nasu},
  \citenamefont {Motome}, \citenamefont {Kwon}, \citenamefont {Adroja},
  \citenamefont {Voneshen}, \citenamefont {Kim}, \citenamefont {Jang},
  \citenamefont {Park}, \citenamefont {Choi},\ and\ \citenamefont
  {Ji}}]{Do2017}%
  \BibitemOpen
  \bibfield  {author} {\bibinfo {author} {\bibfnamefont {S.-H.}\ \bibnamefont
  {Do}}, \bibinfo {author} {\bibfnamefont {S.-Y.}\ \bibnamefont {Park}},
  \bibinfo {author} {\bibfnamefont {J.}~\bibnamefont {Yoshitake}}, \bibinfo
  {author} {\bibfnamefont {J.}~\bibnamefont {Nasu}}, \bibinfo {author}
  {\bibfnamefont {Y.}~\bibnamefont {Motome}}, \bibinfo {author} {\bibfnamefont
  {Y.~S.}\ \bibnamefont {Kwon}}, \bibinfo {author} {\bibfnamefont {D.~T.}\
  \bibnamefont {Adroja}}, \bibinfo {author} {\bibfnamefont {D.~J.}\
  \bibnamefont {Voneshen}}, \bibinfo {author} {\bibfnamefont {K.}~\bibnamefont
  {Kim}}, \bibinfo {author} {\bibfnamefont {T.-H.}\ \bibnamefont {Jang}},
  \bibinfo {author} {\bibfnamefont {J.-H.}\ \bibnamefont {Park}}, \bibinfo
  {author} {\bibfnamefont {K.-Y.}\ \bibnamefont {Choi}}, \ and\ \bibinfo
  {author} {\bibfnamefont {S.}~\bibnamefont {Ji}},\ }\bibfield  {title}
  {\enquote {\bibinfo {title} {Majorana fermions in the $\textrm{K}$itaev
  quantum spin system $\alpha$-$\textrm{RuCl}{}_3$},}\ }\href {\doibase
  10.1038/nphys4264} {\bibfield  {journal} {\bibinfo  {journal} {Nat. Phys.}\
  }\textbf {\bibinfo {volume} {13}},\ \bibinfo {pages} {1079--1084} (\bibinfo
  {year} {2017})}\BibitemShut {NoStop}%
\bibitem [{\citenamefont {Pezz\`{e}}\ and\ \citenamefont
  {Smerzi}(2009)}]{Pezze2009}%
  \BibitemOpen
  \bibfield  {author} {\bibinfo {author} {\bibfnamefont {L.}~\bibnamefont
  {Pezz\`{e}}}\ and\ \bibinfo {author} {\bibfnamefont {A.}~\bibnamefont
  {Smerzi}},\ }\bibfield  {title} {\enquote {\bibinfo {title} {Entanglement,
  nonlinear dynamics, and the $\textrm{H}$eisenberg limit},}\ }\href {\doibase
  10.1103/PhysRevLett.102.100401} {\bibfield  {journal} {\bibinfo  {journal}
  {Phys. Rev. Lett.}\ }\textbf {\bibinfo {volume} {102}},\ \bibinfo {pages}
  {100401} (\bibinfo {year} {2009})}\BibitemShut {NoStop}%
\bibitem [{\citenamefont {Hyllus}\ \emph {et~al.}(2012)\citenamefont {Hyllus},
  \citenamefont {Laskowski}, \citenamefont {Krischek}, \citenamefont
  {Schwemmer}, \citenamefont {Wieczorek}, \citenamefont {Weinfurter},
  \citenamefont {Pezz\`{e}},\ and\ \citenamefont {Smerzi}}]{Hyllus2012}%
  \BibitemOpen
  \bibfield  {author} {\bibinfo {author} {\bibfnamefont {P.}~\bibnamefont
  {Hyllus}}, \bibinfo {author} {\bibfnamefont {W.}~\bibnamefont {Laskowski}},
  \bibinfo {author} {\bibfnamefont {R.}~\bibnamefont {Krischek}}, \bibinfo
  {author} {\bibfnamefont {C.}~\bibnamefont {Schwemmer}}, \bibinfo {author}
  {\bibfnamefont {W.}~\bibnamefont {Wieczorek}}, \bibinfo {author}
  {\bibfnamefont {H.}~\bibnamefont {Weinfurter}}, \bibinfo {author}
  {\bibfnamefont {L.}~\bibnamefont {Pezz\`{e}}}, \ and\ \bibinfo {author}
  {\bibfnamefont {A.}~\bibnamefont {Smerzi}},\ }\bibfield  {title} {\enquote
  {\bibinfo {title} {Fisher information and multiparticle entanglement},}\
  }\href {\doibase 10.1103/PhysRevA.85.022321} {\bibfield  {journal} {\bibinfo
  {journal} {Phys. Rev. A}\ }\textbf {\bibinfo {volume} {85}},\ \bibinfo
  {pages} {022321} (\bibinfo {year} {2012})}\BibitemShut {NoStop}%
\bibitem [{\citenamefont {T\'{o}th}(2012)}]{Toth2012}%
  \BibitemOpen
  \bibfield  {author} {\bibinfo {author} {\bibfnamefont {G.}~\bibnamefont
  {T\'{o}th}},\ }\bibfield  {title} {\enquote {\bibinfo {title} {Multipartite
  entanglement and high-precision metrology},}\ }\href {\doibase
  10.1103/PhysRevA.85.022322} {\bibfield  {journal} {\bibinfo  {journal} {Phys.
  Rev. A}\ }\textbf {\bibinfo {volume} {85}},\ \bibinfo {pages} {022322}
  (\bibinfo {year} {2012})}\BibitemShut {NoStop}%
\bibitem [{\citenamefont {Strobel}\ \emph {et~al.}(2014)\citenamefont
  {Strobel}, \citenamefont {Muessel}, \citenamefont {Linnemann}, \citenamefont
  {Zibold}, \citenamefont {Hume}, \citenamefont {Pezz\`{e}}, \citenamefont
  {Smerzi},\ and\ \citenamefont {Oberthaler}}]{Strobel2014}%
  \BibitemOpen
  \bibfield  {author} {\bibinfo {author} {\bibfnamefont {H.}~\bibnamefont
  {Strobel}}, \bibinfo {author} {\bibfnamefont {W.}~\bibnamefont {Muessel}},
  \bibinfo {author} {\bibfnamefont {D.}~\bibnamefont {Linnemann}}, \bibinfo
  {author} {\bibfnamefont {T.}~\bibnamefont {Zibold}}, \bibinfo {author}
  {\bibfnamefont {D.~B.}\ \bibnamefont {Hume}}, \bibinfo {author}
  {\bibfnamefont {L.}~\bibnamefont {Pezz\`{e}}}, \bibinfo {author}
  {\bibfnamefont {A.}~\bibnamefont {Smerzi}}, \ and\ \bibinfo {author}
  {\bibfnamefont {M.~K.}\ \bibnamefont {Oberthaler}},\ }\bibfield  {title}
  {\enquote {\bibinfo {title} {Fisher information and entanglement of
  non-$\textrm{G}$aussian spin states},}\ }\href {\doibase
  10.1126/science.1250147} {\bibfield  {journal} {\bibinfo  {journal}
  {Science}\ }\textbf {\bibinfo {volume} {345}},\ \bibinfo {pages} {424--427}
  (\bibinfo {year} {2014})}\BibitemShut {NoStop}%
\bibitem [{\citenamefont {Braunstein}\ and\ \citenamefont
  {Caves}(1994)}]{BRAUNSTEIN1994}%
  \BibitemOpen
  \bibfield  {author} {\bibinfo {author} {\bibfnamefont {S.~L.}\ \bibnamefont
  {Braunstein}}\ and\ \bibinfo {author} {\bibfnamefont {C.~M.}\ \bibnamefont
  {Caves}},\ }\bibfield  {title} {\enquote {\bibinfo {title} {Statistical
  distance and the geometry of quantum states},}\ }\href {\doibase
  10.1103/PhysRevLett.72.3439} {\bibfield  {journal} {\bibinfo  {journal}
  {Phys. Rev. Lett.}\ }\textbf {\bibinfo {volume} {72}},\ \bibinfo {pages}
  {3439--3443} (\bibinfo {year} {1994})}\BibitemShut {NoStop}%
\bibitem [{\citenamefont {Giovannetti}\ \emph {et~al.}(2006)\citenamefont
  {Giovannetti}, \citenamefont {Lloyd},\ and\ \citenamefont
  {Maccone}}]{Giovannetti2006}%
  \BibitemOpen
  \bibfield  {author} {\bibinfo {author} {\bibfnamefont {V.}~\bibnamefont
  {Giovannetti}}, \bibinfo {author} {\bibfnamefont {S.}~\bibnamefont {Lloyd}},
  \ and\ \bibinfo {author} {\bibfnamefont {L.}~\bibnamefont {Maccone}},\
  }\bibfield  {title} {\enquote {\bibinfo {title} {Quantum metrology},}\ }\href
  {\doibase 10.1103/PhysRevLett.96.010401} {\bibfield  {journal} {\bibinfo
  {journal} {Phys. Rev. Lett.}\ }\textbf {\bibinfo {volume} {96}},\ \bibinfo
  {pages} {010401} (\bibinfo {year} {2006})}\BibitemShut {NoStop}%
\bibitem [{\citenamefont {Giovannetti}\ \emph {et~al.}(2011)\citenamefont
  {Giovannetti}, \citenamefont {Lloyd},\ and\ \citenamefont
  {Maccone}}]{Giovannetti2011}%
  \BibitemOpen
  \bibfield  {author} {\bibinfo {author} {\bibfnamefont {V.}~\bibnamefont
  {Giovannetti}}, \bibinfo {author} {\bibfnamefont {S.}~\bibnamefont {Lloyd}},
  \ and\ \bibinfo {author} {\bibfnamefont {L.}~\bibnamefont {Maccone}},\
  }\bibfield  {title} {\enquote {\bibinfo {title} {Advances in quantum
  metrology},}\ }\href {\doibase 10.1038/NPHOTON.2011.35} {\bibfield  {journal}
  {\bibinfo  {journal} {Nat. Photon.}\ }\textbf {\bibinfo {volume} {5}},\
  \bibinfo {pages} {222--229} (\bibinfo {year} {2011})}\BibitemShut {NoStop}%
\bibitem [{\citenamefont {Pezz\`{e}}\ \emph {et~al.}(2016)\citenamefont
  {Pezz\`{e}}, \citenamefont {Li}, \citenamefont {Li},\ and\ \citenamefont
  {Smerzi}}]{Pezze2016}%
  \BibitemOpen
  \bibfield  {author} {\bibinfo {author} {\bibfnamefont {L.}~\bibnamefont
  {Pezz\`{e}}}, \bibinfo {author} {\bibfnamefont {Y.}~\bibnamefont {Li}},
  \bibinfo {author} {\bibfnamefont {W.~D.}\ \bibnamefont {Li}}, \ and\ \bibinfo
  {author} {\bibfnamefont {A.}~\bibnamefont {Smerzi}},\ }\bibfield  {title}
  {\enquote {\bibinfo {title} {Witnessing entanglement without entanglement
  witness operators},}\ }\href {\doibase 10.1073/pnas.1603346113} {\bibfield
  {journal} {\bibinfo  {journal} {PNAS}\ }\textbf {\bibinfo {volume} {113}},\
  \bibinfo {pages} {11459--11464} (\bibinfo {year} {2016})}\BibitemShut
  {NoStop}%
\bibitem [{\citenamefont {Gross}\ \emph {et~al.}(2010)\citenamefont {Gross},
  \citenamefont {Zibold}, \citenamefont {Nicklas}, \citenamefont {Esteve},\
  and\ \citenamefont {Oberthaler}}]{Gross2010}%
  \BibitemOpen
  \bibfield  {author} {\bibinfo {author} {\bibfnamefont {C.}~\bibnamefont
  {Gross}}, \bibinfo {author} {\bibfnamefont {T.}~\bibnamefont {Zibold}},
  \bibinfo {author} {\bibfnamefont {E.}~\bibnamefont {Nicklas}}, \bibinfo
  {author} {\bibfnamefont {J.}~\bibnamefont {Esteve}}, \ and\ \bibinfo {author}
  {\bibfnamefont {M.~K.}\ \bibnamefont {Oberthaler}},\ }\bibfield  {title}
  {\enquote {\bibinfo {title} {Nonlinear atom interferometer surpasses
  classical precision limit},}\ }\href {\doibase 10.1038/nature08919}
  {\bibfield  {journal} {\bibinfo  {journal} {Nature}\ }\textbf {\bibinfo
  {volume} {464}},\ \bibinfo {pages} {1165--1169} (\bibinfo {year}
  {2010})}\BibitemShut {NoStop}%
\bibitem [{\citenamefont {Lucke}\ \emph {et~al.}(2011)\citenamefont {Lucke},
  \citenamefont {Scherer}, \citenamefont {Kruse}, \citenamefont {Pezz\`{e}},
  \citenamefont {Deuretzbacher}, \citenamefont {Hyllus}, \citenamefont {Topic},
  \citenamefont {Peise}, \citenamefont {Ertmer}, \citenamefont {Arlt},
  \citenamefont {Santos}, \citenamefont {Smerzi},\ and\ \citenamefont
  {Klempt}}]{Lucke2011}%
  \BibitemOpen
  \bibfield  {author} {\bibinfo {author} {\bibfnamefont {B.}~\bibnamefont
  {Lucke}}, \bibinfo {author} {\bibfnamefont {M.}~\bibnamefont {Scherer}},
  \bibinfo {author} {\bibfnamefont {J.}~\bibnamefont {Kruse}}, \bibinfo
  {author} {\bibfnamefont {L.}~\bibnamefont {Pezz\`{e}}}, \bibinfo {author}
  {\bibfnamefont {F.}~\bibnamefont {Deuretzbacher}}, \bibinfo {author}
  {\bibfnamefont {P.}~\bibnamefont {Hyllus}}, \bibinfo {author} {\bibfnamefont
  {O.}~\bibnamefont {Topic}}, \bibinfo {author} {\bibfnamefont
  {J.}~\bibnamefont {Peise}}, \bibinfo {author} {\bibfnamefont
  {W.}~\bibnamefont {Ertmer}}, \bibinfo {author} {\bibfnamefont
  {J.}~\bibnamefont {Arlt}}, \bibinfo {author} {\bibfnamefont {L.}~\bibnamefont
  {Santos}}, \bibinfo {author} {\bibfnamefont {A.}~\bibnamefont {Smerzi}}, \
  and\ \bibinfo {author} {\bibfnamefont {C.}~\bibnamefont {Klempt}},\
  }\bibfield  {title} {\enquote {\bibinfo {title} {Twin matter waves for
  interferometry beyond the classical limit},}\ }\href {\doibase
  10.1126/science.1208798} {\bibfield  {journal} {\bibinfo  {journal}
  {Science}\ }\textbf {\bibinfo {volume} {334}},\ \bibinfo {pages} {773--776}
  (\bibinfo {year} {2011})}\BibitemShut {NoStop}%
\bibitem [{\citenamefont {Pezz\`{e}}\ \emph {et~al.}(2017)\citenamefont
  {Pezz\`{e}}, \citenamefont {Gabbrielli}, \citenamefont {Lepori},\ and\
  \citenamefont {Smerzi}}]{Pezze2017}%
  \BibitemOpen
  \bibfield  {author} {\bibinfo {author} {\bibfnamefont {L.}~\bibnamefont
  {Pezz\`{e}}}, \bibinfo {author} {\bibfnamefont {M.}~\bibnamefont
  {Gabbrielli}}, \bibinfo {author} {\bibfnamefont {L.}~\bibnamefont {Lepori}},
  \ and\ \bibinfo {author} {\bibfnamefont {A.}~\bibnamefont {Smerzi}},\
  }\bibfield  {title} {\enquote {\bibinfo {title} {Multipartite entanglement in
  topological quantum phases},}\ }\href {\doibase
  10.1103/PhysRevLett.119.250401} {\bibfield  {journal} {\bibinfo  {journal}
  {Phys. Rev. Lett.}\ }\textbf {\bibinfo {volume} {119}},\ \bibinfo {pages}
  {250401} (\bibinfo {year} {2017})}\BibitemShut {NoStop}%
\bibitem [{\citenamefont {Chiu}\ \emph {et~al.}(2016)\citenamefont {Chiu},
  \citenamefont {Teo}, \citenamefont {Schnyder},\ and\ \citenamefont
  {Ryu}}]{Chiu2016}%
  \BibitemOpen
  \bibfield  {author} {\bibinfo {author} {\bibfnamefont {C.~K.}\ \bibnamefont
  {Chiu}}, \bibinfo {author} {\bibfnamefont {J.~C.~Y.}\ \bibnamefont {Teo}},
  \bibinfo {author} {\bibfnamefont {A.~P.}\ \bibnamefont {Schnyder}}, \ and\
  \bibinfo {author} {\bibfnamefont {S.}~\bibnamefont {Ryu}},\ }\bibfield
  {title} {\enquote {\bibinfo {title} {Classification of topological quantum
  matter with symmetries},}\ }\href {\doibase 10.1103/RevModPhys.88.035005}
  {\bibfield  {journal} {\bibinfo  {journal} {Rev. Mod. Phys.}\ }\textbf
  {\bibinfo {volume} {88}},\ \bibinfo {pages} {035005} (\bibinfo {year}
  {2016})}\BibitemShut {NoStop}%
\bibitem [{\citenamefont {Li}\ \emph {et~al.}(2016)\citenamefont {Li},
  \citenamefont {Yang},\ and\ \citenamefont {Chen}}]{Li2016}%
  \BibitemOpen
  \bibfield  {author} {\bibinfo {author} {\bibfnamefont {L.~H.}\ \bibnamefont
  {Li}}, \bibinfo {author} {\bibfnamefont {C.}~\bibnamefont {Yang}}, \ and\
  \bibinfo {author} {\bibfnamefont {S.}~\bibnamefont {Chen}},\ }\bibfield
  {title} {\enquote {\bibinfo {title} {Topological invariants for phase
  transition points of one-dimensional $\textrm{Z}_2$ topological systems},}\
  }\href {\doibase 10.1140/epjb/e2016-70325-x} {\bibfield  {journal} {\bibinfo
  {journal} {Eur. Phys. J. B}\ }\textbf {\bibinfo {volume} {89}},\ \bibinfo
  {pages} {195} (\bibinfo {year} {2016})}\BibitemShut {NoStop}%
\bibitem [{\citenamefont {Tewari}\ and\ \citenamefont
  {Sau}(2012)}]{Tewari2012}%
  \BibitemOpen
  \bibfield  {author} {\bibinfo {author} {\bibfnamefont {S.}~\bibnamefont
  {Tewari}}\ and\ \bibinfo {author} {\bibfnamefont {J.~D.}\ \bibnamefont
  {Sau}},\ }\bibfield  {title} {\enquote {\bibinfo {title} {Topological
  invariants for spin-orbit coupled superconductor nanowires},}\ }\href
  {\doibase 10.1103/PhysRevLett.109.150408} {\bibfield  {journal} {\bibinfo
  {journal} {Phys. Rev. Lett.}\ }\textbf {\bibinfo {volume} {109}},\ \bibinfo
  {pages} {150408} (\bibinfo {year} {2012})}\BibitemShut {NoStop}%
\bibitem [{\citenamefont {Fradkin}\ and\ \citenamefont
  {Susskind}(1978)}]{FRADKIN1978}%
  \BibitemOpen
  \bibfield  {author} {\bibinfo {author} {\bibfnamefont {E.}~\bibnamefont
  {Fradkin}}\ and\ \bibinfo {author} {\bibfnamefont {L.}~\bibnamefont
  {Susskind}},\ }\bibfield  {title} {\enquote {\bibinfo {title} {Order and
  disorder in gauge systems and magnets},}\ }\href {\doibase
  10.1103/PhysRevD.17.2637} {\bibfield  {journal} {\bibinfo  {journal} {Phys.
  Rev. D}\ }\textbf {\bibinfo {volume} {17}},\ \bibinfo {pages} {2637--2658}
  (\bibinfo {year} {1978})}\BibitemShut {NoStop}%
\bibitem [{\citenamefont {Feng}\ \emph {et~al.}(2007)\citenamefont {Feng},
  \citenamefont {Zhang},\ and\ \citenamefont {Xiang}}]{Feng2007}%
  \BibitemOpen
  \bibfield  {author} {\bibinfo {author} {\bibfnamefont {X.~Y.}\ \bibnamefont
  {Feng}}, \bibinfo {author} {\bibfnamefont {G.~M.}\ \bibnamefont {Zhang}}, \
  and\ \bibinfo {author} {\bibfnamefont {T.}~\bibnamefont {Xiang}},\ }\bibfield
   {title} {\enquote {\bibinfo {title} {Topological characterization of quantum
  phase transitions in a spin-1/2 model},}\ }\href {\doibase
  10.1103/PhysRevLett.98.087204} {\bibfield  {journal} {\bibinfo  {journal}
  {Phys. Rev. Lett.}\ }\textbf {\bibinfo {volume} {98}},\ \bibinfo {pages}
  {087204} (\bibinfo {year} {2007})}\BibitemShut {NoStop}%
\bibitem [{\citenamefont {Qin}\ \emph {et~al.}(2017)\citenamefont {Qin},
  \citenamefont {He}, \citenamefont {You}, \citenamefont {Lu}, \citenamefont
  {Sen}, \citenamefont {Sandvik}, \citenamefont {Xu},\ and\ \citenamefont
  {Meng}}]{Qin2017}%
  \BibitemOpen
  \bibfield  {author} {\bibinfo {author} {\bibfnamefont {Y.~Q.}\ \bibnamefont
  {Qin}}, \bibinfo {author} {\bibfnamefont {Y.~Y.}\ \bibnamefont {He}},
  \bibinfo {author} {\bibfnamefont {Y.~Z.}\ \bibnamefont {You}}, \bibinfo
  {author} {\bibfnamefont {Z.~Y.}\ \bibnamefont {Lu}}, \bibinfo {author}
  {\bibfnamefont {A.}~\bibnamefont {Sen}}, \bibinfo {author} {\bibfnamefont
  {A.~W.}\ \bibnamefont {Sandvik}}, \bibinfo {author} {\bibfnamefont {C.~K.}\
  \bibnamefont {Xu}}, \ and\ \bibinfo {author} {\bibfnamefont {Z.~Y.}\
  \bibnamefont {Meng}},\ }\bibfield  {title} {\enquote {\bibinfo {title}
  {Duality between the deconfined quantum-critical point and the bosonic
  topological transition},}\ }\href {\doibase 10.1103/PhysRevX.7.031052}
  {\bibfield  {journal} {\bibinfo  {journal} {Phys. Rev. X}\ }\textbf {\bibinfo
  {volume} {7}},\ \bibinfo {pages} {031052} (\bibinfo {year}
  {2017})}\BibitemShut {NoStop}%
\bibitem [{\citenamefont {Smacchia}\ \emph {et~al.}(2011)\citenamefont
  {Smacchia}, \citenamefont {Amico}, \citenamefont {Facchi}, \citenamefont
  {Fazio}, \citenamefont {Florio}, \citenamefont {Pascazio},\ and\
  \citenamefont {Vedral}}]{Smacchia2011}%
  \BibitemOpen
  \bibfield  {author} {\bibinfo {author} {\bibfnamefont {P.}~\bibnamefont
  {Smacchia}}, \bibinfo {author} {\bibfnamefont {L.}~\bibnamefont {Amico}},
  \bibinfo {author} {\bibfnamefont {P.}~\bibnamefont {Facchi}}, \bibinfo
  {author} {\bibfnamefont {R.}~\bibnamefont {Fazio}}, \bibinfo {author}
  {\bibfnamefont {G.}~\bibnamefont {Florio}}, \bibinfo {author} {\bibfnamefont
  {S.}~\bibnamefont {Pascazio}}, \ and\ \bibinfo {author} {\bibfnamefont
  {V.}~\bibnamefont {Vedral}},\ }\bibfield  {title} {\enquote {\bibinfo {title}
  {Statistical mechanics of the cluster $\textrm{I}$sing model},}\ }\href
  {\doibase 10.1103/PhysRevA.84.022304} {\bibfield  {journal} {\bibinfo
  {journal} {Phys. Rev. A}\ }\textbf {\bibinfo {volume} {84}},\ \bibinfo
  {pages} {022304} (\bibinfo {year} {2011})}\BibitemShut {NoStop}%
\bibitem [{\citenamefont {Venuti}\ and\ \citenamefont
  {Roncaglia}(2005)}]{Venuti2005}%
  \BibitemOpen
  \bibfield  {author} {\bibinfo {author} {\bibfnamefont {L.~C.}\ \bibnamefont
  {Venuti}}\ and\ \bibinfo {author} {\bibfnamefont {M.}~\bibnamefont
  {Roncaglia}},\ }\bibfield  {title} {\enquote {\bibinfo {title} {Analytic
  relations between localizable entanglement and string correlations in spin
  systems},}\ }\href {\doibase 10.1103/PhysRevLett.94.207207} {\bibfield
  {journal} {\bibinfo  {journal} {Phys. Rev. Lett.}\ }\textbf {\bibinfo
  {volume} {94}},\ \bibinfo {pages} {207207} (\bibinfo {year}
  {2005})}\BibitemShut {NoStop}%
\bibitem [{\citenamefont {Cui}\ \emph {et~al.}(2013)\citenamefont {Cui},
  \citenamefont {Amico}, \citenamefont {Fan}, \citenamefont {Gu}, \citenamefont
  {Hamma},\ and\ \citenamefont {Vedral}}]{Cui2013}%
  \BibitemOpen
  \bibfield  {author} {\bibinfo {author} {\bibfnamefont {J.}~\bibnamefont
  {Cui}}, \bibinfo {author} {\bibfnamefont {L.}~\bibnamefont {Amico}}, \bibinfo
  {author} {\bibfnamefont {H.}~\bibnamefont {Fan}}, \bibinfo {author}
  {\bibfnamefont {M.}~\bibnamefont {Gu}}, \bibinfo {author} {\bibfnamefont
  {A.}~\bibnamefont {Hamma}}, \ and\ \bibinfo {author} {\bibfnamefont
  {V.}~\bibnamefont {Vedral}},\ }\bibfield  {title} {\enquote {\bibinfo {title}
  {Local characterization of one-dimensional topologically ordered states},}\
  }\href {\doibase 10.1103/PhysRevB.88.125117} {\bibfield  {journal} {\bibinfo
  {journal} {Phys. Rev. B}\ }\textbf {\bibinfo {volume} {88}},\ \bibinfo
  {pages} {125117} (\bibinfo {year} {2013})}\BibitemShut {NoStop}%
\bibitem [{\citenamefont {Kitaev}(2006)}]{Kitaev2006a}%
  \BibitemOpen
  \bibfield  {author} {\bibinfo {author} {\bibfnamefont {A.}~\bibnamefont
  {Kitaev}},\ }\bibfield  {title} {\enquote {\bibinfo {title} {Anyons in an
  exactly solved model and beyond},}\ }\href {\doibase
  10.1016/j.aop.2005.10.005} {\bibfield  {journal} {\bibinfo  {journal} {Ann.
  Phys.}\ }\textbf {\bibinfo {volume} {321}},\ \bibinfo {pages} {2--111}
  (\bibinfo {year} {2006})}\BibitemShut {NoStop}%
\bibitem [{\citenamefont {Suzuki}(1971)}]{Suzuki1971}%
  \BibitemOpen
  \bibfield  {author} {\bibinfo {author} {\bibfnamefont {M.}~\bibnamefont
  {Suzuki}},\ }\bibfield  {title} {\enquote {\bibinfo {title} {Relationship
  among exactly soluble models of critical phenomena .1. 2{D} {I}sing model,
  dimer problem and generalized {XY}-model},}\ }\href {\doibase
  10.1143/PTP.46.1337} {\bibfield  {journal} {\bibinfo  {journal} {Prog. Theor.
  Phys.}\ }\textbf {\bibinfo {volume} {46}},\ \bibinfo {pages} {1337} (\bibinfo
  {year} {1971})}\BibitemShut {NoStop}%
\bibitem [{\citenamefont {Niu}\ \emph {et~al.}(2012)\citenamefont {Niu},
  \citenamefont {Chung}, \citenamefont {Hsu}, \citenamefont {Mandal},
  \citenamefont {Raghu},\ and\ \citenamefont {Chakravarty}}]{Niu2012}%
  \BibitemOpen
  \bibfield  {author} {\bibinfo {author} {\bibfnamefont {Y.~Z.}\ \bibnamefont
  {Niu}}, \bibinfo {author} {\bibfnamefont {S.~B.}\ \bibnamefont {Chung}},
  \bibinfo {author} {\bibfnamefont {C.~H.}\ \bibnamefont {Hsu}}, \bibinfo
  {author} {\bibfnamefont {I.}~\bibnamefont {Mandal}}, \bibinfo {author}
  {\bibfnamefont {S.}~\bibnamefont {Raghu}}, \ and\ \bibinfo {author}
  {\bibfnamefont {S.}~\bibnamefont {Chakravarty}},\ }\bibfield  {title}
  {\enquote {\bibinfo {title} {Majorana zero modes in a quantum
  $\textrm{I}$sing chain with longer-ranged interactions},}\ }\href {\doibase
  10.1103/PhysRevB.85.035110} {\bibfield  {journal} {\bibinfo  {journal} {Phys.
  Rev. B}\ }\textbf {\bibinfo {volume} {85}},\ \bibinfo {pages} {035110}
  (\bibinfo {year} {2012})}\BibitemShut {NoStop}%
\bibitem [{\citenamefont {Zhang}\ and\ \citenamefont {Song}(2015)}]{Zhang2015}%
  \BibitemOpen
  \bibfield  {author} {\bibinfo {author} {\bibfnamefont {G.}~\bibnamefont
  {Zhang}}\ and\ \bibinfo {author} {\bibfnamefont {Z.}~\bibnamefont {Song}},\
  }\bibfield  {title} {\enquote {\bibinfo {title} {Topological characterization
  of extended quantum $\textrm{I}$sing models},}\ }\href {\doibase
  10.1103/PhysRevLett.115.177204} {\bibfield  {journal} {\bibinfo  {journal}
  {Phys. Rev. Lett.}\ }\textbf {\bibinfo {volume} {115}},\ \bibinfo {pages}
  {177204} (\bibinfo {year} {2015})}\BibitemShut {NoStop}%
\bibitem [{\citenamefont {Zhang}\ \emph {et~al.}(2017)\citenamefont {Zhang},
  \citenamefont {Li},\ and\ \citenamefont {Song}}]{Zhang2017a}%
  \BibitemOpen
  \bibfield  {author} {\bibinfo {author} {\bibfnamefont {G.}~\bibnamefont
  {Zhang}}, \bibinfo {author} {\bibfnamefont {C.}~\bibnamefont {Li}}, \ and\
  \bibinfo {author} {\bibfnamefont {Z.}~\bibnamefont {Song}},\ }\bibfield
  {title} {\enquote {\bibinfo {title} {Majorana charges, winding numbers and
  $\textrm{C}$hern numbers in quantum $\textrm{I}$sing models},}\ }\href
  {\doibase 10.1038/s41598-017-08323-0} {\bibfield  {journal} {\bibinfo
  {journal} {Sci. Rep.}\ }\textbf {\bibinfo {volume} {7}},\ \bibinfo {pages}
  {8176} (\bibinfo {year} {2017})}\BibitemShut {NoStop}%
\bibitem [{\citenamefont {Zhang}\ and\ \citenamefont {Guo}(2017)}]{Zhang2017}%
  \BibitemOpen
  \bibfield  {author} {\bibinfo {author} {\bibfnamefont {X.~Z.}\ \bibnamefont
  {Zhang}}\ and\ \bibinfo {author} {\bibfnamefont {J.~L.}\ \bibnamefont
  {Guo}},\ }\bibfield  {title} {\enquote {\bibinfo {title} {Quantum correlation
  and quantum phase transition in the one-dimensional extended $\textrm{I}$sing
  model},}\ }\href {\doibase 10.1007/s11128-017-1670-3} {\bibfield  {journal}
  {\bibinfo  {journal} {Quantum Inf. Process.}\ }\textbf {\bibinfo {volume}
  {16}},\ \bibinfo {pages} {223} (\bibinfo {year} {2017})}\BibitemShut
  {NoStop}%
\bibitem [{SM()}]{SM}%
  \BibitemOpen
  \href@noop {} {\emph {\bibinfo {title} {\emph{Supplementary Material is
  available at http://}}}}\BibitemShut {NoStop}%
\bibitem [{\citenamefont {Ahlfors}(1953)}]{Ahlfors1953}%
  \BibitemOpen
  \bibfield  {author} {\bibinfo {author} {\bibfnamefont {L.~V.}\ \bibnamefont
  {Ahlfors}},\ }\href@noop {} {\emph {\bibinfo {title} {Complex analysis: an
  introduction to the theory of analytic functions of one complex variable}}}\
  (\bibinfo  {publisher} {New York, London},\ \bibinfo {year}
  {1953})\BibitemShut {NoStop}%
\bibitem [{\citenamefont {Fendley}(2012)}]{Fendley2012}%
  \BibitemOpen
  \bibfield  {author} {\bibinfo {author} {\bibfnamefont {P.}~\bibnamefont
  {Fendley}},\ }\bibfield  {title} {\enquote {\bibinfo {title} {Parafermionic
  edge zero modes in {Z}(n)-invariant spin chains},}\ }\href {\doibase
  10.1088/1742-5468/2012/11/P11020} {\bibfield  {journal} {\bibinfo  {journal}
  {J. Stat. Mech.}\ }\textbf {\bibinfo {volume} {2012}},\ \bibinfo {pages}
  {P11020} (\bibinfo {year} {2012})}\BibitemShut {NoStop}%
\bibitem [{\citenamefont {Viyuela}\ \emph {et~al.}(2016)\citenamefont
  {Viyuela}, \citenamefont {Vodola}, \citenamefont {Pupillo},\ and\
  \citenamefont {Martin-Delgado}}]{Viyuela2016}%
  \BibitemOpen
  \bibfield  {author} {\bibinfo {author} {\bibfnamefont {O.}~\bibnamefont
  {Viyuela}}, \bibinfo {author} {\bibfnamefont {D.}~\bibnamefont {Vodola}},
  \bibinfo {author} {\bibfnamefont {G.}~\bibnamefont {Pupillo}}, \ and\
  \bibinfo {author} {\bibfnamefont {M.~A.}\ \bibnamefont {Martin-Delgado}},\
  }\bibfield  {title} {\enquote {\bibinfo {title} {Topological massive {D}irac
  edge modes and long-range superconducting {H}amiltonians},}\ }\href {\doibase
  10.1103/PhysRevB.94.125121} {\bibfield  {journal} {\bibinfo  {journal} {Phys.
  Rev. B}\ }\textbf {\bibinfo {volume} {94}},\ \bibinfo {pages} {125121}
  (\bibinfo {year} {2016})}\BibitemShut {NoStop}%
\bibitem [{\citenamefont {Kitagawa}\ and\ \citenamefont
  {Ueda}(1993)}]{KITAGAWA1993}%
  \BibitemOpen
  \bibfield  {author} {\bibinfo {author} {\bibfnamefont {M.}~\bibnamefont
  {Kitagawa}}\ and\ \bibinfo {author} {\bibfnamefont {M.}~\bibnamefont
  {Ueda}},\ }\bibfield  {title} {\enquote {\bibinfo {title} {Squeezed spin
  states},}\ }\href {\doibase 10.1103/PhysRevA.47.5138} {\bibfield  {journal}
  {\bibinfo  {journal} {Phys. Rev. A}\ }\textbf {\bibinfo {volume} {47}},\
  \bibinfo {pages} {5138--5143} (\bibinfo {year} {1993})}\BibitemShut {NoStop}%
\bibitem [{\citenamefont {Ma}\ \emph {et~al.}(2011)\citenamefont {Ma},
  \citenamefont {Wang}, \citenamefont {Sun},\ and\ \citenamefont
  {Nori}}]{Ma2011}%
  \BibitemOpen
  \bibfield  {author} {\bibinfo {author} {\bibfnamefont {J.}~\bibnamefont
  {Ma}}, \bibinfo {author} {\bibfnamefont {X.~G.}\ \bibnamefont {Wang}},
  \bibinfo {author} {\bibfnamefont {C.~P.}\ \bibnamefont {Sun}}, \ and\
  \bibinfo {author} {\bibfnamefont {F.}~\bibnamefont {Nori}},\ }\bibfield
  {title} {\enquote {\bibinfo {title} {Quantum spin squeezing},}\ }\href
  {\doibase 10.1016/j.physrep.2011.08.003} {\bibfield  {journal} {\bibinfo
  {journal} {Phys. Rep.}\ }\textbf {\bibinfo {volume} {509}},\ \bibinfo {pages}
  {89--165} (\bibinfo {year} {2011})}\BibitemShut {NoStop}%
\bibitem [{\citenamefont {Hauke}\ \emph {et~al.}(2016)\citenamefont {Hauke},
  \citenamefont {Heyl}, \citenamefont {Tagliacozzo},\ and\ \citenamefont
  {Zoller}}]{Hauke2016}%
  \BibitemOpen
  \bibfield  {author} {\bibinfo {author} {\bibfnamefont {P.}~\bibnamefont
  {Hauke}}, \bibinfo {author} {\bibfnamefont {M.}~\bibnamefont {Heyl}},
  \bibinfo {author} {\bibfnamefont {L.}~\bibnamefont {Tagliacozzo}}, \ and\
  \bibinfo {author} {\bibfnamefont {P.}~\bibnamefont {Zoller}},\ }\bibfield
  {title} {\enquote {\bibinfo {title} {Measuring multipartite entanglement
  through dynamic susceptibilities},}\ }\href {\doibase 10.1038/NPHYS3700}
  {\bibfield  {journal} {\bibinfo  {journal} {Nat. Phys.}\ }\textbf {\bibinfo
  {volume} {12}},\ \bibinfo {pages} {778--782} (\bibinfo {year}
  {2016})}\BibitemShut {NoStop}%
\bibitem [{\citenamefont {Suzuki}\ \emph {et~al.}(2012)\citenamefont {Suzuki},
  \citenamefont {Inoue},\ and\ \citenamefont {Chakrabarti}}]{Suzuki2012}%
  \BibitemOpen
  \bibfield  {author} {\bibinfo {author} {\bibfnamefont {S.}~\bibnamefont
  {Suzuki}}, \bibinfo {author} {\bibfnamefont {J.-I.}\ \bibnamefont {Inoue}}, \
  and\ \bibinfo {author} {\bibfnamefont {B.~K.}\ \bibnamefont {Chakrabarti}},\
  }\href@noop {} {\emph {\bibinfo {title} {Quantum {I}sing phases and
  transitions in transverse {I}sing models}}},\ Vol.\ \bibinfo {volume} {862}\
  (\bibinfo  {publisher} {Springer},\ \bibinfo {year} {2012})\BibitemShut
  {NoStop}%
\bibitem [{\citenamefont {Cobanera}\ \emph {et~al.}(2011)\citenamefont
  {Cobanera}, \citenamefont {Ortiz},\ and\ \citenamefont
  {Nussinov}}]{Cobanera2011}%
  \BibitemOpen
  \bibfield  {author} {\bibinfo {author} {\bibfnamefont {E.}~\bibnamefont
  {Cobanera}}, \bibinfo {author} {\bibfnamefont {G.}~\bibnamefont {Ortiz}}, \
  and\ \bibinfo {author} {\bibfnamefont {Z.}~\bibnamefont {Nussinov}},\
  }\bibfield  {title} {\enquote {\bibinfo {title} {The bond-algebraic approach
  to dualities},}\ }\href {\doibase 10.1080/00018732.2011.619814} {\bibfield
  {journal} {\bibinfo  {journal} {Adv. Phys.}\ }\textbf {\bibinfo {volume}
  {60}},\ \bibinfo {pages} {679--798} (\bibinfo {year} {2011})}\BibitemShut
  {NoStop}%
\bibitem [{\citenamefont {Fidkowski}\ and\ \citenamefont
  {Kitaev}(2011)}]{Fidkowski2011}%
  \BibitemOpen
  \bibfield  {author} {\bibinfo {author} {\bibfnamefont {L.}~\bibnamefont
  {Fidkowski}}\ and\ \bibinfo {author} {\bibfnamefont {A.}~\bibnamefont
  {Kitaev}},\ }\bibfield  {title} {\enquote {\bibinfo {title} {Topological
  phases of fermions in one dimension},}\ }\href {\doibase
  10.1103/PhysRevB.83.075103} {\bibfield  {journal} {\bibinfo  {journal} {Phys.
  Rev. B}\ }\textbf {\bibinfo {volume} {83}},\ \bibinfo {pages} {075103}
  (\bibinfo {year} {2011})}\BibitemShut {NoStop}%
\bibitem [{\citenamefont {Kitaev}(2001)}]{Kitaev2001}%
  \BibitemOpen
  \bibfield  {author} {\bibinfo {author} {\bibfnamefont {A.~Yu}\ \bibnamefont
  {Kitaev}},\ }\bibfield  {title} {\enquote {\bibinfo {title} {Unpaired
  $\textrm{M}$ajorana fermions in quantum wires},}\ }\href@noop {} {\bibfield
  {journal} {\bibinfo  {journal} {Phys. Usp.}\ }\textbf {\bibinfo {volume}
  {44}},\ \bibinfo {pages} {131} (\bibinfo {year} {2001})}\BibitemShut
  {NoStop}%
\bibitem [{\citenamefont {Stoferle}\ \emph {et~al.}(2004)\citenamefont
  {Stoferle}, \citenamefont {Moritz}, \citenamefont {Schori}, \citenamefont
  {Kohl},\ and\ \citenamefont {Esslinger}}]{Stoferle2004}%
  \BibitemOpen
  \bibfield  {author} {\bibinfo {author} {\bibfnamefont {T.}~\bibnamefont
  {Stoferle}}, \bibinfo {author} {\bibfnamefont {H.}~\bibnamefont {Moritz}},
  \bibinfo {author} {\bibfnamefont {C.}~\bibnamefont {Schori}}, \bibinfo
  {author} {\bibfnamefont {M.}~\bibnamefont {Kohl}}, \ and\ \bibinfo {author}
  {\bibfnamefont {T.}~\bibnamefont {Esslinger}},\ }\bibfield  {title} {\enquote
  {\bibinfo {title} {Transition from a strongly interacting 1$\textrm{D}$
  superfluid to a {M}ott insulator},}\ }\href {\doibase
  10.1103/PhysRevLett.92.130403} {\bibfield  {journal} {\bibinfo  {journal}
  {Phys. Rev. Lett.}\ }\textbf {\bibinfo {volume} {92}},\ \bibinfo {pages}
  {130403} (\bibinfo {year} {2004})}\BibitemShut {NoStop}%
\bibitem [{\citenamefont {Ernst}\ \emph {et~al.}(2010)\citenamefont {Ernst},
  \citenamefont {Gotze}, \citenamefont {Krauser}, \citenamefont {Pyka},
  \citenamefont {Luhmann}, \citenamefont {Pfannkuche},\ and\ \citenamefont
  {Sengstock}}]{Ernst2010}%
  \BibitemOpen
  \bibfield  {author} {\bibinfo {author} {\bibfnamefont {P.~T.}\ \bibnamefont
  {Ernst}}, \bibinfo {author} {\bibfnamefont {S.}~\bibnamefont {Gotze}},
  \bibinfo {author} {\bibfnamefont {J.~S.}\ \bibnamefont {Krauser}}, \bibinfo
  {author} {\bibfnamefont {K.}~\bibnamefont {Pyka}}, \bibinfo {author}
  {\bibfnamefont {D.~S.}\ \bibnamefont {Luhmann}}, \bibinfo {author}
  {\bibfnamefont {D.}~\bibnamefont {Pfannkuche}}, \ and\ \bibinfo {author}
  {\bibfnamefont {K.}~\bibnamefont {Sengstock}},\ }\bibfield  {title} {\enquote
  {\bibinfo {title} {Probing superfluids in optical lattices by
  momentum-resolved $\textrm{B}$ragg spectroscopy},}\ }\href {\doibase
  10.1038/NPHYS1476} {\bibfield  {journal} {\bibinfo  {journal} {Nat. Phys.}\
  }\textbf {\bibinfo {volume} {6}},\ \bibinfo {pages} {56--61} (\bibinfo {year}
  {2010})}\BibitemShut {NoStop}%
\bibitem [{\citenamefont {Shirane}\ \emph {et~al.}(2002)\citenamefont
  {Shirane}, \citenamefont {Shapiro},\ and\ \citenamefont
  {Tranquada}}]{Shirane2002}%
  \BibitemOpen
  \bibfield  {author} {\bibinfo {author} {\bibfnamefont {G.}~\bibnamefont
  {Shirane}}, \bibinfo {author} {\bibfnamefont {S.~M.}\ \bibnamefont
  {Shapiro}}, \ and\ \bibinfo {author} {\bibfnamefont {J.~M.}\ \bibnamefont
  {Tranquada}},\ }\href@noop {} {\emph {\bibinfo {title} {Neutron Scattering
  with a Triple-Axis Spectrometer, Basic Techniques}}}\ (\bibinfo  {publisher}
  {Cambridge Univ. Press},\ \bibinfo {year} {2002})\BibitemShut {NoStop}%
\bibitem [{\citenamefont {Yang}\ \emph {et~al.}(2008)\citenamefont {Yang},
  \citenamefont {Gu}, \citenamefont {Sun},\ and\ \citenamefont
  {Lin}}]{Yang2008}%
  \BibitemOpen
  \bibfield  {author} {\bibinfo {author} {\bibfnamefont {S.}~\bibnamefont
  {Yang}}, \bibinfo {author} {\bibfnamefont {S.~J.}\ \bibnamefont {Gu}},
  \bibinfo {author} {\bibfnamefont {C.~P.}\ \bibnamefont {Sun}}, \ and\
  \bibinfo {author} {\bibfnamefont {H.~Q.}\ \bibnamefont {Lin}},\ }\bibfield
  {title} {\enquote {\bibinfo {title} {Fidelity susceptibility and long-range
  correlation in the {K}itaev honeycomb model},}\ }\href {\doibase
  10.1103/PhysRevA.78.012304} {\bibfield  {journal} {\bibinfo  {journal} {Phys.
  Rev. A}\ }\textbf {\bibinfo {volume} {78}},\ \bibinfo {pages} {012304}
  (\bibinfo {year} {2008})}\BibitemShut {NoStop}%
\bibitem [{\citenamefont {Abasto}\ and\ \citenamefont
  {Zanardi}(2009)}]{Abasto2009}%
  \BibitemOpen
  \bibfield  {author} {\bibinfo {author} {\bibfnamefont {D.~F.}\ \bibnamefont
  {Abasto}}\ and\ \bibinfo {author} {\bibfnamefont {P.}~\bibnamefont
  {Zanardi}},\ }\bibfield  {title} {\enquote {\bibinfo {title} {Thermal states
  of the {K}itaev honeycomb model: Bures metric analysis},}\ }\href {\doibase
  10.1103/PhysRevA.79.012321} {\bibfield  {journal} {\bibinfo  {journal} {Phys.
  Rev. A}\ }\textbf {\bibinfo {volume} {79}},\ \bibinfo {pages} {012321}
  (\bibinfo {year} {2009})}\BibitemShut {NoStop}%
\bibitem [{\citenamefont {Shi}\ \emph {et~al.}(2009)\citenamefont {Shi},
  \citenamefont {Yu}, \citenamefont {You},\ and\ \citenamefont
  {Nori}}]{Shi2009}%
  \BibitemOpen
  \bibfield  {author} {\bibinfo {author} {\bibfnamefont {X.~F.}\ \bibnamefont
  {Shi}}, \bibinfo {author} {\bibfnamefont {Y.}~\bibnamefont {Yu}}, \bibinfo
  {author} {\bibfnamefont {J.~Q.}\ \bibnamefont {You}}, \ and\ \bibinfo
  {author} {\bibfnamefont {F.}~\bibnamefont {Nori}},\ }\bibfield  {title}
  {\enquote {\bibinfo {title} {Topological quantum phase transition in the
  extended $\textrm{K}$itaev spin model},}\ }\href {\doibase
  10.1103/PhysRevB.79.134431} {\bibfield  {journal} {\bibinfo  {journal} {Phys.
  Rev. B}\ }\textbf {\bibinfo {volume} {79}},\ \bibinfo {pages} {134431}
  (\bibinfo {year} {2009})}\BibitemShut {NoStop}%
\bibitem [{\citenamefont {Chen}\ \emph {et~al.}(2016)\citenamefont {Chen},
  \citenamefont {Cui}, \citenamefont {Zhang},\ and\ \citenamefont
  {Fan}}]{Chen2016}%
  \BibitemOpen
  \bibfield  {author} {\bibinfo {author} {\bibfnamefont {J.~J.}\ \bibnamefont
  {Chen}}, \bibinfo {author} {\bibfnamefont {J.}~\bibnamefont {Cui}}, \bibinfo
  {author} {\bibfnamefont {Y.~R.}\ \bibnamefont {Zhang}}, \ and\ \bibinfo
  {author} {\bibfnamefont {H.}~\bibnamefont {Fan}},\ }\bibfield  {title}
  {\enquote {\bibinfo {title} {Coherence susceptibility as a probe of quantum
  phase transitions},}\ }\href {\doibase 10.1103/PhysRevA.94.022112} {\bibfield
   {journal} {\bibinfo  {journal} {Phys. Rev. A}\ }\textbf {\bibinfo {volume}
  {94}},\ \bibinfo {pages} {022112} (\bibinfo {year} {2016})}\BibitemShut
  {NoStop}%
\bibitem [{\citenamefont {Lieb}(1994)}]{Lieb1994}%
  \BibitemOpen
  \bibfield  {author} {\bibinfo {author} {\bibfnamefont {E.~H.}\ \bibnamefont
  {Lieb}},\ }\bibfield  {title} {\enquote {\bibinfo {title} {Flux phase of the
  half-filled band},}\ }\href {\doibase 10.1103/PhysRevLett.73.2158} {\bibfield
   {journal} {\bibinfo  {journal} {Phys. Rev. Lett.}\ }\textbf {\bibinfo
  {volume} {73}},\ \bibinfo {pages} {2158--2161} (\bibinfo {year}
  {1994})}\BibitemShut {NoStop}%
\bibitem [{\citenamefont {Stormer}\ \emph {et~al.}(1999)\citenamefont
  {Stormer}, \citenamefont {Tsui},\ and\ \citenamefont
  {Gossard}}]{Stormer1999}%
  \BibitemOpen
  \bibfield  {author} {\bibinfo {author} {\bibfnamefont {H.~L.}\ \bibnamefont
  {Stormer}}, \bibinfo {author} {\bibfnamefont {D.~C.}\ \bibnamefont {Tsui}}, \
  and\ \bibinfo {author} {\bibfnamefont {A.~C.}\ \bibnamefont {Gossard}},\
  }\bibfield  {title} {\enquote {\bibinfo {title} {The fractional quantum
  $\textrm{H}$all effect},}\ }\href {\doibase 10.1103/RevModPhys.71.S298}
  {\bibfield  {journal} {\bibinfo  {journal} {Rev. Mod. Phys.}\ }\textbf
  {\bibinfo {volume} {71}},\ \bibinfo {pages} {S298--S305} (\bibinfo {year}
  {1999})}\BibitemShut {NoStop}%
\bibitem [{\citenamefont {Leykam}\ \emph {et~al.}(2017)\citenamefont {Leykam},
  \citenamefont {Bliokh}, \citenamefont {Huang}, \citenamefont {Chong},\ and\
  \citenamefont {Nori}}]{Leykam2017}%
  \BibitemOpen
  \bibfield  {author} {\bibinfo {author} {\bibfnamefont {D.}~\bibnamefont
  {Leykam}}, \bibinfo {author} {\bibfnamefont {K.~Y.}\ \bibnamefont {Bliokh}},
  \bibinfo {author} {\bibfnamefont {C.~L.}\ \bibnamefont {Huang}}, \bibinfo
  {author} {\bibfnamefont {Y.~D.}\ \bibnamefont {Chong}}, \ and\ \bibinfo
  {author} {\bibfnamefont {F.}~\bibnamefont {Nori}},\ }\bibfield  {title}
  {\enquote {\bibinfo {title} {Edge modes, degeneracies, and topological
  numbers in non-$\textrm{H}$ermitian systems},}\ }\href {\doibase
  10.1103/PhysRevLett.118.040401} {\bibfield  {journal} {\bibinfo  {journal}
  {Phys. Rev. Lett.}\ }\textbf {\bibinfo {volume} {118}},\ \bibinfo {pages}
  {040401} (\bibinfo {year} {2017})}\BibitemShut {NoStop}%
\end{thebibliography}%


\begin{thebibliography}{32}%
\makeatletter
\providecommand \@ifxundefined [1]{%
 \@ifx{#1\undefined}
}%
\providecommand \@ifnum [1]{%
 \ifnum #1\expandafter \@firstoftwo
 \else \expandafter \@secondoftwo
 \fi
}%
\providecommand \@ifx [1]{%
 \ifx #1\expandafter \@firstoftwo
 \else \expandafter \@secondoftwo
 \fi
}%
\providecommand \natexlab [1]{#1}%
\providecommand \enquote  [1]{``#1''}%
\providecommand \bibnamefont  [1]{#1}%
\providecommand \bibfnamefont [1]{#1}%
\providecommand \citenamefont [1]{#1}%
\providecommand \href@noop [0]{\@secondoftwo}%
\providecommand \href [0]{\begingroup \@sanitize@url \@href}%
\providecommand \@href[1]{\@@startlink{#1}\@@href}%
\providecommand \@@href[1]{\endgroup#1\@@endlink}%
\providecommand \@sanitize@url [0]{\catcode `\\12\catcode `\$12\catcode
  `\&12\catcode `\#12\catcode `\^12\catcode `\_12\catcode `\%12\relax}%
\providecommand \@@startlink[1]{}%
\providecommand \@@endlink[0]{}%
\providecommand \url  [0]{\begingroup\@sanitize@url \@url }%
\providecommand \@url [1]{\endgroup\@href {#1}{\urlprefix }}%
\providecommand \urlprefix  [0]{URL }%
\providecommand \Eprint [0]{\href }%
\providecommand \doibase [0]{http://dx.doi.org/}%
\providecommand \selectlanguage [0]{\@gobble}%
\providecommand \bibinfo  [0]{\@secondoftwo}%
\providecommand \bibfield  [0]{\@secondoftwo}%
\providecommand \translation [1]{[#1]}%
\providecommand \BibitemOpen [0]{}%
\providecommand \bibitemStop [0]{}%
\providecommand \bibitemNoStop [0]{.\EOS\space}%
\providecommand \EOS [0]{\spacefactor3000\relax}%
\providecommand \BibitemShut  [1]{\csname bibitem#1\endcsname}%
\let\auto@bib@innerbib\@empty
\bibitem [{\citenamefont {Alecce}\ and\ \citenamefont
  {Dell'Anna}(2017)}]{Alecce2017}%
  \BibitemOpen
  \bibfield  {author} {\bibinfo {author} {\bibfnamefont {A.}~\bibnamefont
  {Alecce}}\ and\ \bibinfo {author} {\bibfnamefont {L.}~\bibnamefont
  {Dell'Anna}},\ }\bibfield  {title} {\enquote {\bibinfo {title} {Extended
  $\textrm{K}$itaev chain with longer-range hopping and pairing},}\ }\href
  {\doibase 10.1103/PhysRevB.95.195160} {\bibfield  {journal} {\bibinfo
  {journal} {Phys. Rev. B}\ }\textbf {\bibinfo {volume} {95}},\ \bibinfo
  {pages} {195160} (\bibinfo {year} {2017})}\BibitemShut {NoStop}%
\bibitem [{\citenamefont {Chiu}\ \emph {et~al.}(2016)\citenamefont {Chiu},
  \citenamefont {Teo}, \citenamefont {Schnyder},\ and\ \citenamefont
  {Ryu}}]{Chiu2016}%
  \BibitemOpen
  \bibfield  {author} {\bibinfo {author} {\bibfnamefont {C.~K.}\ \bibnamefont
  {Chiu}}, \bibinfo {author} {\bibfnamefont {J.~C.~Y.}\ \bibnamefont {Teo}},
  \bibinfo {author} {\bibfnamefont {A.~P.}\ \bibnamefont {Schnyder}}, \ and\
  \bibinfo {author} {\bibfnamefont {S.}~\bibnamefont {Ryu}},\ }\bibfield
  {title} {\enquote {\bibinfo {title} {Classification of topological quantum
  matter with symmetries},}\ }\href {\doibase 10.1103/RevModPhys.88.035005}
  {\bibfield  {journal} {\bibinfo  {journal} {Rev. Mod. Phys.}\ }\textbf
  {\bibinfo {volume} {88}},\ \bibinfo {pages} {035005} (\bibinfo {year}
  {2016})}\BibitemShut {NoStop}%
\bibitem [{\citenamefont {Li}\ \emph {et~al.}(2016)\citenamefont {Li},
  \citenamefont {Yang},\ and\ \citenamefont {Chen}}]{Li2016}%
  \BibitemOpen
  \bibfield  {author} {\bibinfo {author} {\bibfnamefont {L.~H.}\ \bibnamefont
  {Li}}, \bibinfo {author} {\bibfnamefont {C.}~\bibnamefont {Yang}}, \ and\
  \bibinfo {author} {\bibfnamefont {S.}~\bibnamefont {Chen}},\ }\bibfield
  {title} {\enquote {\bibinfo {title} {Topological invariants for phase
  transition points of one-dimensional $\textrm{Z}_2$ topological systems},}\
  }\href {\doibase 10.1140/epjb/e2016-70325-x} {\bibfield  {journal} {\bibinfo
  {journal} {Eur. Phys. J. B}\ }\textbf {\bibinfo {volume} {89}},\ \bibinfo
  {pages} {195} (\bibinfo {year} {2016})}\BibitemShut {NoStop}%
\bibitem [{\citenamefont {Zhang}\ and\ \citenamefont {Song}(2015)}]{Zhang2015}%
  \BibitemOpen
  \bibfield  {author} {\bibinfo {author} {\bibfnamefont {G.}~\bibnamefont
  {Zhang}}\ and\ \bibinfo {author} {\bibfnamefont {Z.}~\bibnamefont {Song}},\
  }\bibfield  {title} {\enquote {\bibinfo {title} {Topological characterization
  of extended quantum $\textrm{I}$sing models},}\ }\href {\doibase
  10.1103/PhysRevLett.115.177204} {\bibfield  {journal} {\bibinfo  {journal}
  {Phys. Rev. Lett.}\ }\textbf {\bibinfo {volume} {115}},\ \bibinfo {pages}
  {177204} (\bibinfo {year} {2015})}\BibitemShut {NoStop}%
\bibitem [{\citenamefont {Ahlfors}(1953)}]{Ahlfors1953}%
  \BibitemOpen
  \bibfield  {author} {\bibinfo {author} {\bibfnamefont {L.~V.}\ \bibnamefont
  {Ahlfors}},\ }\href@noop {} {\emph {\bibinfo {title} {Complex analysis: an
  introduction to the theory of analytic functions of one complex variable}}}\
  (\bibinfo  {publisher} {New York, London},\ \bibinfo {year}
  {1953})\BibitemShut {NoStop}%
\bibitem [{\citenamefont {Fendley}(2012)}]{Fendley2012}%
  \BibitemOpen
  \bibfield  {author} {\bibinfo {author} {\bibfnamefont {P.}~\bibnamefont
  {Fendley}},\ }\bibfield  {title} {\enquote {\bibinfo {title} {Parafermionic
  edge zero modes in {Z}(n)-invariant spin chains},}\ }\href {\doibase
  10.1088/1742-5468/2012/11/P11020} {\bibfield  {journal} {\bibinfo  {journal}
  {J. Stat. Mech.}\ }\textbf {\bibinfo {volume} {2012}},\ \bibinfo {pages}
  {P11020} (\bibinfo {year} {2012})}\BibitemShut {NoStop}%
\bibitem [{\citenamefont {Sarma}\ \emph {et~al.}(2015)\citenamefont {Sarma},
  \citenamefont {Freedman},\ and\ \citenamefont {Nayak}}]{Sarma2015}%
  \BibitemOpen
  \bibfield  {author} {\bibinfo {author} {\bibfnamefont {S.~D.}\ \bibnamefont
  {Sarma}}, \bibinfo {author} {\bibfnamefont {M.}~\bibnamefont {Freedman}}, \
  and\ \bibinfo {author} {\bibfnamefont {C.}~\bibnamefont {Nayak}},\ }\bibfield
   {title} {\enquote {\bibinfo {title} {Majorana zero modes and topological
  quantum computation},}\ }\href {\doibase 10.1038/npjqi.2015.1} {\bibfield
  {journal} {\bibinfo  {journal} {npj Quantum Inf.}\ }\textbf {\bibinfo
  {volume} {1}},\ \bibinfo {pages} {15001} (\bibinfo {year}
  {2015})}\BibitemShut {NoStop}%
\bibitem [{\citenamefont {Elliott}\ and\ \citenamefont
  {Franz}(2015)}]{Elliott2015}%
  \BibitemOpen
  \bibfield  {author} {\bibinfo {author} {\bibfnamefont {S.~R.}\ \bibnamefont
  {Elliott}}\ and\ \bibinfo {author} {\bibfnamefont {M.}~\bibnamefont
  {Franz}},\ }\bibfield  {title} {\enquote {\bibinfo {title} {Colloquium:
  $\textrm{M}$ajorana fermions in nuclear, particle, and solid-state
  physics},}\ }\href {\doibase 10.1103/RevModPhys.87.137} {\bibfield  {journal}
  {\bibinfo  {journal} {Rev. Mod. Phys.}\ }\textbf {\bibinfo {volume} {87}},\
  \bibinfo {pages} {137--163} (\bibinfo {year} {2015})}\BibitemShut {NoStop}%
\bibitem [{\citenamefont {Niu}\ \emph {et~al.}(2012)\citenamefont {Niu},
  \citenamefont {Chung}, \citenamefont {Hsu}, \citenamefont {Mandal},
  \citenamefont {Raghu},\ and\ \citenamefont {Chakravarty}}]{Niu2012}%
  \BibitemOpen
  \bibfield  {author} {\bibinfo {author} {\bibfnamefont {Y.~Z.}\ \bibnamefont
  {Niu}}, \bibinfo {author} {\bibfnamefont {S.~B.}\ \bibnamefont {Chung}},
  \bibinfo {author} {\bibfnamefont {C.~H.}\ \bibnamefont {Hsu}}, \bibinfo
  {author} {\bibfnamefont {I.}~\bibnamefont {Mandal}}, \bibinfo {author}
  {\bibfnamefont {S.}~\bibnamefont {Raghu}}, \ and\ \bibinfo {author}
  {\bibfnamefont {S.}~\bibnamefont {Chakravarty}},\ }\bibfield  {title}
  {\enquote {\bibinfo {title} {Majorana zero modes in a quantum
  $\textrm{I}$sing chain with longer-ranged interactions},}\ }\href {\doibase
  10.1103/PhysRevB.85.035110} {\bibfield  {journal} {\bibinfo  {journal} {Phys.
  Rev. B}\ }\textbf {\bibinfo {volume} {85}},\ \bibinfo {pages} {035110}
  (\bibinfo {year} {2012})}\BibitemShut {NoStop}%
\bibitem [{\citenamefont {Kitaev}(2001)}]{Kitaev2001}%
  \BibitemOpen
  \bibfield  {author} {\bibinfo {author} {\bibfnamefont {A.~Yu}\ \bibnamefont
  {Kitaev}},\ }\bibfield  {title} {\enquote {\bibinfo {title} {Unpaired
  $\textrm{M}$ajorana fermions in quantum wires},}\ }\href@noop {} {\bibfield
  {journal} {\bibinfo  {journal} {Phys. Usp.}\ }\textbf {\bibinfo {volume}
  {44}},\ \bibinfo {pages} {131} (\bibinfo {year} {2001})}\BibitemShut
  {NoStop}%
\bibitem [{\citenamefont {Lepori}\ and\ \citenamefont
  {Dell'Anna}(2017)}]{Lepori2017}%
  \BibitemOpen
  \bibfield  {author} {\bibinfo {author} {\bibfnamefont {L.}~\bibnamefont
  {Lepori}}\ and\ \bibinfo {author} {\bibfnamefont {L.}~\bibnamefont
  {Dell'Anna}},\ }\bibfield  {title} {\enquote {\bibinfo {title} {Long-range
  topological insulators and weakened bulk-boundary correspondence},}\ }\href
  {\doibase https://doi.org/10.1088/1367-2630/aa84d0} {\bibfield  {journal}
  {\bibinfo  {journal} {New J. Phys.}\ }\textbf {\bibinfo {volume} {19}},\
  \bibinfo {pages} {103030} (\bibinfo {year} {2017})}\BibitemShut {NoStop}%
\bibitem [{\citenamefont {Braunstein}\ and\ \citenamefont
  {Caves}(1994)}]{BRAUNSTEIN1994}%
  \BibitemOpen
  \bibfield  {author} {\bibinfo {author} {\bibfnamefont {S.~L.}\ \bibnamefont
  {Braunstein}}\ and\ \bibinfo {author} {\bibfnamefont {C.~M.}\ \bibnamefont
  {Caves}},\ }\bibfield  {title} {\enquote {\bibinfo {title} {Statistical
  distance and the geometry of quantum states},}\ }\href {\doibase
  10.1103/PhysRevLett.72.3439} {\bibfield  {journal} {\bibinfo  {journal}
  {Phys. Rev. Lett.}\ }\textbf {\bibinfo {volume} {72}},\ \bibinfo {pages}
  {3439--3443} (\bibinfo {year} {1994})}\BibitemShut {NoStop}%
\bibitem [{\citenamefont {Pezz\'{e}}\ and\ \citenamefont
  {Smerzi}(2009)}]{Pezze2009}%
  \BibitemOpen
  \bibfield  {author} {\bibinfo {author} {\bibfnamefont {L.}~\bibnamefont
  {Pezz\'{e}}}\ and\ \bibinfo {author} {\bibfnamefont {A.}~\bibnamefont
  {Smerzi}},\ }\bibfield  {title} {\enquote {\bibinfo {title} {Entanglement,
  nonlinear dynamics, and the $\textrm{H}$eisenberg limit},}\ }\href {\doibase
  10.1103/PhysRevLett.102.100401} {\bibfield  {journal} {\bibinfo  {journal}
  {Phys. Rev. Lett.}\ }\textbf {\bibinfo {volume} {102}},\ \bibinfo {pages}
  {100401} (\bibinfo {year} {2009})}\BibitemShut {NoStop}%
\bibitem [{\citenamefont {Giovannetti}\ \emph {et~al.}(2011)\citenamefont
  {Giovannetti}, \citenamefont {Lloyd},\ and\ \citenamefont
  {Maccone}}]{Giovannetti2011}%
  \BibitemOpen
  \bibfield  {author} {\bibinfo {author} {\bibfnamefont {V.}~\bibnamefont
  {Giovannetti}}, \bibinfo {author} {\bibfnamefont {S.}~\bibnamefont {Lloyd}},
  \ and\ \bibinfo {author} {\bibfnamefont {L.}~\bibnamefont {Maccone}},\
  }\bibfield  {title} {\enquote {\bibinfo {title} {Advances in quantum
  metrology},}\ }\href {\doibase 10.1038/NPHOTON.2011.35} {\bibfield  {journal}
  {\bibinfo  {journal} {Nat. Photonics}\ }\textbf {\bibinfo {volume} {5}},\
  \bibinfo {pages} {222--229} (\bibinfo {year} {2011})}\BibitemShut {NoStop}%
\bibitem [{\citenamefont {Ma}\ \emph {et~al.}(2011)\citenamefont {Ma},
  \citenamefont {Wang}, \citenamefont {Sun},\ and\ \citenamefont
  {Nori}}]{Ma2011}%
  \BibitemOpen
  \bibfield  {author} {\bibinfo {author} {\bibfnamefont {J.}~\bibnamefont
  {Ma}}, \bibinfo {author} {\bibfnamefont {X.~G.}\ \bibnamefont {Wang}},
  \bibinfo {author} {\bibfnamefont {C.~P.}\ \bibnamefont {Sun}}, \ and\
  \bibinfo {author} {\bibfnamefont {F.}~\bibnamefont {Nori}},\ }\bibfield
  {title} {\enquote {\bibinfo {title} {Quantum spin squeezing},}\ }\href
  {\doibase 10.1016/j.physrep.2011.08.003} {\bibfield  {journal} {\bibinfo
  {journal} {Phys. Rep.}\ }\textbf {\bibinfo {volume} {509}},\ \bibinfo {pages}
  {89--165} (\bibinfo {year} {2011})}\BibitemShut {NoStop}%
\bibitem [{\citenamefont {Pezz\`{e}}\ \emph {et~al.}(2017)\citenamefont
  {Pezz\`{e}}, \citenamefont {Gabbrielli}, \citenamefont {Lepori},\ and\
  \citenamefont {Smerzi}}]{Pezze2017}%
  \BibitemOpen
  \bibfield  {author} {\bibinfo {author} {\bibfnamefont {L.}~\bibnamefont
  {Pezz\`{e}}}, \bibinfo {author} {\bibfnamefont {M.}~\bibnamefont
  {Gabbrielli}}, \bibinfo {author} {\bibfnamefont {L.}~\bibnamefont {Lepori}},
  \ and\ \bibinfo {author} {\bibfnamefont {A.}~\bibnamefont {Smerzi}},\
  }\bibfield  {title} {\enquote {\bibinfo {title} {Multipartite entanglement in
  topological quantum phases},}\ }\href {\doibase
  10.1103/PhysRevLett.119.250401} {\bibfield  {journal} {\bibinfo  {journal}
  {Phys. Rev. Lett.}\ }\textbf {\bibinfo {volume} {119}},\ \bibinfo {pages}
  {250401} (\bibinfo {year} {2017})}\BibitemShut {NoStop}%
\bibitem [{\citenamefont {Barouch}\ and\ \citenamefont
  {Mccoy}(1971)}]{BAROUCH1971}%
  \BibitemOpen
  \bibfield  {author} {\bibinfo {author} {\bibfnamefont {E.}~\bibnamefont
  {Barouch}}\ and\ \bibinfo {author} {\bibfnamefont {B.~M.}\ \bibnamefont
  {Mccoy}},\ }\bibfield  {title} {\enquote {\bibinfo {title} {Statistical
  mechanics of $\textrm{XY}$-model .2. spin-correlation functions},}\ }\href
  {\doibase 10.1103/PhysRevA.3.786} {\bibfield  {journal} {\bibinfo  {journal}
  {Phys. Rev. A}\ }\textbf {\bibinfo {volume} {3}},\ \bibinfo {pages}
  {786--804} (\bibinfo {year} {1971})}\BibitemShut {NoStop}%
\bibitem [{\citenamefont {Fradkin}\ and\ \citenamefont
  {Susskind}(1978)}]{FRADKIN1978}%
  \BibitemOpen
  \bibfield  {author} {\bibinfo {author} {\bibfnamefont {E.}~\bibnamefont
  {Fradkin}}\ and\ \bibinfo {author} {\bibfnamefont {L.}~\bibnamefont
  {Susskind}},\ }\bibfield  {title} {\enquote {\bibinfo {title} {Order and
  disorder in gauge systems and magnets},}\ }\href {\doibase
  10.1103/PhysRevD.17.2637} {\bibfield  {journal} {\bibinfo  {journal} {Phys.
  Rev. D}\ }\textbf {\bibinfo {volume} {17}},\ \bibinfo {pages} {2637--2658}
  (\bibinfo {year} {1978})}\BibitemShut {NoStop}%
\bibitem [{\citenamefont {Smacchia}\ \emph {et~al.}(2011)\citenamefont
  {Smacchia}, \citenamefont {Amico}, \citenamefont {Facchi}, \citenamefont
  {Fazio}, \citenamefont {Florio}, \citenamefont {Pascazio},\ and\
  \citenamefont {Vedral}}]{Smacchia2011}%
  \BibitemOpen
  \bibfield  {author} {\bibinfo {author} {\bibfnamefont {P.}~\bibnamefont
  {Smacchia}}, \bibinfo {author} {\bibfnamefont {L.}~\bibnamefont {Amico}},
  \bibinfo {author} {\bibfnamefont {P.}~\bibnamefont {Facchi}}, \bibinfo
  {author} {\bibfnamefont {R.}~\bibnamefont {Fazio}}, \bibinfo {author}
  {\bibfnamefont {G.}~\bibnamefont {Florio}}, \bibinfo {author} {\bibfnamefont
  {S.}~\bibnamefont {Pascazio}}, \ and\ \bibinfo {author} {\bibfnamefont
  {V.}~\bibnamefont {Vedral}},\ }\bibfield  {title} {\enquote {\bibinfo {title}
  {Statistical mechanics of the cluster $\textrm{I}$sing model},}\ }\href
  {\doibase 10.1103/PhysRevA.84.022304} {\bibfield  {journal} {\bibinfo
  {journal} {Phys. Rev. A}\ }\textbf {\bibinfo {volume} {84}},\ \bibinfo
  {pages} {022304} (\bibinfo {year} {2011})}\BibitemShut {NoStop}%
\bibitem [{\citenamefont {Feng}\ \emph {et~al.}(2007)\citenamefont {Feng},
  \citenamefont {Zhang},\ and\ \citenamefont {Xiang}}]{Feng2007}%
  \BibitemOpen
  \bibfield  {author} {\bibinfo {author} {\bibfnamefont {X.~Y.}\ \bibnamefont
  {Feng}}, \bibinfo {author} {\bibfnamefont {G.~M.}\ \bibnamefont {Zhang}}, \
  and\ \bibinfo {author} {\bibfnamefont {T.}~\bibnamefont {Xiang}},\ }\bibfield
   {title} {\enquote {\bibinfo {title} {Topological characterization of quantum
  phase transitions in a spin-1/2 model},}\ }\href {\doibase
  10.1103/PhysRevLett.98.087204} {\bibfield  {journal} {\bibinfo  {journal}
  {Phys. Rev. Lett.}\ }\textbf {\bibinfo {volume} {98}},\ \bibinfo {pages}
  {087204} (\bibinfo {year} {2007})}\BibitemShut {NoStop}%
\bibitem [{\citenamefont {Qin}\ \emph {et~al.}(2017)\citenamefont {Qin},
  \citenamefont {He}, \citenamefont {You}, \citenamefont {Lu}, \citenamefont
  {Sen}, \citenamefont {Sandvik}, \citenamefont {Xu},\ and\ \citenamefont
  {Meng}}]{Qin2017}%
  \BibitemOpen
  \bibfield  {author} {\bibinfo {author} {\bibfnamefont {Y.~Q.}\ \bibnamefont
  {Qin}}, \bibinfo {author} {\bibfnamefont {Y.~Y.}\ \bibnamefont {He}},
  \bibinfo {author} {\bibfnamefont {Y.~Z.}\ \bibnamefont {You}}, \bibinfo
  {author} {\bibfnamefont {Z.~Y.}\ \bibnamefont {Lu}}, \bibinfo {author}
  {\bibfnamefont {A.}~\bibnamefont {Sen}}, \bibinfo {author} {\bibfnamefont
  {A.~W.}\ \bibnamefont {Sandvik}}, \bibinfo {author} {\bibfnamefont {C.~K.}\
  \bibnamefont {Xu}}, \ and\ \bibinfo {author} {\bibfnamefont {Z.~Y.}\
  \bibnamefont {Meng}},\ }\bibfield  {title} {\enquote {\bibinfo {title}
  {Duality between the deconfined quantum-critical point and the bosonic
  topological transition},}\ }\href {\doibase 10.1103/PhysRevX.7.031052}
  {\bibfield  {journal} {\bibinfo  {journal} {Phys. Rev. X}\ }\textbf {\bibinfo
  {volume} {7}},\ \bibinfo {pages} {031052} (\bibinfo {year}
  {2017})}\BibitemShut {NoStop}%
\bibitem [{\citenamefont {Cui}\ \emph {et~al.}(2013)\citenamefont {Cui},
  \citenamefont {Amico}, \citenamefont {Fan}, \citenamefont {Gu}, \citenamefont
  {Hamma},\ and\ \citenamefont {Vedral}}]{Cui2013}%
  \BibitemOpen
  \bibfield  {author} {\bibinfo {author} {\bibfnamefont {J.}~\bibnamefont
  {Cui}}, \bibinfo {author} {\bibfnamefont {L.}~\bibnamefont {Amico}}, \bibinfo
  {author} {\bibfnamefont {H.}~\bibnamefont {Fan}}, \bibinfo {author}
  {\bibfnamefont {M.}~\bibnamefont {Gu}}, \bibinfo {author} {\bibfnamefont
  {A.}~\bibnamefont {Hamma}}, \ and\ \bibinfo {author} {\bibfnamefont
  {V.}~\bibnamefont {Vedral}},\ }\bibfield  {title} {\enquote {\bibinfo {title}
  {Local characterization of one-dimensional topologically ordered states},}\
  }\href {\doibase 10.1103/PhysRevB.88.125117} {\bibfield  {journal} {\bibinfo
  {journal} {Phys. Rev. B}\ }\textbf {\bibinfo {volume} {88}},\ \bibinfo
  {pages} {125117} (\bibinfo {year} {2013})}\BibitemShut {NoStop}%
\bibitem [{\citenamefont {Venuti}\ and\ \citenamefont
  {Roncaglia}(2005)}]{Venuti2005}%
  \BibitemOpen
  \bibfield  {author} {\bibinfo {author} {\bibfnamefont {L.~C.}\ \bibnamefont
  {Venuti}}\ and\ \bibinfo {author} {\bibfnamefont {M.}~\bibnamefont
  {Roncaglia}},\ }\bibfield  {title} {\enquote {\bibinfo {title} {Analytic
  relations between localizable entanglement and string correlations in spin
  systems},}\ }\href {\doibase 10.1103/PhysRevLett.94.207207} {\bibfield
  {journal} {\bibinfo  {journal} {Phys. Rev. Lett.}\ }\textbf {\bibinfo
  {volume} {94}},\ \bibinfo {pages} {207207} (\bibinfo {year}
  {2005})}\BibitemShut {NoStop}%
\bibitem [{\citenamefont {Cobanera}\ \emph {et~al.}(2011)\citenamefont
  {Cobanera}, \citenamefont {Ortiz},\ and\ \citenamefont
  {Nussinov}}]{Cobanera2011}%
  \BibitemOpen
  \bibfield  {author} {\bibinfo {author} {\bibfnamefont {E.}~\bibnamefont
  {Cobanera}}, \bibinfo {author} {\bibfnamefont {G.}~\bibnamefont {Ortiz}}, \
  and\ \bibinfo {author} {\bibfnamefont {Z.}~\bibnamefont {Nussinov}},\
  }\bibfield  {title} {\enquote {\bibinfo {title} {The bond-algebraic approach
  to dualities},}\ }\href {\doibase 10.1080/00018732.2011.619814} {\bibfield
  {journal} {\bibinfo  {journal} {Adv. Phys.}\ }\textbf {\bibinfo {volume}
  {60}},\ \bibinfo {pages} {679--798} (\bibinfo {year} {2011})}\BibitemShut
  {NoStop}%
\bibitem [{\citenamefont {Fidkowski}\ and\ \citenamefont
  {Kitaev}(2011)}]{Fidkowski2011}%
  \BibitemOpen
  \bibfield  {author} {\bibinfo {author} {\bibfnamefont {L.}~\bibnamefont
  {Fidkowski}}\ and\ \bibinfo {author} {\bibfnamefont {A.}~\bibnamefont
  {Kitaev}},\ }\bibfield  {title} {\enquote {\bibinfo {title} {Topological
  phases of fermions in one dimension},}\ }\href {\doibase
  10.1103/PhysRevB.83.075103} {\bibfield  {journal} {\bibinfo  {journal} {Phys.
  Rev. B}\ }\textbf {\bibinfo {volume} {83}},\ \bibinfo {pages} {075103}
  (\bibinfo {year} {2011})}\BibitemShut {NoStop}%
\bibitem [{\citenamefont {Viyuela}\ \emph {et~al.}(2016)\citenamefont
  {Viyuela}, \citenamefont {Vodola}, \citenamefont {Pupillo},\ and\
  \citenamefont {Martin-Delgado}}]{Viyuela2016}%
  \BibitemOpen
  \bibfield  {author} {\bibinfo {author} {\bibfnamefont {O.}~\bibnamefont
  {Viyuela}}, \bibinfo {author} {\bibfnamefont {D.}~\bibnamefont {Vodola}},
  \bibinfo {author} {\bibfnamefont {G.}~\bibnamefont {Pupillo}}, \ and\
  \bibinfo {author} {\bibfnamefont {M.~A.}\ \bibnamefont {Martin-Delgado}},\
  }\bibfield  {title} {\enquote {\bibinfo {title} {Topological massive {D}irac
  edge modes and long-range superconducting {H}amiltonians},}\ }\href {\doibase
  10.1103/PhysRevB.94.125121} {\bibfield  {journal} {\bibinfo  {journal} {Phys.
  Rev. B}\ }\textbf {\bibinfo {volume} {94}},\ \bibinfo {pages} {125121}
  (\bibinfo {year} {2016})}\BibitemShut {NoStop}%
\bibitem [{\citenamefont {Vodola}\ \emph {et~al.}(2014)\citenamefont {Vodola},
  \citenamefont {Lepori}, \citenamefont {Ercolessi}, \citenamefont {Gorshkov},\
  and\ \citenamefont {Pupillo}}]{Vodola2014}%
  \BibitemOpen
  \bibfield  {author} {\bibinfo {author} {\bibfnamefont {D.}~\bibnamefont
  {Vodola}}, \bibinfo {author} {\bibfnamefont {L.}~\bibnamefont {Lepori}},
  \bibinfo {author} {\bibfnamefont {E.}~\bibnamefont {Ercolessi}}, \bibinfo
  {author} {\bibfnamefont {A.~V.}\ \bibnamefont {Gorshkov}}, \ and\ \bibinfo
  {author} {\bibfnamefont {G.}~\bibnamefont {Pupillo}},\ }\bibfield  {title}
  {\enquote {\bibinfo {title} {Kitaev chains with long-range pairing},}\ }\href
  {\doibase 10.1103/PhysRevLett.113.156402} {\bibfield  {journal} {\bibinfo
  {journal} {Phys. Rev. Lett.}\ }\textbf {\bibinfo {volume} {113}},\ \bibinfo
  {pages} {156402} (\bibinfo {year} {2014})}\BibitemShut {NoStop}%
\bibitem [{\citenamefont {Vodola}\ \emph {et~al.}(2015)\citenamefont {Vodola},
  \citenamefont {Lepori}, \citenamefont {Ercolessi},\ and\ \citenamefont
  {Pupillo}}]{Vodola2015}%
  \BibitemOpen
  \bibfield  {author} {\bibinfo {author} {\bibfnamefont {D.}~\bibnamefont
  {Vodola}}, \bibinfo {author} {\bibfnamefont {L.}~\bibnamefont {Lepori}},
  \bibinfo {author} {\bibfnamefont {E.}~\bibnamefont {Ercolessi}}, \ and\
  \bibinfo {author} {\bibfnamefont {G.}~\bibnamefont {Pupillo}},\ }\bibfield
  {title} {\enquote {\bibinfo {title} {Long-range $\textrm{I}$sing and
  $\textrm{K}$itaev models: phases, correlations and edge modes},}\ }\href
  {\doibase 10.1088/1367-2630/18/1/015001} {\bibfield  {journal} {\bibinfo
  {journal} {New J. Phys.}\ }\textbf {\bibinfo {volume} {18}},\ \bibinfo
  {pages} {015001} (\bibinfo {year} {2015})}\BibitemShut {NoStop}%
\bibitem [{\citenamefont {Kitaev}(2006)}]{Kitaev2006a}%
  \BibitemOpen
  \bibfield  {author} {\bibinfo {author} {\bibfnamefont {A.}~\bibnamefont
  {Kitaev}},\ }\bibfield  {title} {\enquote {\bibinfo {title} {Anyons in an
  exactly solved model and beyond},}\ }\href {\doibase
  10.1016/j.aop.2005.10.005} {\bibfield  {journal} {\bibinfo  {journal} {Ann.
  Phys.}\ }\textbf {\bibinfo {volume} {321}},\ \bibinfo {pages} {2--111}
  (\bibinfo {year} {2006})}\BibitemShut {NoStop}%
\bibitem [{\citenamefont {Lieb}(1994)}]{Lieb1994}%
  \BibitemOpen
  \bibfield  {author} {\bibinfo {author} {\bibfnamefont {E.~H.}\ \bibnamefont
  {Lieb}},\ }\bibfield  {title} {\enquote {\bibinfo {title} {Flux phase of the
  half-filled band},}\ }\href {\doibase 10.1103/PhysRevLett.73.2158} {\bibfield
   {journal} {\bibinfo  {journal} {Phys. Rev. Lett.}\ }\textbf {\bibinfo
  {volume} {73}},\ \bibinfo {pages} {2158--2161} (\bibinfo {year}
  {1994})}\BibitemShut {NoStop}%
\bibitem [{\citenamefont {Chen}\ \emph {et~al.}(2016)\citenamefont {Chen},
  \citenamefont {Cui}, \citenamefont {Zhang},\ and\ \citenamefont
  {Fan}}]{Chen2016}%
  \BibitemOpen
  \bibfield  {author} {\bibinfo {author} {\bibfnamefont {J.~J.}\ \bibnamefont
  {Chen}}, \bibinfo {author} {\bibfnamefont {J.}~\bibnamefont {Cui}}, \bibinfo
  {author} {\bibfnamefont {Y.~R.}\ \bibnamefont {Zhang}}, \ and\ \bibinfo
  {author} {\bibfnamefont {H.}~\bibnamefont {Fan}},\ }\bibfield  {title}
  {\enquote {\bibinfo {title} {Coherence susceptibility as a probe of quantum
  phase transitions},}\ }\href {\doibase 10.1103/PhysRevA.94.022112} {\bibfield
   {journal} {\bibinfo  {journal} {Phys. Rev. A}\ }\textbf {\bibinfo {volume}
  {94}},\ \bibinfo {pages} {022112} (\bibinfo {year} {2016})}\BibitemShut
  {NoStop}%
\bibitem [{\citenamefont {Zhang}\ \emph {et~al.}()\citenamefont {Zhang},
  \citenamefont {Fan}, \citenamefont {You},\ and\ \citenamefont
  {Nori}}]{Zhang2018}%
  \BibitemOpen
  \bibfield  {author} {\bibinfo {author} {\bibfnamefont {Y.~R.}\ \bibnamefont
  {Zhang}}, \bibinfo {author} {\bibfnamefont {H.}~\bibnamefont {Fan}}, \bibinfo
  {author} {\bibfnamefont {J.~Q.}\ \bibnamefont {You}}, \ and\ \bibinfo
  {author} {\bibfnamefont {F.}~\bibnamefont {Nori}},\ }\href@noop {} {\enquote
  {\bibinfo {title} {Dual multipartite entanglement as a resource in
  topological quantum computation with anyons},}\ }\bibinfo {note} {{i}n
  preparation}\BibitemShut {NoStop}%
\end{thebibliography}%

\end{document}


\title{SUPPLEMENTAL MATERIAL:\\Characterization of topological states via dual multipartite entanglement}
\author{Yu-Ran Zhang}
\affiliation{Beijing Computational Science Research Center, Beijing 100094, China}
\affiliation{Theoretical Quantum Physics Laboratory, RIKEN, Saitama 351-0198, Japan}
\author{Yu Zeng}
\affiliation{Institute of Physics, Chinese Academy of Sciences, Beijing 100190, China}
\author{Heng Fan}
\affiliation{Institute of Physics, Chinese Academy of Sciences, Beijing 100190, China}
\affiliation{CAS Central of Excellence for Topological Quantum Computation, University of Chinese Academy of Sciences, Beijing 100190, China}
\author{J. Q. You}
\email{jqyou@zju.edu.cn}
\affiliation{Beijing Computational Science Research Center, Beijing 100094, China}
\affiliation{Department of Physics, Zhejiang University, Hangzhou 310027, China}
\author{Franco Nori}
\email{fnori@riken.jp}
\affiliation{Theoretical Quantum Physics Laboratory, RIKEN, Saitama 351-0198, Japan}
\affiliation{Physics Department, University of Michigan, Ann Arbor, Michigan 48109-1040, USA}
\date{\today}

\begin{abstract}
\end{abstract}

\maketitle

\renewcommand{\theequation}{S\arabic{equation}}
\setcounter{equation}{0}  

\renewcommand{\thefigure}{S\arabic{figure}}
\setcounter{figure}{0}  

\section{Mapping to the extended Ising model and Exact solutions}
We start from the extended quantum Ising model with longer-range interactions in a transverse
field, with the Hamiltonian
\begin{equation}
H=\sum_{n=1}^{N_f}\sum_{j=1}^L\left(\frac{{J}_n^x}{2}\sigma_j^x\sigma_{j+n}^x\!+\frac{{J}_n^y}{2}\sigma_j^y\sigma_{j+n}^y\right)\!\!\prod_{l=j+1}^{j+n-1}\!\!\!\sigma_l^z+\!\sum_{j=1}^L\frac{\mu}{2}\sigma_j^z,\label{eq1}
\end{equation}
where $\sigma_j^{x,y,z}$ are Pauli matrices for the spin at site $j$, and $L$ (assumed even)
is the total number of sites. By the Jordan-Wigner transformation
\begin{equation}
c_1=-\sigma_1^+=-(\sigma^x_1+i\sigma^y_1)/2,\hspace*{0.2in}c_j=-\sigma_j^+\prod_{i=1}^{j-1}\sigma^z_i,
\end{equation}
we can obtain a spinless fermion Hamiltonian with longer-range pairing
and hopping terms with fermion parity $(-1)^{N_p}$ of the number of fermions
\begin{equation}
N_p=\sum_{j=1}^Lc_j^\dag c_j,
\end{equation}
as $H=H_{{o}}+H_{{b}}$, where the open chain part is
\begin{align}
H_{{o}}=&\sum_{n=1}^{N_f}\sum_{j=1}^{L-n}\left(\frac{{J}_n^+}{2}c^\dag_jc_{j+n}+\frac{{J}_n^-}{2}c^\dag_jc_{j+n}^\dag+\textrm{h.c.}\right)\nonumber\\
&-\sum_{j=1}^L\mu\left(c^\dag_jc_j-\frac{1}{2}\right), \label{open}
\end{align}
and the boundary part reads
\begin{equation}
H_{{b}}=\frac{(-1)^{N_p}}{2}\sum_{n=1}^{N_f}\!\sum_{~j=L-n+1}^L\!\!\!\!\!({J}_n^+c^\dag_jc_{j+n}+{J}_n^-c^\dag_jc_{j+n}^\dag+\textrm{h.c.}),\label{eq4}
\end{equation}
with ${J}_n^\pm\equiv {J}_n^x\pm {J}_n^y$. Thus, given a definite even fermion parity $(-1)^{N_p}=1$,
this extended Kitaev fermion chain  \cite{Alecce2017} has an antiperiodic boundary condition $c_{j+L}=-c_j$.
Here we choose all the hopping and pairing parameters as real, which make the Hamiltonian preserve
time-reversal symmetry and belong to the {BDI} class ($\mathbf{Z}$ type) characterized by the winding
numbers \cite{Chiu2016,Li2016}.

\begin{figure*}[t]
\centering
\includegraphics[width=0.98\textwidth]{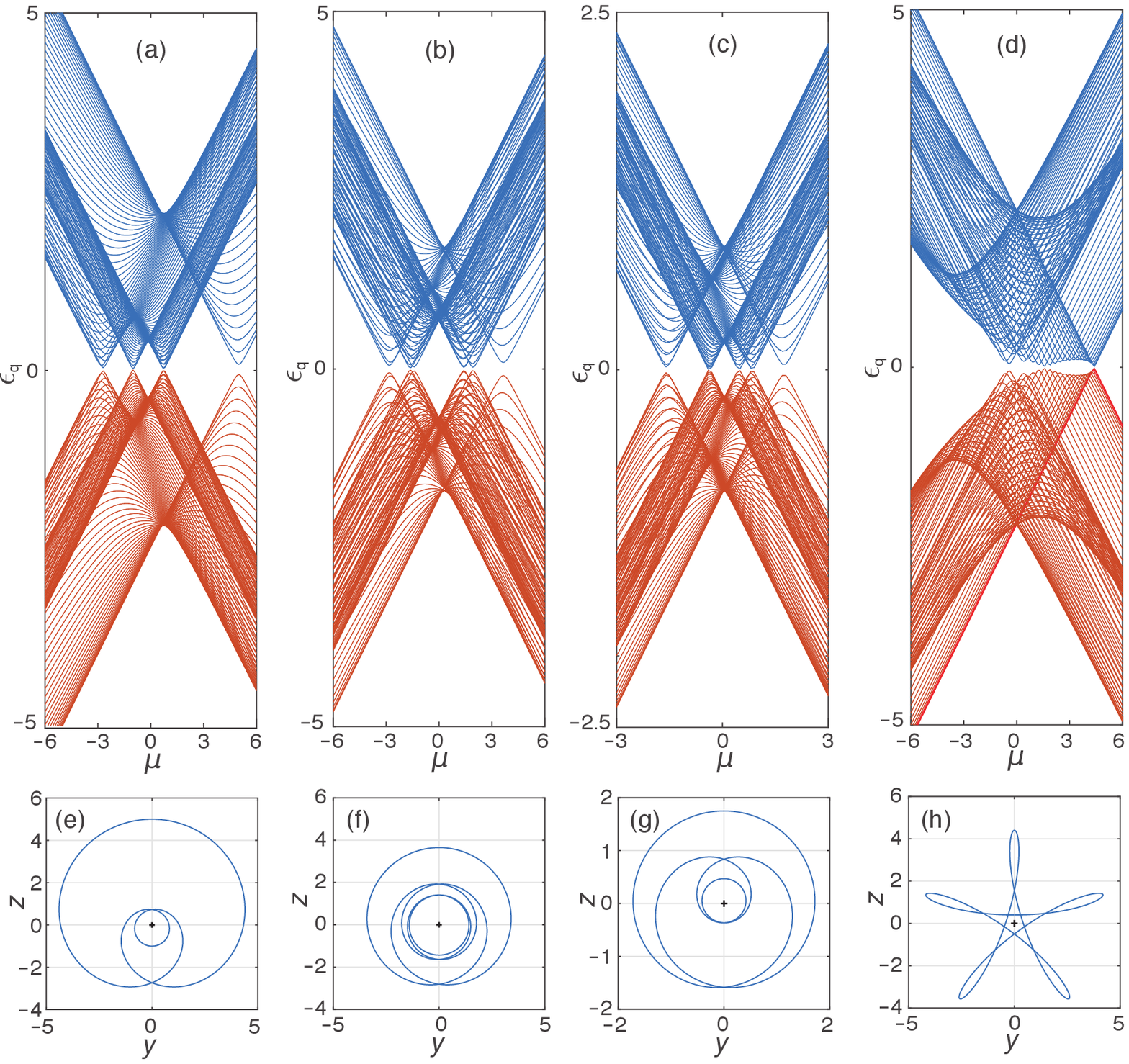}\\
\caption{(color online) (a-d) Energy spectra for $L=200$  and (e-h)
trajectories of winding vectors for an extended Kitaev fermion chain with parameters:
(a,e) ${J}_1^+={J}_1^-=1$, ${J}_{2}^+={J}_{2}^-=2$, ${J}_{3}^+={J}_{3}^-=2$ ($N_f=3$); 
(b,f) ${J}_1^+={J}_1^-=0.1$, ${J}_{2}^+={J}_{2}^-=0.21$, ${J}_{3}^+={J}_{3}^-=0.44$, ${J}_4^+={J}_4^-=0.9$, ${J}_{5}^+={J}_{5}^-=2$ ($N_f=5$); 
(c,g) ${J}_1^+={J}_1^-=0.1$, ${J}_{2}^+={J}_{2}^-=0.21$, ${J}_{3}^+={J}_{3}^-=-0.74$, ${J}_4^+={J}_4^-=0.9$ ($N_f=4$);
and (d,h) ${J}_{2}^+={J}_{2}^-=2.4$, $J_3^+=2$, $J_3^-=-2$ ($N_f=3$).
}\label{fig:S4}
\end{figure*}

For the thermodynamic limit $L\gg N_f\geq1$, we  use the Fourier transformation,
\begin{equation}
c_j=\frac{1}{\sqrt{L}}\sum_{q}\exp({-iqj})\;c_q,
\end{equation}
 to express the Bogoliubov-de Gennes Hamiltonian as
 \begin{equation}
H=\sum_{q}(c_q^\dag,c_{-q})\mathcal{H}_q\left(\begin{array}{c}c_q\\c_{-q}^\dag\end{array}\right),
\end{equation}
where the complete set of wavevectors is $q=2\pi m/L$ with
\begin{equation}
m=-\frac{L-1}{2},-\frac{L-3}{2},\cdots,\frac{L-3}{2},\frac{L-1}{2}.
\end{equation}
Here, we can write
\begin{equation}
\mathcal{H}_q=\frac{1}{2}\bm{r}(q)\cdot \bm{\sigma},
\end{equation}
with the vector $\bm{r}(q)=(0,y(q),z(q))$ in the auxiliary two-dimensional $y$-$z$ space,
\begin{align}
&y(q)=\sum_{n=1}^{N_f}{J}_n^-\sin(nq),\\
&z(q)=\sum_{n=1}^{N_f}{J}_n^+\cos(nq)-\mu,
\end{align}
and $\bm{\sigma}=(\sigma^x,\sigma^y,\sigma^z)$.
Using the Bogoliubov transformation
\begin{equation}
c_q=\cos\frac{\Theta}{2}\eta_q+i\sin\frac{\Theta}{2}\eta_{-q}^\dag,
\end{equation}
with $\tan\Theta=y(q)/z(q)$, we can diagonalize the Hamiltonian as
\begin{equation}
{H}=\sum_q\epsilon_q\left(\eta_q^\dag\eta_q-\frac{1}{2}\right),
\end{equation}
and obtain the ground state
\begin{equation}
|\mathcal{G}\rangle=\prod_{q}[\cos\frac{\Theta}{2}+i\sin\frac{\Theta}{2}\eta_q^{\dag}\eta_{-q}^\dag]|0\rangle,
\end{equation}
where the energy spectra are
\begin{equation}
\epsilon_q=\pm\frac{1}{2}\sqrt{y(q)^2+z(q)^2}.
\end{equation}
In Fig.~\ref{fig:S4}, we plot the energy spectra for $L = 200$ and trajectories of
winding vectors for four different extended Kitaev fermion chain models \cite{Alecce2017} considered
in the main text.

\section{Winding numbers}
For the {BDI} symmetry class Kitaev chain fermion systems, the winding number in the auxiliary space
of momentum behaves as a $\mathbf{Z}$ topological invariant \cite{Chiu2016,Zhang2015}, which is a
fundamental concept in geometric topology. The winding number of the closed loop in auxiliary $y$-$z$
plane around the origin can be written as
\begin{equation}
\nu=\frac{1}{2\pi}\oint\frac{ydz-zdy}{|\bm{r}|^2}.
\end{equation}
Via the substitution $\zeta(q)\equiv\exp(iq)$, we can rewrite in complex space that
\begin{equation}
y(q)=\sum_{n=1}^{N_f}\frac{{J}^-_n{(\zeta^n-\zeta^{-n})}}{2i}\equiv Y(\zeta),
\end{equation}
and
\begin{equation}
z(q)=\sum_{n=1}^{N_f}\frac{{J}^+_n{(\zeta^n+\zeta^{-n})}}{2}-\mu\equiv Z(\zeta).
\end{equation}
By defining a complex characteristic function
\begin{eqnarray}
g(\zeta)&\equiv& Z(\zeta)+iY(\zeta)\\
&=&\sum_{n=1}^{N_f}({J}_n^x\zeta^n+{J}_n^y\zeta^{-n})-\mu,\label{eq5}
\end{eqnarray}
we obtain the winding number by calculating the logarithmic residue of $g(\zeta)$
in accordance with the Cauchy's argument principle \cite{Ahlfors1953}
\begin{equation}
\nu=\frac{1}{2\pi i}\oint_{{}_{|\zeta|=1}}\!\!\!\!\!\!\!\!\!d\zeta\;\frac{g'(\zeta)}{g(\zeta)}=\mathcal{N}-\mathcal{P},
\end{equation}
where in the complex region $|\zeta|<1$, $\mathcal{N}$ is the number of zeros for
$g(\zeta)=0$, and $\mathcal{P}$ is the number of poles for $g(\zeta)=\infty$.
For two special cases: ${J}_n^y=0$  $\forall n$, we have
\begin{equation}
g(\zeta)=\sum_{n=1}^{N_f}J_n^x\zeta^n+\mu,
\end{equation}
and only zeros exist; while ${J}_n^x=0$ there only poles exist.

\section{Majorana zero modes}
We can write the open-chain Hamiltonian (\ref{open}) in terms of Majorana fermion
operators:
\begin{equation}
a_j=c_j^\dag+c_j,\hspace{0.2in}b_j=i(c^\dag_j-c_j),
\end{equation}
 with relations $\{a_i,a_j\}=\{b_i,b_j\}=2\delta_{ij}$,
$\{a_i,b_j\}=0$ as
\begin{equation}
H_{{o}}=-\frac{i}{2}\sum_{n=1}^{N_f}\sum_{j=1}^{L-n}({J}_n^xb_{j}a_{j+n}+{J}_n^yb_{j+n}a_{j})+\frac{i\mu}{2}\sum_{j=1}^La_{j}b_{j}.
\end{equation}
We can assume an ansatz wave function as a linear combination
of Majorana operators $a_j$ \cite{Fendley2012}:
\begin{equation}
\phi=\sum_{j=1}^{L}\alpha_ja_j,
\end{equation}
and calculate the commutation to satisfy the condition $[H,\phi]=0$ for the existence
of Majorana zero modes \cite{Sarma2015,Elliott2015}. Then, the coefficients
are given by the recursion relations
\begin{equation}
\sum_{n=1}^{N_f}({J}_n^x\alpha_{j+n}+{J}_n^y\alpha_{j-n})-\mu\alpha_j=0,
\end{equation}
for $j=n+1,n+2\cdots,L-n$. These recursion equations can be solved with the solutions of
characteristic equations $g(\zeta)=0$ \cite{Niu2012} given $g(\zeta)$ in Eq.~(\ref{eq5}).
If $\mathcal{N}\geq\mathcal{P}$, we should require Majorana zero modes at the left end satisfying
$|\alpha_{L}|\rightarrow0$, for the thermodynamic limit $L\gg1$, and
only in the range $|\zeta|<1$ should the zeros $\{\zeta_{l}\}$ be considered. Thus, we have
$\mathcal{N}$ independent solutions
\begin{equation}
\alpha_{j}=\sum_{l=1}^{\mathcal{N}}\omega_{l}(\zeta_l)^j,
\end{equation}
with $\{\omega_l\}$ undetermined coefficients, and for $j\leq \mathcal{P}$, we
have $\mathcal{P}$ constraint conditions
\begin{eqnarray}
\sum_{n=1}^{N_f}{J}_n^x\alpha_{j+n}+\mu\alpha_j+\sum_{n=1}^{j-1}{J}_n^y\alpha_{j-n}=0.\label{cc}
\end{eqnarray}
Thus, we have $(\mathcal{N}-\mathcal{P})$ independent normalized left zero modes
$\phi_{\textrm{L}}^1,...,\phi_{\textrm{L}}^{(\mathcal{N}-\mathcal{P})}$ with coefficients
$\{\alpha_j^1\},...,\{\alpha_j^{(\mathcal{N}-\mathcal{P})}\}$, where the orthogonal Majorana zero modes
can be obtained by using the Schmidt orthogonalization with conditions
$\{\phi^i,{\phi^j}^\dag\}=2\delta_{ij}$. These considerations also hold for linear combinations
of Majorana operators $\{b_j\}$ with the form
\begin{equation}
\psi^i=\sum_{j=1}^{L}\beta_j^i b_j,
\end{equation}
and
\begin{equation}
\beta_j^i=\alpha_{L-j+1}^i,
\end{equation}
because Majorana zero modes appear in pairs \cite{Kitaev2001}. For the other case $\mathcal{N}<\mathcal{P}$,
we should consider right Majorana zero modes that require $|\alpha_{1}|\rightarrow0$ for $L\gg1$ and
the characteristic equation $\bar{g}(\zeta)=g(1/\zeta)=0$, with $\bar{\mathcal{N}}$ zeros
and $\bar{\mathcal{P}}$ poles in $|\zeta|<1$, where we can obtain that
\begin{equation}
\mathcal{N}+\bar{\mathcal{N}}=\bar{\mathcal{P}}+\mathcal{P},
\end{equation}
and have $(\mathcal{P}-\mathcal{N})$ right Majorana zero modes $\phi_{\textrm{R}}^1,\phi_{\textrm{R}}^2,\cdots,\phi_{\textrm{R}}^{(\mathcal{P}-\mathcal{N})}$.
Therefore, we derive that in the thermodynamic limit $L\gg N_f\geq1$, the number of
Majorana zero modes at each end of the extended Kitaev open chain, defined as $\mathcal{M}_{0}$,
equals the absolute value of the winding number:
\begin{eqnarray}
\mathcal{M}_{0}=|\mathcal{N}-\mathcal{P}|=|\nu|.
\end{eqnarray}
Here, we should note that there exist special cases when degenerate
solutions of Majorana zero modes might occur for some choices of parameters and
could be averted as we consider the perturbation of characteristic
functions.

Moreover, while the coefficients $\{\alpha_j\}$ are not real, the zero modes $\phi$ and $\psi$, with
conditions $\{\phi^i,{\phi^j}^\dag\}=\{\psi^i,{\psi^j}^\dag\}=2\delta_{ij}$ and
$\{\phi^i,{\psi^j}^\dag\}=\{\phi^i,{\psi^j}\}=0$, are not Majorana  operators \cite{Lepori2017}.
Fortunately, for $\mathcal{N}\geq\mathcal{P}$, left and right Majorana zero modes can be combined as
$(\mathcal{N}-\mathcal{P})$ fermion modes $d^1,d^2,\cdots,d^{(\mathcal{N}-\mathcal{P})}$ with
\begin{equation}
d^i={(\phi^i_{\textrm{L}}+i\psi^i_{\textrm{R}})}/{2},
\end{equation}
that commute with the Hamiltonian in the thermodynamic limit. Conversely, for $\mathcal{P}\geq\mathcal{N}$,
there exist $(\mathcal{P}-\mathcal{N})$ fermion zero modes with operators $\bar{d}^1,\bar{d}^2,\cdots,\bar{d}^{(\mathcal{P}-\mathcal{N})}$, where
\begin{equation}
\bar{d}^i={(\phi^i_{\textrm{R}}+i\psi^i_{\textrm{L}})}/{2}.
\end{equation}

Our discussions also provide an effective method for finding the distribution of Majorana zero modes
by finding the zeros and poles of the characteristic functions $g(\zeta)$ in momentum space.
Moreover, the topological phase transitions occur when the parameters satisfy the existence
of zeros of the characteristic functions on the critical contour $|\zeta|=1$, see Sec.~\ref{VI}
for details.

\section{Quantum Fisher information of topological states}
Given a generator $\mathcal{O}$ with respect to the parameter $t$, the quantum
Fisher information of the pure ground state $|\mathcal{G}\rangle$ can be written as
\cite{BRAUNSTEIN1994,Pezze2009,Giovannetti2011,Ma2011}
\begin{eqnarray}
F_Q[\mathcal{O},|\mathcal{G}\rangle]\ =\ 4(\Delta\mathcal{O})^2\ =\ 4(\langle\mathcal{O}^2\rangle_{\mathcal{G}}-\langle\mathcal{O}\rangle_{\mathcal{G}}^2).
\end{eqnarray}
For critical systems with $L$ sites, we consider the quantum Fisher information density
with the form
\begin{eqnarray}
f_Q[\mathcal{O},|\mathcal{G}\rangle]=\frac{F_{Q}[\mathcal{O},|\mathcal{G}\rangle]}{L},
\end{eqnarray}
and the violation of the inequality
$f_Q\leq\kappa$ signals $(\kappa+1)$-partite entanglement ($1\leq\kappa\leq L-1$).

For instance, we consider a Kitaev chain which is a tight-binding
model with strengths of tunneling $J$ and superconducting pairing $\Delta$ \cite{Kitaev2001}:
\begin{equation}
H=\sum_{j=1}^{L-1}\left(\frac{\Delta}{2}c_jc_{j+1}-\frac{J}{2}c_j^\dag c_{j+1}+\textrm{h.c.}\right)-{\mu}\sum_{j=1}^L\left(n_j-\frac{1}{2}\right),
\end{equation}
with the fermion number operator $n_j\equiv c_j^\dag c_j$. For $J=\Delta$ and
zero chemical potentials $\mu=0$, we have one Majorana zero mode at each end, and the Hamiltonian
may be written in terms of Majorana operators and Dirac fermion
operators
\begin{equation}
d_{j,1}=(b_j+ia_{j+1})/2
\end{equation}
as a diagonal form
\begin{eqnarray}
H=i\frac{J}{2}\sum_{j=1}^{L-1}b_{j}a_{j+1}=\sum_{j=1}^{L-1}J\left(d_{j,1}^\dag d_{j,1}-\frac{1}{2}\right),\label{ge}
\end{eqnarray}
where we have a winding number $\nu=1$. Here, to detect multipartite entanglement,
it requires to choose a pair of nonlocal generators \cite{Pezze2017}
\begin{eqnarray}
\mathcal{O}_{\nu=1}=\sum_{j=1}^L\sigma^{x}_j/2,\hspace*{0.2in}\mathcal{O}_{\nu=1}^{(\textrm{st})}=\sum_{j=1}^L(-)^j\sigma^{x}_j/2.\ \ \ \ 
\end{eqnarray}
Using the Jordan-Wigner transformation as
\begin{equation}
-\sigma^{x}_j=c_j^\dag \exp\left({i\pi\sum_{l=1}^{j-l}c_l^\dag c_l}\right)+\exp\left({-i\pi\sum_{l=1}^{j-l}c_l^\dag c_l}\right)c_j,
\end{equation}
the quantum Fisher information density of the ground state of the Kitaev chain can
be written in terms of longitudinal spin-spin correlation functions:
\begin{align}
&f_{Q}[\mathcal{O}_{\nu=1},|\mathcal{G}\rangle]=1+\sum_{r=1}^{L-1}C_{\nu=1}(r),\\
&f_{Q}[\mathcal{O}_{\nu=1}^{(\textrm{st})},|\mathcal{G}\rangle]=1+\sum_{r=1}^{L-1}(-)^rC_{\nu=1}(r),
\end{align}
with respect to the generators $\mathcal{O}_{\nu=1}$ and $\mathcal{O}_{\nu=1}^{(\textrm{st})}$,
respectively. Here, we have used the fact that $\langle\sigma_j^{x}\rangle_{\mathcal{G}}=0$
and considered a closed chain for $L\gg1$. Moreover, the $x$-directional longitudinal
correlation function can be written
as
\begin{equation}
C_{\nu=1}(r)=\left\langle\prod_{l=i}^{j-1}(- ib_{l}a_{l+1})\right\rangle_{\!\!\!\mathcal{G}}
=\left\langle\prod_{l=i}^{j-1}(1-2d_{l,1}^\dag d_{l,1})\right\rangle_{\!\!\!\mathcal{G}},
\end{equation}
which represents the average of the Majorana parity from site $i$ to $j$ ($j-i=r$) and
does not include the edge modes. For $J>0$, we have
\begin{equation}
\langle d_{l,1}^\dag d_{l,1}\rangle_{\mathcal{G}}=0,
\end{equation}
so the Majorana zero modes give
\begin{equation}
f_{Q}[\mathcal{O}_{\nu=1},|\mathcal{G}\rangle] = L,
\end{equation}
which signals the maximal $L$-partite
entanglement with the generator $\mathcal{O}_{\nu=1}$. On the contrary, for
$J<0$, we have
\begin{equation}
\langle d_{l,1}^\dag d_{l,1}\rangle_{\mathcal{G}}=1,
\end{equation}
 such that the
edge Majorana zero modes lead to the fact that
\begin{equation}
f_{Q}[\mathcal{O}_{\nu=1}^{(\textrm{st})},|\mathcal{G}\rangle] = L,
\end{equation}
with respect to the generator $\mathcal{O}_{\nu=1}^{(\textrm{st})}$. Therefore, the choice of generators between
the  operator $\mathcal{O}_{\nu=1}$ and the staggered operator $\mathcal{O}_{\nu=1}^{(\textrm{st})}$
depends on the sign of the direct interaction between the chain ends as discussed
in Ref.~\cite{Kitaev2001}. These results also hold for the open chain, because the
correlation function does not include the fermion edge modes. For the other case,
we choose $J=-\Delta$ and $\mu=0$, where the winding number is $\nu=-1$. Then,
the quantum Fisher information
density $f_Q$ of the ground state $|\mathcal{G}\rangle$ with respect to the generators:
\begin{eqnarray}
\mathcal{O}_{\nu=-1}=\sum_{j=1}^L\sigma^{y}_j/2,\hspace*{0.2in}\mathcal{O}_{\nu=-1}^{(\textrm{st})}=\sum_{j=1}^L(-)^j\sigma^{y}_j/2.\ \ \ \ 
\end{eqnarray}
can detect symmetry-protected topological order and Majorana zero modes with $\nu=-1$.

\begin{figure*}[t]
\centering
\includegraphics[width=0.97\textwidth]{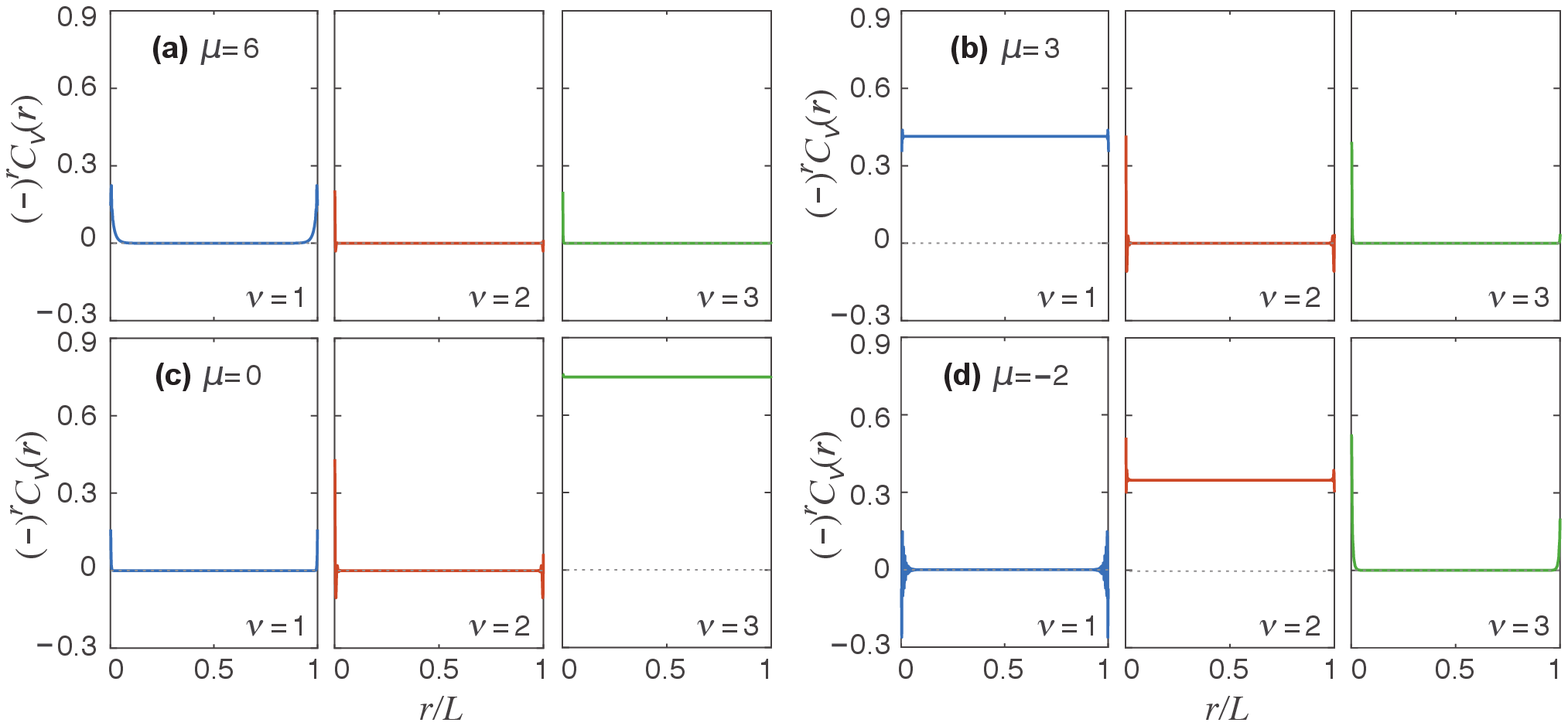}\\
\caption{(color online) The staggered string correlation functions $(-)^rC_\nu(r)$ versus the normalized distance $r/L$ for the extended
Kitaev fermion chain  with a system size $L=600$, third neighbor interactions ($N_f=3$) and nonzero parameters: ${J}_1^+={J}_1^-=1$, ${J}_{2}^+={J}_{2}^-=2$, ${J}_{3}^+={J}_{3}^-=2$.}\label{fig:A1}
\end{figure*}

The interchange between the quantum phases with positive and negative winding numbers $\nu=\pm1$
\begin{align}
\mathcal{O}_{\nu=1}^{(\textrm{st})}\leftrightarrow\mathcal{O}_{\nu=-1}^{(\textrm{st})},&\hspace*{0.2in}\mathcal{O}_{\nu=1}\leftrightarrow\mathcal{O}_{\nu=-1}\\
{f}_Q[\mathcal{O}_{\nu=1}^{(\textrm{st})}]\leftrightarrow{f}_Q[\mathcal{O}_{\nu=-1}^{(\textrm{st})}],&\hspace*{0.2in}{f}_Q[\mathcal{O}_{\nu=1}]\leftrightarrow{f}_Q[\mathcal{O}_{\nu=-1}]
\end{align}
can be realized by a phase redefinition $c_j \rightarrow \pm i c_j$.
Another interchange between the staggered operator $\mathcal{O}_{\nu=1}^{(\textrm{st})}$ and the operator $\mathcal{O}_{\nu=1}$, for the positive and negative signs of the interaction between Dirac fermions localized
at the chain ends, respectively,
\begin{align}
\mathcal{O}_{\nu=1}^{(\textrm{st})}\leftrightarrow\mathcal{O}_{\nu=1},&\hspace*{0.2in}\mathcal{O}^{(\textrm{st})}_{\nu=-1}\leftrightarrow\mathcal{O}_{\nu=-1}\\
{f}_Q[\mathcal{O}_{\nu=1}^{(\textrm{st})}]\leftrightarrow{f}_Q[\mathcal{O}_{\nu=1}],&\hspace*{0.2in}{f}_Q[\mathcal{O}_{\nu=-1}^{(\textrm{st})}]\leftrightarrow{f}_Q[\mathcal{O}_{\nu=-1}]
\end{align}
can be realized by a Hermitian conjugate transformation $c_j \rightarrow c_j^\dag$.

\begin{figure}[t]
\centering
\includegraphics[width=0.47\textwidth]{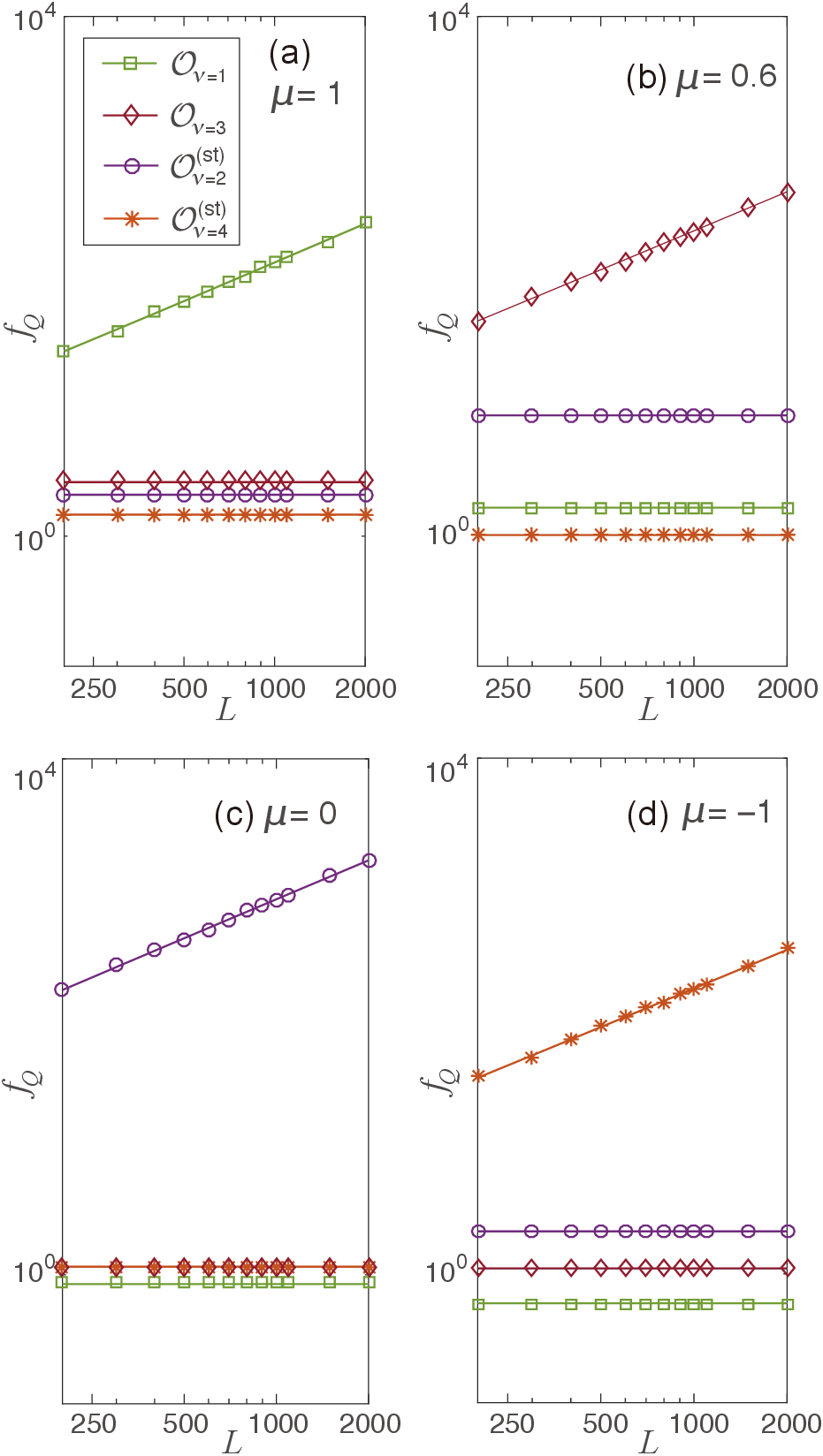}\\
\caption{(color online) Dual quantum Fisher information density $f_Q$ of the ground state $|\mathcal{G}\rangle$ with respect to the dual generators
$\mathcal{O}_{\nu}$ and $\mathcal{O}_{\nu}^{(\textrm{st})}$ as a
function of $L$ for the extended Kitaev fermion chain with longer-range interactions and with nonzero
parameters: ${J}_1^+={J}_1^-=0.1$, ${J}_{2}^+={J}_{2}^-=0.21$, ${J}_{3}^+={J}_{3}^-=-0.74$, ${J}_4^+={J}_4^-=0.9$ ($N_f=4$), in different topological phases. (a)
For $\mu=1$, the winding number $\nu=1$, and the fitting nontrivial scaling topological index $\lambda_{\nu=1}=0.9837$. (b) For $\mu=0.6$, $\nu=3$, and $\lambda_{\nu=3}=0.9941$. (c) For $\mu=0$, $\nu=2$, and  $\lambda_{\nu=2}^{(\textrm{st})}=1.0051$. (d) For $\mu=4$, $\nu=4$, and $\lambda_{\nu=4}^{(\textrm{st})}=0.9933$.
}\label{fig:S1}
\end{figure}

Generally for $\mu\neq0$, we can calculate the longitudinal correlation function by defining
\begin{eqnarray}
A_l=c_l^\dag+c_l=a_l,\hspace*{0.2in}B_{l}=c_l^\dag-c_l=-ib_l.
\end{eqnarray}
The correlation functions in the $x$ and $y$ directions can be written as
\begin{align}
C_{\nu=1}(r)&=\langle\mathcal{G}|B_iA_{i+1}...A_{j-1}B_{j-1}A_j|\mathcal{G}\rangle,\\
C_{\nu=-1}(r)&=-\langle\mathcal{G}|A_iB_{i+1}...B_{j-1}A_{j-1}B_j|\mathcal{G}\rangle,
\end{align}
where $j-i=r$.
Using Wick's theorem, we can write the $x$-directional spin correlation function into a
determinant of size $r$ \cite{BAROUCH1971}
\begin{eqnarray}
C_{\nu=1}(r)=\left|\begin{array}{c c c c}
G_{-1}&G_{-2}&\cdots&G_{-r}\\
G_0&G_{-1}&\cdots&G_{-r+1}\\
G_{1}&G_0&\cdots&G_{-r+2}\\
\vdots&\vdots&\vdots&\vdots\\
G_{r-2}&G_{r-3}&\cdots&G_{-1}
\end{array}\right|,
\end{eqnarray}
and similarly, we have the $y$-directional spin correlation function as
\begin{eqnarray}
C_{\nu=-1}(r)=\left|\begin{array}{c c c c}
G_{1}&G_{0}&\cdots&G_{-r+2}\\
G_2&G_{1}&\cdots&G_{-r+3}\\
G_{3}&G_2&\cdots&G_{-r+4}\\
\vdots&\vdots&\vdots&\vdots\\
G_{r}&G_{r-1}&\cdots&G_{1}
\end{array}\right|,
\end{eqnarray}
where we have 
\begin{align}
G_{-r}\equiv\langle\mathcal{G}|B_iA_{i+r}|\mathcal{G}\rangle
\end{align} 
and $\langle\mathcal{G}|A_iA_{j}|\mathcal{G}\rangle=\langle\mathcal{G}|B_iB_{j}|\mathcal{G}\rangle=\delta_{ij}$.

\begin{table*}[t]
\begin{tabular}{c|c c |c c| c c| c c}
\hline
\hline
$~~~\mu~~~$ & $\lambda_{\nu=1}^{(\textrm{st})}$ & $\lambda_{\nu=1}$  & $\lambda_{\nu=2}^{(\textrm{st})}$ & $\lambda_{\nu=2}$ & $\lambda_{\nu=3}^{(\textrm{st})}$  & $\lambda_{\nu=3}$ & $\lambda_{\nu=4}^{(\textrm{st})}$  & $\lambda_{\nu=4}$\\
\hline
1&
$4.8\times10^{-7}$ &
\textcolor{blue}{0.9837}&
$-2.0\times10^{-6}$&
$2.1\times10^{-6}$ &
$-8.0\times10^{-7}$&
$4.4\times10^{-5}$ &
$1.9\times10^{-6}$&
$5.2\times10^{-7}$\\
0.6&
$-8.6\times10^{-8}$ &
$8.0\times10^{-8}$&
$-1.3\times10^{-7}$&
$3.3\times10^{-8}$&
$-6.9\times10^{-8}$ &
\textcolor{blue}{0.9941}&
$7.4\times10^{-7}$&
$-3.5\times10^{-8}$\\
$0$ &
$5.8\times10^{-14}$  &
$-6.7\times10^{-14}$ &
$3.1\times10^{-14}$ &
$1.5\times10^{-14}$ &
$6.1\times10^{-14}$ &
$-5.5\times10^{-14}$ &
\textcolor{blue}{1.0051}&
$2.5\times10^{-13}$
\\
$-1$ &
$9.5\times10^{-14}$  &
$-2.4\times10^{-13}$ &
\textcolor{blue}{0.9933} &
$2.1\times10^{-14}$ &
$-2.2\times10^{-14}$ &
$-1.6\times10^{-13}$ &
$3.3\times10^{-14}$&
$3.8\times10^{-14}$ \\
\hline
\hline
\end{tabular}
\caption{Fitting of the scaling coefficients $\lambda_{\nu}$ and 
$\lambda_{\nu}^{(\textrm{st})}$ with respect to the
dual generators $\mathcal{O}_{\nu}$ and $\mathcal{O}_{\nu}^{(\textrm{st})}$, respectively, for the different
topological phases for the  extended  Kitaev fermion chain with parameters
${J}_1^+={J}_1^-=0.1$, ${J}_{2}^+={J}_{2}^-=0.21$, ${J}_{3}^+={J}_{3}^-=-0.74$, ${J}_4^+={J}_4^-=0.9$ ($N_f=4$), and chain length up to $L=2000$.
The four essentially non-zero scaling coefficients are shown in blue font, and all
four are close to $1$.}\label{tab:1}
\end{table*}

\section{Duality Transformation}
The duality transformation connects different  but equivalent mathematical descriptions
of a system or a state of matter through a mapping by the change of variables in quantum physics \cite{FRADKIN1978,Smacchia2011,Feng2007,Qin2017}.
For example, an Ising chain with an external field $h$ has a self-duality symmetry, mapping between
the ordered and disordered phases, expressed as
\begin{align}
H_{\textrm{Ising}}=\sum_j(\sigma_j^x\sigma_{j+1}^x+h\sigma_j^z)=h\sum_j(s_j^xs_{j+1}^x+h^{-1}s_j^z)
\end{align}
with the duality transformation
\begin{align}
s_j^x=\prod_{k\leq j}\sigma_k^z,\hspace{0.2in}s_j^z=\sigma_j^x\sigma_{j+1}^x, \hspace{0.2in}s_j^y=-is_j^zs_j^x,
\end{align}
where both $\sigma$ and $s$ satisfy the same algebra.
By this duality transformation, the cluster Ising model \cite{Smacchia2011,Cui2013} can
be mapped to  an anisotropic $XY$ model
\begin{align}
H_{\textrm{cluster}}&=\sum_j(\sigma_{j-1}^x\sigma_{j}^z\sigma_{j+1}^x+h\sigma_j^z)\\
&=\sum_j(-s_j^ys_{j+1}^y+hs_j^xs_{j+1}^x),\label{dI}
\end{align}
of which the ordered phase can help to characterize the symmetry-protected topological phase by
a $\mathbf{Z}_2\times\mathbf{Z}_2$ symmetry of the cluster Ising model.
Therefore, as shown in \cite{Smacchia2011,Cui2013}, this symmetry-protected topological phase
can be characterized by the \emph{unlocal} string correlation function \cite{Venuti2005} equal to a \emph{local} correlator  in the dual
lattice of the Ising model with the form
\begin{align}
(-)^rC_{\nu=2}(r)&=(-)^r\langle s_j^ys_{j+r}^y\rangle_{\mathcal{G}}\\
=&(-)^r\left\langle\sigma_j^x\sigma_{j+1}^y\left(\prod_{k=2}^{r-1}\sigma_{j+k}^z\right)\sigma_{j+r}^y\sigma_{j+r+1}^x\right\rangle_{\mathcal{G}},\label{unlocal}
\end{align}
from site $j$ to $(j+r)$ in the dual lattice.
It is shown in Ref.~\cite{Cobanera2011} that the Jordan-Wigner transformation mapping between a 
one-dimensional spin-$\frac{1}{2}$
model and free fermion chain  can also be regarded as a dual transformation with a bond-algebraic
approach. Through the Jordan-Wigner transformation, the cluster Ising model corresponds to an extended
Kitaev chain with a $\textbf{Z}_4$ symmetry.
Thus, the self-duality properties of the Ising model (\ref{dI}) can help to study
topological phases and multipartite entanglement in the symmetry-protected phase with a winding number $\nu=2$ in
the extended Kitaev chain.
Generally, we find that for the extended Kitaev chain, the string correlation function can be written
as a spin correlation function with respect to the spin operators from the self-duality symmetry of
the extended Ising model.

The duality transformation for topological phases with a winding number $\nu=2$ can be written as
\begin{align}
&\mathbb{Z}_j^{{(2)}}=\sigma_j^x\sigma_{j+1}^x,\hspace*{0.2in}\mathbb{X}_j^{{(2)}}=\prod_{l=1}^{j}\sigma_l^z,\\
&\mathbb{Y}_j^{{(2)}}=-i\mathbb{Z}_j^{{(2)}}\mathbb{X}_j^{{(2)}}=-\left(\prod_{l=1}^{j-1}\sigma_l^z\right)\sigma_j^y\sigma_{j+1}^x
\end{align}
which implies that
\begin{equation}
\mathbb{X}_j^{{(2)}} \mathbb{X}_{j+1}^{{(2)}}=\sigma_{j+1}^z.
\end{equation}
Therefore, the duality transformation connects two Ising models as
\begin{eqnarray}
\sum_{j=1}^L\sigma_j^x\sigma_{j+1}^x+\mu\sigma_j^z=\sum_{j=1}^L\mathbb{Z}_j^{{(2)}}+\mu\mathbb{X}_j^{{(2)}}\mathbb{X}_{j+1}^{{(2)}}.
\end{eqnarray}
The spin correlation function with dual $y$-directional spin operators between sites $i$ and $j{=i+r}$  equals to the string correlation function:
\begin{align}
{C_{\nu=2}(r)}=\left\langle\mathbb{Y}_i^{{(2)}}\mathbb{Y}_j^{{(2)}}\right\rangle_{\mathcal{G}}=\left\langle\prod_{l=i}^{j-1}\sigma_l^{x}\sigma_{l+1}^{z}\sigma_{l+2}^x\right\rangle_{\!\!\!\mathcal{G}}.
\end{align}
Similarly, the duality transformation for topological phases with $\nu=-2$ can be written as
\begin{align}
&{\mathbb{Z}}_j^{{(-2)}}=\sigma_j^y\sigma_{j+1}^y,\hspace*{0.2in}{\mathbb{Y}}_j^{{(-2)}}=\prod_{l=1}^{j}\sigma_k^z,\\
&{\mathbb{X}}_j^{{(-2)}}=-i{\mathbb{Y}}_j^{{(-2)}}{\mathbb{Z}}_j^{{(-2)}}=-\left(\prod_{l=1}^{j-1}\sigma_l^z\right)\sigma_j^x\sigma_{j+1}^y
\end{align}
which implies that
\begin{equation}
{\mathbb{X}}_j^{{(-2)}} {\mathbb{X}}_{j+1}^{{(-2)}}=\sigma_{j+1}^z,
\end{equation}
and
\begin{equation}
\sum_{j=1}^L\sigma_j^y\sigma_{j+1}^y+\mu\sigma_j^z=\sum_{j=1}^L\mathbb{Z}_j^{{(-2)}}+\mu\mathbb{Y}_j^{{(-2)}}\mathbb{Y}_{j+1}^{{(-2)}}.
\end{equation}
The dual $x$-directional correlation function between sites $i$ and $j{=i+r}$ equals to the string correlation function
\begin{align}
{C_{\nu=-2}(r)}=\left\langle{\mathbb{X}}_i^{{(-2)}}{\mathbb{X}}_j^{{(-2)}}\right\rangle_{\mathcal{G}}
=\left\langle\prod_{l=i}^{j-1}\sigma_l^{y}\sigma_{l+1}^{z}\sigma_{l+2}^y\right\rangle_{\!\!\!\mathcal{G}}.
\end{align}
We can therefore define the dual spin operators as
\begin{eqnarray}\left\{
\begin{array}{ll}
\tau_j^{{(2)}}=\mathbb{Y}_j^{{(2)}},&\textrm{for}\ \nu=2,\\
\tau_j^{{(-2)}}=\mathbb{X}_j^{{(-2)}},&\textrm{for}\ \nu=-2.
\end{array}\right.
\end{eqnarray}

The duality transformation for $\nu=3$ can be written as
\begin{align}
&\mathbb{Z}_j^{{(3)}}=\sigma_j^x\sigma_{j+1}^z\sigma_{j+2}^x,\hspace*{0.2in}\mathbb{X}_j^{{(3)}}=\sigma_{j+1}^x,\\
&\mathbb{Y}_j^{{(3)}}=-i\mathbb{Z}_j^{{(3)}}\mathbb{X}_j^{{(3)}}=\sigma_j^x\sigma_{j+1}^y\sigma_{j+2}^x
\end{align}
which implies that
\begin{equation}
\mathbb{X}_j^{{(3)}}\mathbb{Z}_{j+1}^{{(3)}}\mathbb{X}_{j+2}^{{(3)}}=\sigma_{j+2}^z.
\end{equation}
The duality transformation for $\nu=-3$ can be written as
\begin{align}
&\mathbb{Z}_j^{{(-3)}}=\sigma_j^y\sigma_{j+1}^z\sigma_{j+2}^y,\hspace*{0.2in}\mathbb{Y}_j^{{(-3)}}=\sigma_{j+1}^y,\\
&\mathbb{X}_j^{{(-3)}}=-i\mathbb{Y}_j^{{(-3)}}\mathbb{Z}_j^{{(-3)}}=\sigma_j^y\sigma_{j+1}^x\sigma_{j+2}^y
\end{align}
which implies that
\begin{equation}
\mathbb{Y}_j^{{(-3)}}\mathbb{Z}_{j+1}^{{(-3)}}\mathbb{Y}_{j+2}^{{(-3)}}=\sigma_{j+2}^z.
\end{equation}
Thus, we can  define the dual spin operators as
\begin{eqnarray}\left\{
\begin{array}{ll}
\tau_j^{{(3)}}=\mathbb{Y}_j^{{(3)}},&\textrm{for}\ \nu=3,\\
\tau_j^{{(-3)}}=\mathbb{X}_j^{{(-3)}},&\textrm{for}\ \nu=-3.
\end{array}\right.
\end{eqnarray}

Generally, the formalism  of string correlation functions and dual
spin operators depend on the parity of the winding numbers \cite{Fidkowski2011}.
We first consider
the odd winding numbers with $p>1$:
For positive odd winding numbers $\nu=2p-1$, we have
\begin{align}
&\mathbb{Z}_j^{(2p-1)}=\sigma_j^x\left(\prod_{l=1}^{2p-3}\sigma_{j+l}^z\right)\sigma_{j+2p-2}^x,\\
&\mathbb{X}_j^{(2p-1)}=\left(\prod_{l=1}^{p-2}\sigma_{j+2l-1}^x\sigma_{j+2l}^y\right)\sigma_{j+2p-3}^x,\\
&\mathbb{Y}_j^{(2p-1)}=\sigma_j^x\left(\prod_{l=1}^{p-1}\sigma_{j+2l-1}^y\sigma_{j+2l}^x\right),
\end{align}
which implies
\begin{equation}
\mathbb{X}_j^{(2p-1)}\left(\prod_{l=1}^{2p-3}\mathbb{Z}_{j+l}^{(2p-1)}\right)\mathbb{X}_{j+2p-2}^{(2p-1)}=\sigma_{j+2p-2}^z.
\end{equation}
For negative odd winding numbers $\nu=1-2p$, we have
\begin{align}
&{\mathbb{Z}}_j^{(1-2p)}=\sigma_j^y\left(\prod_{l=1}^{2p-3}\sigma_{j+l}^z\right)\sigma_{j+2p-2}^y,\\
&{\mathbb{Y}}_j^{(1-2p)}=\left(\prod_{l=1}^{p-2}\sigma_{j+2l-1}^y\sigma_{j+2l}^x\right)\sigma_{j+2p-3}^y,\\
&{\mathbb{X}}_j^{(1-2p)}=\sigma_j^y\left(\prod_{l=1}^{p-1}\sigma_{j+2l-1}^x\sigma_{j+2l}^y\right),
\end{align}
which implies
\begin{equation}
\mathbb{Y}_j^{(1-2p)}\left(\prod_{l=1}^{2p-3}\mathbb{Z}_{j+l}^{(1-2p)}\right)\mathbb{Y}_{j+2p-2}^{(1-2p)}=\sigma_{j+2p-2}^z.
\end{equation}
Thus, we can  define the dual spin operators as
\begin{eqnarray}
\left\{
\begin{array}{ll}
\tau_j^{(2p-1)}=\mathbb{Y}_j^{(2p-1)},&\textrm{for}\ \nu=2p-1,\\
\tau_j^{(1-2p)}=\mathbb{X}_j^{(1-2p)},&\textrm{for}\ \nu=1-2p.
\end{array}\right.
\end{eqnarray}

We then consider the even winding numbers with $p>1$:
For positive even winding numbers $\nu=2p$, we have
\begin{align}
&\mathbb{Z}_j^{{(2p)}}=\sigma_j^x\left(\prod_{l=1}^{2p-2}\sigma_{j+l}^z\right)\sigma_{j+2p-1}^x,\\
&\mathbb{X}_j^{{(2p)}}=\left(\prod_{k=1}^{j}\sigma_k^z\right)\left(\prod_{l=1}^{p-1}\sigma_{j+2l-1}^y\sigma_{j+2l}^x\right)\\
&\mathbb{Y}_j^{{(2p)}}=-\left(\prod_{k=1}^{j-1}\sigma_k^z\right)\left(\prod_{l=1}^{p}\sigma_{j+2l-2}^y\sigma_{j+2l-1}^x\right)
\end{align}
which implies
\begin{equation}
\mathbb{X}_j^{{(2p)}}\left(\prod_{l=1}^{2p-2}\mathbb{Z}_{j+l}^{{(2p)}}\right)\mathbb{X}_{j+2p-1}^{{(2p)}}=\sigma_{j+2p-1}^z.
\end{equation}
For negative even winding numbers $\nu=-2p$, we have
\begin{align}
&{\mathbb{Z}}_j^{{(-2p)}}=\sigma_j^y\left(\prod_{l=1}^{2p-2}\sigma_{j+l}^z\right)\sigma_{j+2p-1}^y,\\
&{\mathbb{Y}}_j^{{(-2p)}}=\left(\prod_{k=1}^{j}\sigma_k^z\right)\left(\prod_{l=1}^{p-1}\sigma_{j+2l-1}^x\sigma_{j+2l}^y\right)\\
&{\mathbb{X}}_j^{{(-2p)}}=-\left(\prod_{k=1}^{j-1}\sigma_k^z\right)\left(\prod_{l=1}^{p}\sigma_{j+2l-2}^x\sigma_{j+2l-1}^y\right)
\end{align}
which implies
\begin{equation}
\mathbb{Y}_j^{{(-2p)}}\left(\prod_{l=1}^{2p-2}\mathbb{Z}_{j+l}^{{(-2p)}}\right)\mathbb{Y}_{j+2p-1}^{{(-2p)}}=\sigma_{j+2p-1}^z.
\end{equation}
Thus, we can write the dual spin operators as
\begin{eqnarray}\left\{
\begin{array}{ll}
\tau_j^{{(2p)}}=\mathbb{Y}_j^{{(2p)}},&\textrm{for}\ \nu=2p,\\
\tau_j^{{(-2p)}}=\mathbb{X}_j^{{(-2p)}},&\textrm{for}\ \nu=-2p.
\end{array}\right.
\end{eqnarray}

 \begin{figure}[t]
\centering
\includegraphics[width=0.46\textwidth]{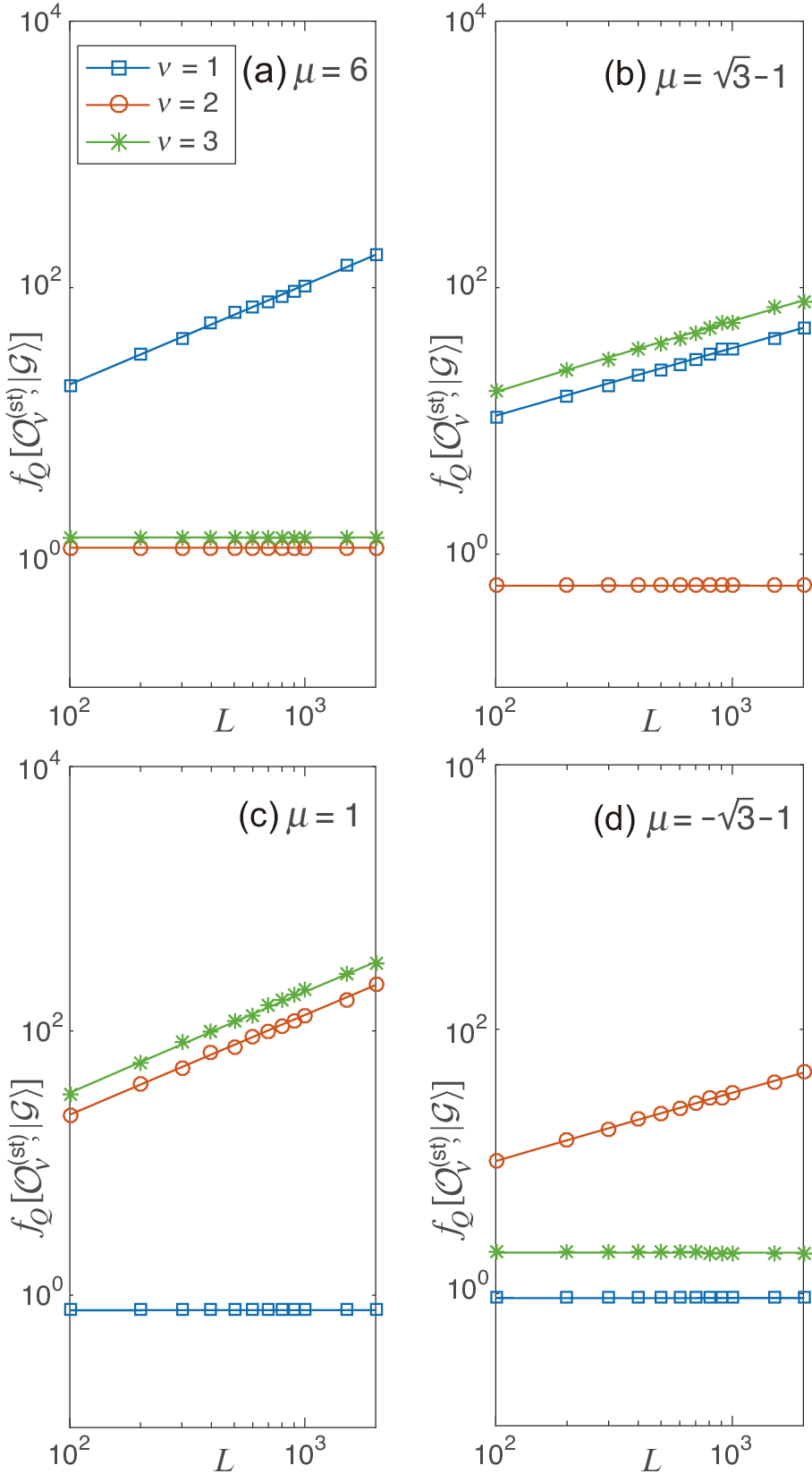}\\
\caption{(color online)  Quantum Fisher information density $f_Q[\mathcal{O}_\nu^{(\textrm{st})},|\mathcal{G}\rangle]$
as a function of $L$ for the extended Kitaev fermion chain
with nonzero parameters ${J}_1^+={J}_1^-=1$, ${J}_{2}^+={J}_{2}^-=2$, ${J}_{3}^+={J}_{3}^-=2$ ($N_f=3$) on the contour
between different topological phases for (a) $\mu=5$, (b) $\mu=\sqrt{3}-1$, (c) $\mu=-1$, and (d) $\mu=-\sqrt{3}-1$.
The scaling coefficients $\lambda_\nu^{(\textrm{st})}$ are shown in Tab.~\ref{tab:2}.}\label{fig:S2}
\end{figure}

\begin{figure}[t]
\centering
\includegraphics[width=0.467\textwidth]{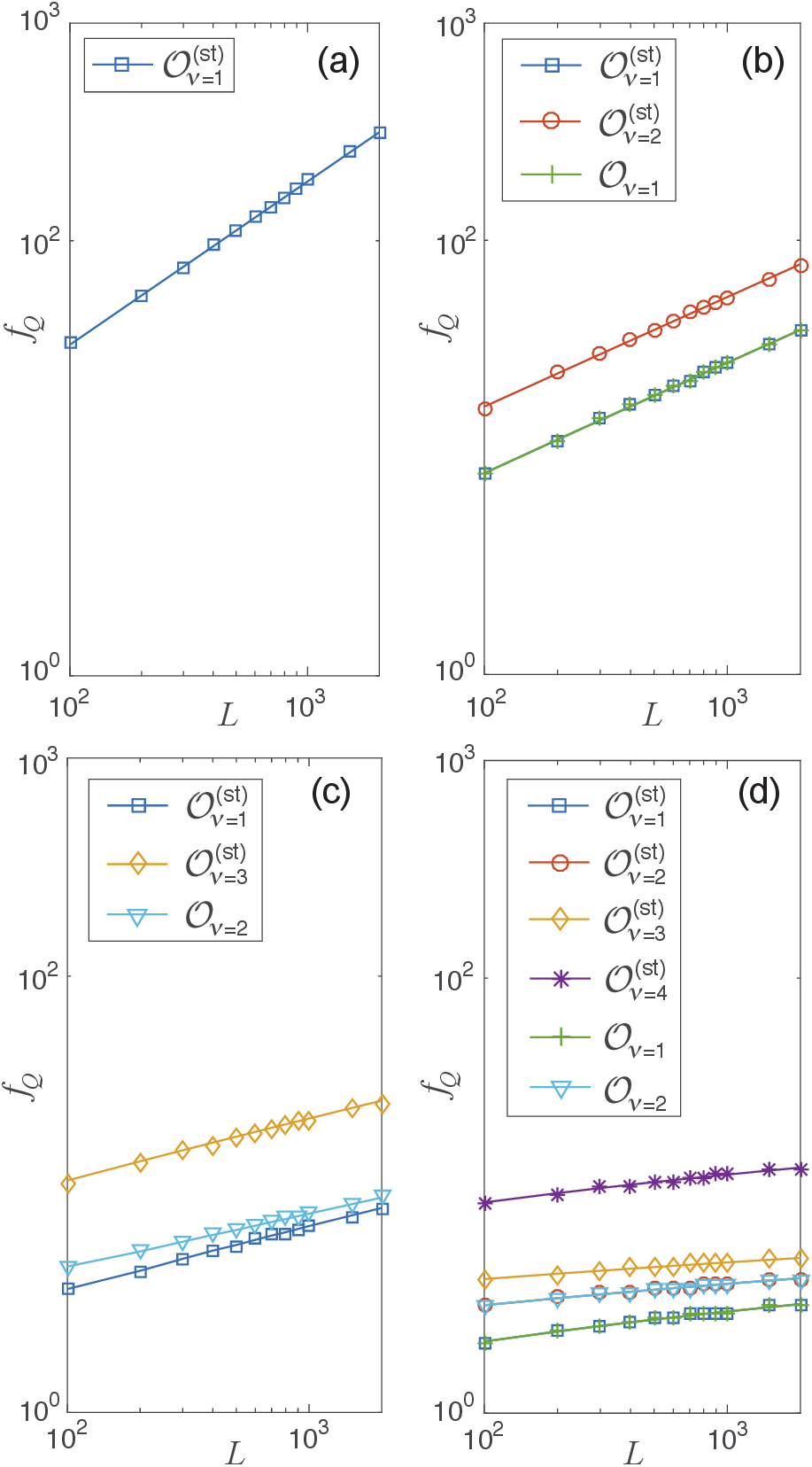}\\
\caption{(color online) Quantum Fisher information density
$f_Q$ of the ground state $|\mathcal{G}\rangle$ with respect to the dual generators
$\mathcal{O}_{\nu}$ and $\mathcal{O}_{\nu}^{(\textrm{st})}$ as a function of $L$ for the extended Kitaev fermion chain when $\mu=1$
with nonzero parameters: (a) ${J}_1^+={J}_1^-=1$ ($N_f=1$); (b) ${J}_2^+={J}_2^-=1$ ($N_f=2$); (c) ${J}_3^+={J}_3^-=1$ ($N_f=3$); and (d) ${J}_4^+={J}_4^-=1$ ($N_f=4$).
The scaling coefficients $\lambda_{\nu}$ and $\lambda_{\nu}^{(\textrm{st})}$ are shown in Tab.~\ref{tab:3}.
}\label{fig:S3}
\end{figure}

\section{Quantum Fisher information density and string correlation functions}
For higher winding numbers $\nu=\pm2,\pm3,\cdots$, the quantum Fisher information with respect to
the dual generators
\begin{equation}
{\mathcal{O}_\nu}=\sum_{j=1}^{M}\tau_j^{{(\nu)}},\hspace*{0.2in}{\mathcal{O}_\nu^{(\textrm{st})}}=\sum_{j=1}^{M}(-)^j\tau_j^{{(\nu)}}
\end{equation}
can be written as
\begin{align}
&{F_{Q}[\mathcal{O}_{\nu},|\mathcal{G}\rangle]}=M+M\sum_{r=1}^{M-1}\langle\tau_i^{{{(\nu)}}}\tau_{i+r}^{{{(\nu)}}}\rangle_{\mathcal{G}}\\
&{F_{Q}[\mathcal{O}^{(\textrm{st})}_{\nu},|\mathcal{G}\rangle]}=M+M\sum_{r=1}^{M-1}(-)^{r}\langle\tau_i^{{{(\nu)}}}\tau_{i+r}^{{{(\nu)}}}\rangle_{\mathcal{G}}
\end{align}
where $(\tau_{j}^{{(\nu)}})^2=\mathbb{I}$, with $\mathbb{I}$ the identity, and we let
\begin{equation}
M\equiv L-|\nu|+1.
\end{equation}
For the thermodynamic limit $L\gg N_f\geq1$, we can obtain the dual quantum
Fisher information density as
\begin{align}
&{f_{Q}[\mathcal{O}_{\nu},|\mathcal{G}\rangle]}=\frac{{F_{Q}[\mathcal{O}_{\nu},|\mathcal{G}\rangle]}}{L}=1+\sum_{r=1}^{L-|\nu|}{C_\nu}(r),\\
&{f_{Q}[\mathcal{O}^{(\textrm{st})}_{\nu},|\mathcal{G}\rangle]}=\frac{{F_{Q}[\mathcal{O}^{(\textrm{st})}_{\nu},|\mathcal{G}\rangle]}}{L}=1+\sum_{r=1}^{L-|\nu|}{(-)^{r}C_\nu}(r),
\label{scf}
\end{align}
where $M\simeq L$ as $|\nu|\leq N_f$, and
\begin{eqnarray}
C_{\nu}(r)\equiv\langle\tau_i^{{(\nu)}}\tau_{i+r}^{{(\nu)}}\rangle_{\mathcal{G}}
\end{eqnarray}
is the so-called string correlation function \cite{Venuti2005,Cui2013}
from site $i$ to $j=i+r$ in the dual lattice.
The string correlation function is shown able to reveal hidden symmetry-protected order by $\mathbf{Z}$ symmetry in
many topological systems \cite{Venuti2005,Cui2013,Feng2007,Smacchia2011}.
It is easier to rewrite the string correlation function in terms of Majorana operators and
 fermion operators
 \begin{equation}
 d_{l,\nu}=(b_l+ia_{l+\nu})/{2},\hspace{0.2in}d_{l,\nu}^\dag=(b_l-ia_{l+\nu})/{2}
 \end{equation}
as
\begin{equation}\label{cf}
C_{\nu}(r)=\left\langle\prod_{l=i}^{j-1}(- ib_{l}a_{l+\nu})\right\rangle_{\!\!\!\mathcal{G}}\!=\left\langle\prod_{l=i}^{j-1}(1-2d_{l,\nu}^\dag d_{l,\nu})\right\rangle_{\!\!\!\mathcal{G}}.
\end{equation}
Usually, the string correlation function is written in terms of Pauli matrices as
\begin{eqnarray}
{C_{\nu}(r)}=\left\langle\prod_{l=i}^{j-1}\Big(\sigma_l^{\alpha}\sigma_{l+|\nu|}^{\alpha}\prod_{k=l+1}^{l+|\nu|-1}\sigma_k^z\Big)\right\rangle_{\!\!\!\mathcal{G}},
\end{eqnarray}
where $\alpha=x$ for positive $\nu$, and $\alpha=y$ for negative $\nu$.

The interchange between the quantum phases with positive and negative winding numbers $\nu=\pm n$ ($n$
is a positive integer)
\begin{align}
{\mathcal{O}_{\nu=n}^{(\textrm{st})}}\leftrightarrow{\mathcal{O}_{\nu=-n}^{(\textrm{st})}},&\hspace*{0.2in}{\mathcal{O}_{\nu=n}}\leftrightarrow{\mathcal{O}_{\nu=-n}}\\
{{f}_Q[\mathcal{O}_{\nu=n}^{(\textrm{st})}]}\leftrightarrow{{f}_Q[\mathcal{O}_{\nu=-n}^{(\textrm{st})}]},&\hspace*{0.2in}{{f}_Q[\mathcal{O}_{\nu=n}]}\leftrightarrow{{f}_Q[\mathcal{O}_{\nu=-n}]}
\end{align}
can be realized by a phase redefinition $c_j \rightarrow \pm i c_j$.

Another interchange between the staggered operator $\mathcal{O}_{\nu=1}^{(\textrm{st})}$ and the  operator $\mathcal{O}_{\nu=1}$, for the positive and negative signs of the interaction between Dirac fermions localized
at the chain ends, respectively,
\begin{align}
{\mathcal{O}_{\nu=n}^{(\textrm{st})}}\leftrightarrow{\mathcal{O}_{\nu=n}},&\hspace*{0.2in}{\mathcal{O}^{(\textrm{st})}_{\nu=-n}}\leftrightarrow{\mathcal{O}_{\nu=-n}}\\
{{f}_Q[\mathcal{O}_{\nu=n}^{(\textrm{st})}]}\leftrightarrow{{f}_Q[\mathcal{O}_{\nu=n}]},&\hspace*{0.2in}{{f}_Q[\mathcal{O}_{\nu=-n}^{(\textrm{st})}]}\leftrightarrow{{f}_Q[\mathcal{O}_{\nu=-n}]}
\end{align}
can be realized by a Hermitian conjugate transformation $c_j \rightarrow c_j^\dag$.

Following the calculations in previous sections, we can write the string correlation
function into a determinant  of size ($r-|\nu|+1$) as
\begin{equation}
{C_{\nu}(r)}=\left|\begin{array}{c c c c}
G_{-\nu}&G_{-\nu-1}&\cdots&G_{-r}\\
G_{1-\nu}&G_{-\nu}&\cdots&G_{1-r}\\
\vdots&\vdots&\vdots&\vdots\\
G_{r-2\nu}&G_{r-2\nu+1}&\cdots&G_{-\nu}
\end{array}\right|
\end{equation}
for positive $\nu$ and
\begin{equation}
{C_{\nu}(r)}=\left|\begin{array}{c c c c}
G_{-\nu}&G_{-\nu-1}&\cdots&G_{-r-2\nu}\\
G_{1-\nu}&G_{-\nu}&\cdots&G_{1-r-2\nu}\\
\vdots&\vdots&\vdots&\vdots\\
G_{r}&G_{r-1}&\cdots&G_{-\nu}
\end{array}\right|
\end{equation}
for negative $\nu$.

Because the string correlation function decays exponentially versus the distance $r$ when 
breaking the hidden $\mathbf{Z}$
symmetry (see, for example, Fig.~\ref{fig:A1}), the quantum Fisher information density as a function of $L$ has a scaling form in the thermodynamic limit,
\begin{align}
{f}_{Q}[\mathcal{O}_\nu, |\mathcal{G}\rangle]&\simeq1+\gamma_\nu L^{\lambda_\nu},\\
{f}_{Q}[\mathcal{O}^{(\textrm{st})}_\nu, |\mathcal{G}\rangle]&\simeq1+\gamma_\nu^{(\textrm{st})} L^{\lambda_\nu^{(\textrm{st})}}
\end{align}
and becomes linear:
\begin{equation}
\lambda_\nu\ \textrm{or}\  \lambda_\nu^{(\textrm{st})}\simeq1
\end{equation}
in the topological quantum phase with a
winding number $\nu$  and constant:
\begin{equation}
\lambda_\nu\ \textrm{and}\  \lambda_\nu^{(\textrm{st})}\simeq0,
\end{equation}
 in the other phases,
see Fig.~\ref{fig:S1} for example.
Thus, the scaling coefficient $\lambda_\nu$ or $\lambda_\nu^{(\textrm{st})}$ obtained by numerical calculations
can identify the topological phases
with higher winding numbers, see numerical results in Tab.~\ref{tab:1}.

\begin{table}[b]
\begin{tabular}{c|c c c}
\hline
\hline
$\mu$ & {$\lambda_{\nu=1}^{(\textrm{st})}$} & ${\lambda_{\nu=2}^{(\textrm{st})}}$  & ${\lambda_{\nu=3}^{(\textrm{st})}}$ \\
\hline
6\footnote{Inside  topological phases.}& $2.8\times10^{-5}$ & $-4.3\times10^{-7}$& $-1.6\times10^{-6}$\\
$3$ & \textcolor{blue}{0.9965}& $9.4\times10^{-14}$& $2.5\times10^{-13}$\\
$0$ &$-4.2\times10^{-14}$  &$1.4\times10^{-13}$ & \textcolor{blue}{1.0047}\\
$-2$ & $-5.6\times10^{-7}$  & \textcolor{blue}{0.9957} & $2.9\times10^{-7}$\\
\hline
5\footnote{On the critical contour between phases.}& $\textcolor{blue}{0.7492}$ & $4.1\times10^{-7}$& $-1.9\times10^{-6}$\\
$\sqrt{3}-1$ & $\textcolor{blue}{0.5054}$& $-2.8\times10^{-3}$& $\textcolor{blue}{0.5165}$\\
$-1$ &$6.8\times10^{-5}$  &\textcolor{blue}{0.7518} & \textcolor{blue}{0.7547}\\
$-\sqrt{3}-1$ & $1.0\times10^{-3}$  &\textcolor{blue}{0.5088}& $-5.6\times10^{-4}$\\
\hline
\hline
\end{tabular}
\caption{Fitting of the scaling coefficients $\lambda_{\nu}^{(\textrm{st})}$ of the
dual quantum Fisher information density ${f}_Q[\mathcal{O}_{\nu}^{(\textrm{st})},|\mathcal{G}\rangle]$ inside different topological phases
and on the critical contour between phases for the extended Kitaev fermion chain
with nonzero parameters ${J}_1^+={J}_1^-=1$, ${J}_{2}^+={J}_{2}^-=2$, ${J}_{3}^+={J}_{3}^-=2$ ($N_f=3$), and chain length up to $L=2000$. The nine
essentially non-zero scaling  coefficients are show in blue font.
}\label{tab:2}
\end{table}

\begin{table*}[t]
\begin{tabular}{c|c c |c c| c c| c c}
\hline
\hline
~~~~~~~~$g(\zeta)$~~~~~~~~ & ~~~~~~~{$\lambda_{\nu=1}^{(\textrm{st})}$}~~~~~~~ & ~~~~~~~{$\lambda_{\nu=1}$}~~~~~~~ & ~~~~~~~{$\lambda_{\nu=2}^{(\textrm{st})}$}~~~~~~~ & ~~~~~~~{$\lambda_{\nu=2}$}~~~~~~~ & ~~~~~~~{$\lambda_{\nu=3}^{(\textrm{st})}$}~~~~~~~  & ~~~~~~~{$\lambda_{\nu=3}$}~~~~~~~ & ~~~~~~~{$\lambda_{\nu=4}^{(\textrm{st})}$}~~~~~~~ & ~~~~~~~{$\lambda_{\nu=4}$}~~~~~~~\\
\hline
$\zeta-1$&
\textcolor{blue}{0.7506} &
$<10^{-5}$&
$<10^{-5}$&
$<10^{-3}$ &
$<10^{-5}$&
$<10^{-4}$ &
$<10^{-4}$&
$<10^{-4}$\\
$\zeta^2-1$&
\textcolor{blue}{0.5072} &
\textcolor{blue}{0.5072}&
\textcolor{blue}{0.5040}&
$<10^{-4}$ &
$<10^{-3}$ &
$<10^{-16}$&
$<10^{-4}$&
$<10^{-16}$\\
$\zeta^3-1$&
\textcolor{blue}{0.2873}  &
0.0043 &
$<10^{-3}$ &
\textcolor{blue}{0.2441} &
\textcolor{blue}{0.2809} &
$<10^{-16}$ &
$<10^{-3}$ &
$<10^{-16}$
\\
$\zeta^4-1$ &
\textcolor{blue}{0.1313}  &
\textcolor{blue}{0.1313} &
\textcolor{blue}{0.0950} &
\textcolor{blue}{0.0950} &
\textcolor{blue}{0.0745} &
$<10^{-16}$ &
\textcolor{blue}{0.1223} &
$<10^{-16}$  \\
\hline
\hline
\end{tabular}
\caption{Fitting of the scaling coefficients $\lambda_{\nu}$ and $\lambda_{\nu}^{(\textrm{st})}$ 
with respect to the dual generators $\mathcal{O}_{\nu}$ and $\mathcal{O}_{\nu}^{(\textrm{st})}$, 
respectively, on the critical contour between
phases for the extended Kitaev fermion chain with characteristic functions $g(\zeta)$ and
chain length up to $L=2000$. The thirteen essentially non-zero scaling coefficients are shown
in blue font.}\label{tab:3}
\end{table*}

\section{Topological phase transitions and Half-integer winding numbers with zeros on the critical contour}\label{VI}
For completeness, we discuss the case when zeros of the characteristic equation
 appear on the contour $|\zeta|=1$, and interpret the physical
implications of half-integer winding numbers therein. We can find that the topological
phase transitions occur at the critical points satisfying
\begin{eqnarray}
g(\zeta)=\sum_{n=1}^{N_f}({J}_n^x\zeta^n+{J}_n^y\zeta^{-n})-\mu=0\label{supeq}
\end{eqnarray}
for $|\zeta|=1$.

For example, we choose the parameters of the extended Kitaev fermion chain as ${J}_1^+={J}_1^-=1$, ${J}_{2}^+={J}_{2}^-=2$, ${J}_{3}^+={J}_{3}^-=2$ ($N_f=3$),
and calculate the real solutions of the chemical potential $\mu$: for $\zeta=1$, $\mu=5$;
for $\zeta=-1$, $\mu=-1$; for
\begin{equation}
\zeta=\exp\{\pm i\arccos[(-\sqrt{3}-1)/4]\},
\end{equation}
$\mu=\sqrt{3}-1$;
and for
\begin{equation}
\zeta=\exp\{\pm i\arccos[(\sqrt{3}-1)/4]\},
\end{equation}
$\mu=-\sqrt{3}-1$,
where the topological phase transitions occur. For another example, we consider the
parameters of the extended Kitaev fermion chain  as $J_{2}^+=J_{2}^-=\lambda$, $J_1^+=1$, $J_1^-=-1$, $\mu=1$,
and change the value of $\lambda$. We can obtain the critical points of topological
phase transitions by solving the characteristic equation:
\begin{eqnarray}
g(\zeta)=\lambda\zeta^2+\zeta^{-1}-1=0
\end{eqnarray}
where we can obtain the transition points: for $\zeta=1$, $\lambda=0$; for $\zeta=-1$,
$\lambda=2$;  for
\begin{equation}
\zeta=\exp\{\pm i\arccos[(1-\sqrt{5})/4]\},
\end{equation}
$\lambda=(-\sqrt{5}-1)/2$; and for
\begin{equation}
\zeta=\exp\{\pm i\arccos[(1+\sqrt{5})/4]\},
\end{equation}
$\lambda=(\sqrt{5}-1)/2$.

We then consider the critical behaviors of quantum states on the  transition
points. From the viewpoint of geometric topology, we consider the Kitaev closed chain
as $\Delta=J$ and assume an anti-periodic boundary conditions $c_{j+L}=-c_{j}$.
If
\begin{equation}
\Delta=-{\mu}=-1,
\end{equation}
the characteristic function becomes
\begin{eqnarray}
g(\zeta)=\zeta-1,
\end{eqnarray}
and the winding number can be calculated
by the Cauchy principal value:
\begin{align}
&\nu=\frac{1}{2\pi i}\oint_{{}_{|\zeta|=1}}\!\!\!\!\!\!\!\!\!d\zeta\;\frac{1}{\zeta-1}\\
&=\frac{1}{2\pi i}\lim_{\varepsilon\rightarrow0}\left[\int_{-\varepsilon}^{2\pi-\varepsilon}\!\!\!\!\!\!\!d\zeta\;\frac{1}{\zeta-1}\right]\nonumber\\
&~~=\frac{1}{2\pi i}\lim_{\varepsilon\rightarrow0}\left[i\varepsilon\int^{\frac{3\pi}{2}}_{\frac{\pi}{2}}\!\!\!d\theta\;\frac{e^{i\theta}}{(\varepsilon e^{i\theta}+1)-1}\right]\\
&=\frac{1}{2},
\end{align}
where we can only obtain massive Dirac edge modes \cite{Viyuela2016} for the open Kitaev chain.
Moreover, in consideration of the boundary parts for the closed chain,
we can write the Hamiltonian in terms of Majorana fermion operators as
\begin{eqnarray}
i{H}=\sum_{j=1}^{L}a_jb_j+\sum_{j=1}^{L-1}b_ja_{j+1}+(-1)^{N_p}b_La_1,
\end{eqnarray}
where we have that
\begin{equation}
\phi=\frac{1}{\sqrt{L}}\sum_{j=1}^L a_j,\hspace{0.2in}\psi=\frac{1}{\sqrt{L}}\sum_{j=1}^L b_j,
\end{equation}
are a pair of zero modes (obviously not edge modes) for even $N_p$, but there exists
no zero mode for odd parity. Therefore, the half-integer winding number represents
a critical phenomenon when the Majorana zero mode exists or not for different
fermion parities $(-1)^{N_p}$ in consideration of boundary Hamiltonian. Generally,
it can be inferred that if we have even number of zeros on the contour, the winding
number is still an integer for different fermion parities.

In Fig.~\ref{fig:S2}, we plot the quantum Fisher information density as a function of
$L$ in critical cases for the  extended Kitaev fermion chain  with ${J}_1^+={J}_1^-=1$, ${J}_{2}^+={J}_{2}^-=2$, ${J}_{3}^+={J}_{3}^-=2$ ($N_f=3$),
and present the scaling coefficients $\lambda_{\nu}^{(\textrm{st})}$ in Tab.~\ref{tab:2}. Then,
we plot in Fig.~\ref{fig:S3} the quantum Fisher information density as a function of
$L$ for an extended Kitaev fermion chain with characteristic functions:
\begin{itemize}
\item[(a)] $g(\zeta)=\zeta-1$,
\item[(b)] $g(\zeta)=\zeta^2-1$,
\item[(c)] $g(\zeta)=\zeta^3-1$,
\item[(d)] $g(\zeta)=\zeta^4-1$,
\end{itemize}
where the zeros are on the contour $|\zeta|=1$ given $\mu=1$. The scaling coefficients $\lambda_{\nu}$ and $\lambda_{\nu}^{(\textrm{st})}$
are shown in Tab.~\ref{tab:3}.
We should note that our discussions would be inappropriate to discuss the
Dirac sector of the topological phase diagram for the extended Kitaev chain  which would have
a half integer winding number \cite{Vodola2014,Viyuela2016,Vodola2015,Lepori2017}, because
the boundary conditions (anti-periodic and periodic) for finite chain length $L$ would destroy long-range
hopping and pairing terms, and the thermodynamic limit $L\gg N_f\geq1$ could not be satisfied.

\section{Characterization of topological phases in a Kitaev honeycomb model via dual multipartite entanglement}
The Kitaev honeycomb model (i.e., a two-dimensional spin model on a hexagonal lattice with
direction-dependent interactions between adjacent lattice sites)
is an analytically solvable model with topological quantum
phase transitions at zero temperature \cite{Kitaev2006a}. The Hamiltonian is
\begin{align}\label{honey}
H_{\textrm{hc}}=-\sum_{\kappa=x,y,z}\!J_\kappa\!\sum_{\langle ij\rangle_\kappa}\sigma_i^\kappa\sigma_j^\kappa,
\end{align}
where $\langle ij\rangle_\kappa$ denotes the nearest-neighbor bonds in the $\kappa$-direction.
At each site, we define four Majorana operators $a^\alpha$, with $\alpha=0,x,y,z$,
satisfying $(a^\alpha)^\dag=a^\alpha$, $\{a^\alpha,a^\beta\}=2\delta_{\alpha\beta}$,
and $a^xa^ya^za^0 = 1$, and write the Pauli operators as
\begin{equation}
\sigma_j^\kappa=ia_j^\kappa a_j,
\end{equation}
with $\kappa=x,y,z$ and $a_j^0\equiv a_j$.
The Hamiltonian is then rewritten with
\begin{align}
\hat{u}_{{\langle ij\rangle}_\kappa}\equiv i a^\kappa_ia^\kappa_j
\end{align}
as
\begin{equation}
H_{\textrm{hc}}=\frac{i}{2}\sum_{\langle ij\rangle_\kappa}J_{\kappa_{\langle ij\rangle}}\hat{u}_{{\langle ij\rangle}_\kappa} a_ia_j,
\end{equation}
where the factor $\frac{1}{2}$ is due to each lattice being counted twice in the summation.
We have  $\hat{u}_{{\langle ij\rangle}_\kappa}^2=1$ and $[H_{\textrm{hc}},\hat{u}_{{\langle ij\rangle}_\kappa}]=0$.
Here we take $\hat{u}_{{\langle ij\rangle}_\kappa}=1$ for all bonds ($\pi$-flux phase), because this vortex-free
configuration has the lowest energy \cite{Kitaev2006a,Lieb1994}. The system size is $N=2LM$, and at first,
we set $M=L$.

\begin{figure}[t]
\centering
\includegraphics[width=0.38\textwidth]{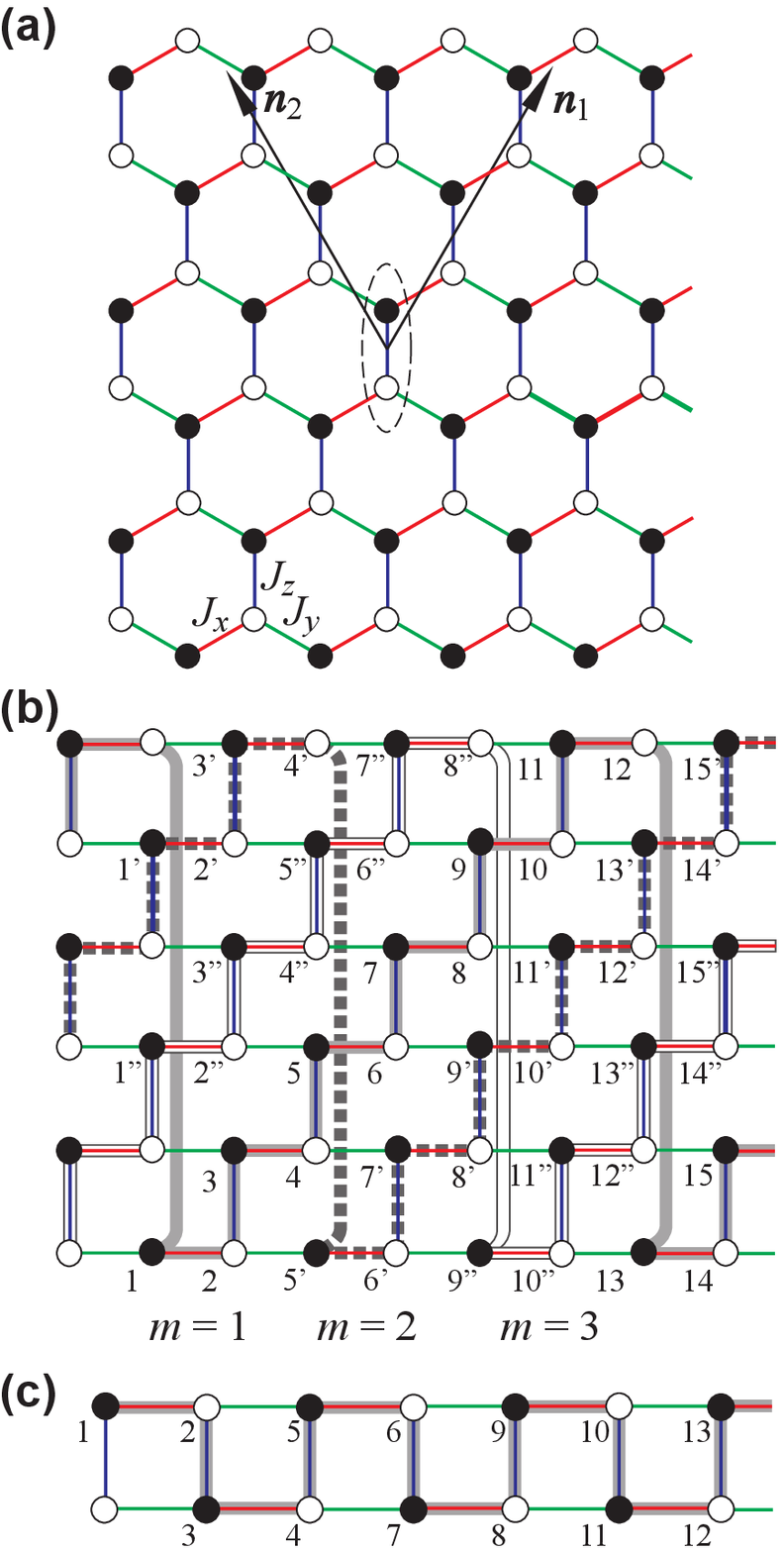}\\
\caption{(color online) (a) A graphic representation of the Kitaev honeycomb model with two
sublattices (empty and full circles). There are three types of bonds labeled by $x,y,z$. (b)
The equivalent brick-wall lattice with three rows ($m=1,2,3$). (c) A single-chain representation
of the two-leg spin ladder.
}\label{fig:S5}
\end{figure}

Using the Fourier transformation, the Hamiltonian in the momentum representation
is \cite{Chen2016}
\begin{align}
H_{\textrm{hc}}=\sum_{\bm{q}}(a_{-\bm{q},1},a_{-\bm{q},2})\ \mathcal{H}_{\bm{q}}\left(\begin{array}{c}a_{\bm{q},1}\\a_{\bm{q},1}\end{array}\right),
\end{align}
where $\bm{q}=(q_1,q_2)$ is the momentum vector and the Bloch matrix of $\mathcal{H}_{\bm{q}}$ is
\begin{align}
\mathcal{H}_{\bm{q}}=-\Delta_{\bm{q}}\sigma^x-\epsilon_{\bm{q}}\sigma^y=\left(\begin{array}{c c}0&i\Upsilon_{\bm{q}}\\-i\Upsilon^*_{\bm{q}}&0\end{array}\right),
\end{align}
with
\begin{align}
\Upsilon_{\bm{q}}&=\epsilon_{\bm{q}}+i\Delta_{\bm{q}},\\
\epsilon_{\bm{q}}&=J_x\cos q_1 +J_y\cos q_2 +J_z,\\
\Delta_{\bm{q}}&=J_x\sin q_1 +J_y\sin q_2.
\end{align}
By choosing the coordinate axes in the $\bm{n}_1$ and $\bm{n}_2$
directions as shown in Fig.~\ref{fig:S5}(a), then the momentum vectors $q_1=\bm{q}\cdot\bm{n}_1$ and $q_2=\bm{q}\cdot\bm{n}_2$
take the values
\begin{align}
q_{1,2}=\frac{2l\pi}{L}, \hspace*{0.2in}l=-\frac{L-1}{2},\cdots,\frac{L-1}{2}.
\end{align}
Using the Bogoliubov transformation
\begin{align}
D_{\bm{q},1}=u_{\bm{q}}a_{\bm{q},1}+v_{\bm{q}}a_{\bm{q},2},\hspace*{0.2in}D_{\bm{q},1}=v^*_{\bm{q}}a_{\bm{q},1}-u^*_{\bm{q}}a_{\bm{q},2}
\end{align}
with $u_{\bm{q}}=1/\sqrt{2}$ and $v_{\bm{q}}=i\Upsilon_{\bm{q}}/(\sqrt{2}|\Upsilon_{\bm{q}}|)$,
the Hamiltonian is diagonalized
\begin{align}
H_{\textrm{hc}}=\sum_{\bm{q}}|f_{\bm{q}}|(1-2D^\dag_{\bm{q},2}D_{\bm{q},2}),
\end{align}
where we have used $\{D^\dag_{\bm{q},\mu},D^\dag_{\bm{q}',\mu'}\}=\delta_{\bm{q},\bm{q}'}\delta_{\mu,\mu'}$,
$D^2_{\bm{q},\mu}=0$, and $D^\dag_{\bm{q},1}D_{\bm{q},1}=1-D^\dag_{\bm{q},2}D_{\bm{q},2}$.
The ground state is
\begin{align}
|\mathcal{G}\rangle=\prod_{\bm{q}}D^\dag_{\bm{q},2}|0\rangle
\end{align}
 and the energy gap is $2\min_{\bm{q}}\{|\Upsilon_{\bm{q}}|\}$.

\begin{figure}[t]
\centering
\includegraphics[width=0.47\textwidth]{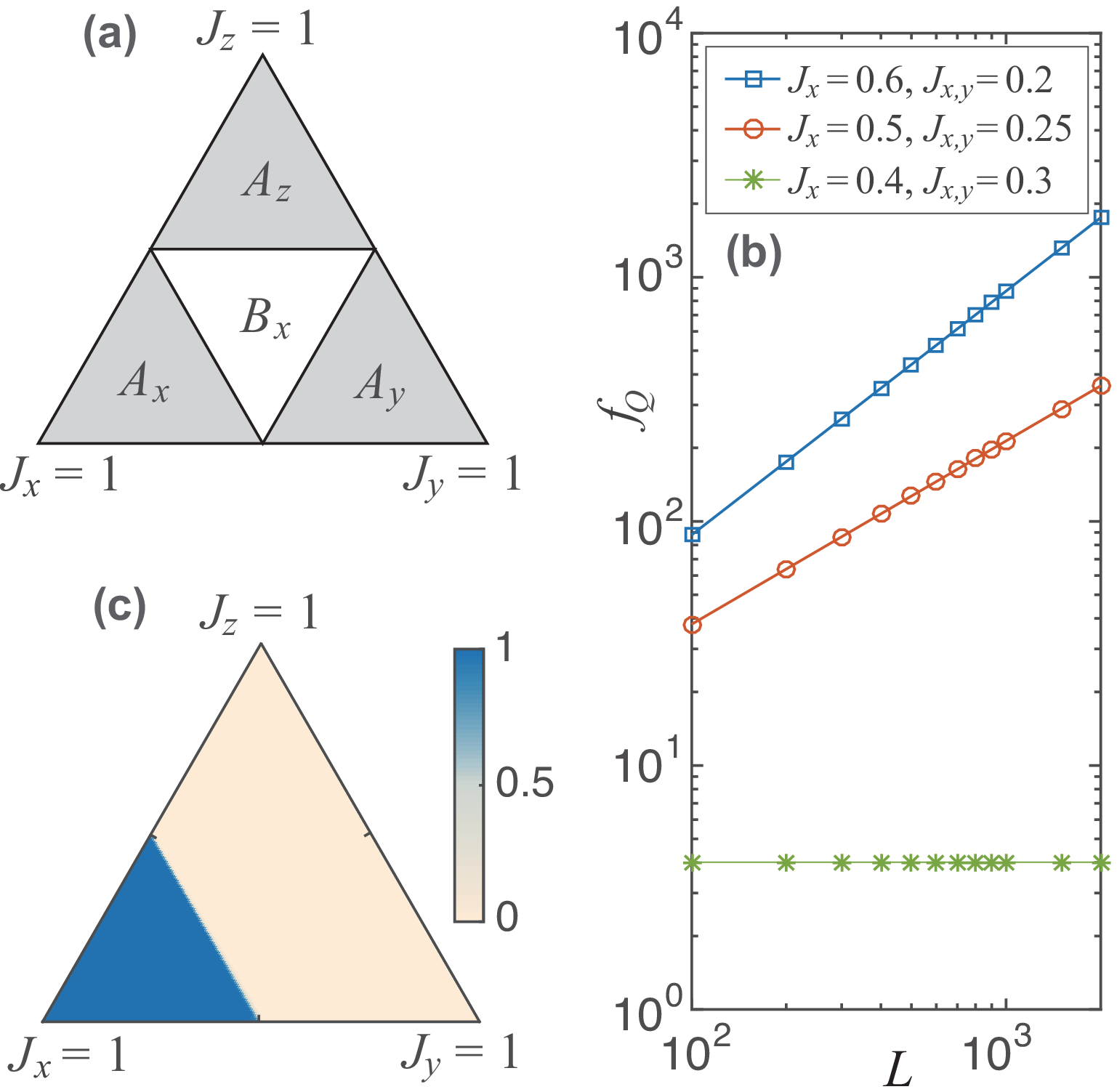}\\
\caption{(color online) (a) The phase diagram of the Kitaev honeycomb model in the
$J_x+J_y+J_z=1$ plane. (b) Quantum Fisher information density in the dual lattice as a function
of $L$ for the two-leg spin ladder. The scaling coefficients are $\lambda_x^{(\textrm{st})}\simeq0.9992$ for
$J_{x,y,z}=0.6,0.2,0.2$, $\lambda_x^{(\textrm{st})}\simeq0.7508$ for $J_{x,y,z}=0.5,0.25,0.25$, and
$\lambda_x^{(\textrm{st})}<10^{-12}$ for $J_{x,y,z}=0.4,0.3,0.3$. (c) Scaling topological index $\lambda_x^{(\textrm{st})}$
with different values of $J_{x,y,z}$ in the $J_x+J_y+J_z=1$ plane versus the system size $2L$
up to 400.
}\label{fig:S6}
\end{figure}

Then, we consider positive bonds, $J_{x,y,z}>0$, and focus on the
$J_x+J_y+J_z=1$ parametric plane. As presented in Fig.~\ref{fig:S6}(a), in the region of
$J_x\leq J_y+J_z$, $J_y\leq J_z+J_x$ and $J_z\leq J_x+J_y$, there is a gapless phase B with non-Abelian
excitations, and in other regions, there are three gapped phases with Abelian anyon
excitations \cite{Kitaev2006a}
\begin{align}
A_x:\ J_x\geq J_y+J_z,\\
A_y:\ J_y\geq J_z+J_x,\\
A_z:\ J_z\geq J_x+J_y.
\end{align}

Following \cite{Feng2007}, we consider a two-leg spin ladder of the Kitaev honeycomb model and relabel all the sites along a special path [as shown in
Fig.~\ref{fig:S5}(c)] and express the Hamiltonian with the third-nearest-neighbor couplings \cite{Feng2007}
\begin{align}
H_{\textrm{2l}}=-\sum_{j=1}^L(J_x\sigma_{2j-1}^x\sigma_{2j}^x+J_y\sigma_{2j}^y\sigma_{2j+3}^y+J_z\sigma_{2j}^z\sigma_{2j+1}^z).
\end{align}
By considering the duality transformation introduced in \cite{Feng2007}
\begin{align}
&\sigma_j^x=\check{s}^x_{j-1}\check{s}^x_{j},\hspace*{0.2in}\sigma_j^z=\prod_{k=j}^{2L}\check{s}^z_k,\\
&\sigma_j^y=-i\sigma_j^z\sigma_j^x=\check{s}^x_{j-1}\check{s}^y_{j}\prod_{k=j+1}^{2L}\check{s}^z_k,
\end{align}
we obtain an anisotropic $XY$ spin chain with a transverse field in the dual space
\begin{align}
H_{\textrm{2l}}=-\sum_{j=1}^L(J_x\check{s}^x_{2j}\check{s}^x_{2j+2}\!+J_y{W}_{j}\check{s}^y_{2j}\check{s}^y_{2j+2}\!+J_z\check{s}^z_{2j}),
\end{align}
where
\begin{align}
{W}_{j}=\check{s}^x_{2j-1}\check{s}^z_{2j+1}\check{s}^x_{2j+3}
\end{align}
is the plaquette operator in the dual lattice and a
good quantum number \cite{Feng2007}. We have ${W}_{j}=-1$ ($\pi$-flux phase \cite{Lieb1994}) for
the ground state. We consider the inverse dual transformation
\begin{align}
&\check{s}^x_j=\prod_{k=1}^{j}\sigma_k^x,\hspace{0.2in}\check{s}_j^z=\sigma^z_{j}\sigma^z_{j+1}\\ &\check{s}_j^y=-i\check{s}_j^z\check{s}_j^x=\sigma^z_{j+1}\sigma^y_{j}\prod_{k=1}^{j-1}\sigma_k^x
\end{align}
and consider the spin correlation function in the dual lattice
\begin{align}
C_x(r)\equiv\langle \check{s}^x_{2i}\check{s}^x_{2j}\rangle_{\mathcal{G}}=\left\langle\prod_{k=2i+1}^{2j}\!\!\!\sigma_{k}^x\right\rangle_{\!\!\mathcal{G}}
\end{align}
where $r=j-i$.
It is shown in Ref.~\cite{Feng2007} that the string correlation order 
\begin{align}
\lim_{r\rightarrow\infty}(-)^rC_x(r)\neq0
\end{align} 
in the phase
$A_x$ ($\ J_x\geq J_y+J_z$) and equals to zero in other regions. Similarly, with respect to the dual generator
\begin{align}
\mathcal{O}_x^{(\textrm{st})}=\sum_{j=1}^{L}(-)^j\check{s}^x_{2j},
\end{align}
the quantum Fisher information density in the dual lattice is
\begin{align}
f_Q[\mathcal{O}_x^{(\textrm{st})},|\mathcal{G}\rangle]&\equiv1+\sum_{r=1}^{L-1}(-)^rC_x(r)\\
&\simeq1+\gamma_x^{(\textrm{st})} L^{\lambda_x^{(\textrm{st})}}.
\end{align}
In the gapped phase $A_x$, the dual QFI density is linear
\begin{align}
\lambda_x^{(\textrm{st})}\simeq1
\end{align}
and constant
\begin{align}
\lambda_x^{(\textrm{st})}\simeq0
\end{align}
in other regions, see Fig.~\ref{fig:S6}(b,c) for example.
Moreover, the gapped phases $A_y$ and $A_z$ as shown in Fig.~\ref{fig:S6}(a)
can be obtained by the substitutions $J_x\rightarrow J_y\rightarrow J_z\rightarrow J_x$ and
$J_x\rightarrow J_z\rightarrow J_y\rightarrow J_x$, respectively.
Therefore, the scaling coefficient of the dual quantum Fisher information density in the dual lattice can
identify three gapped phases $A_x$, $A_y$ and $A_z$ with Abelian
anyon excitations.

Generally, we consider the equivalent brick-wall lattice of the Kitaev honeycomb model as shown in
Fig.~\ref{fig:S5}(b) and rewrite the Hamiltonian (\ref{honey}) as
\begin{align}
H_{\textrm{hc}}=-\sum_{j=1}^{L}\sum_{m=1}^M(&J_x\sigma_{2j-1,m}^x\sigma_{2j,m}^x+J_y\sigma_{2j,m}^y\sigma_{2j+3,m+1}^y\nonumber\\
&+J_z\sigma_{2j,m}^z\sigma_{2j+1,m}^z).
\end{align}
In the two-dimensional limit $M\rightarrow \infty$, the above results for the two-leg spin ladder using
string correlation functions and dual quantum Fisher information density to detect topological phase
transitions can also be extended to the general two-dimensional lattice by transforming the second index $m$
to momentum space \cite{Feng2007,Zhang2018}.

\bibliography{supp_rev}